\documentclass{aa}
\usepackage{natbib}
\bibpunct{(}{)}{;}{a}{}{,} 
\usepackage{graphicx}   
\usepackage{txfonts}
\usepackage[colorlinks=true,allcolors=cyan]{hyperref}

\begin{document}

\titlerunning{Galaxies X events}
\title{Cross-correlation between soft X-rays \\ and galaxies} 
\subtitle{A new benchmark for galaxy evolution models}
\authorrunning{Comparat et al.}
\author{
Johan Comparat\inst{1}\thanks{E-mail: comparat@mpe.mpg.de}, 
Andrea Merloni\inst{1},
Gabriele Ponti\inst{2,3,1},
Soumya Shreeram\inst{1},
Yi Zhang\inst{1},
Thomas H. Reiprich\inst{4},
Ang Liu\inst{1,5},
Riccardo Seppi\inst{6},
Xiaoyuan Zhang\inst{1},
Nicolas Clerc\inst{7},
Andrina Nicola\inst{4},
Kirpal Nandra\inst{1},
Mara Salvato\inst{1},
Nicola Malavasi\inst{1}
}

\institute{
Max-Planck-Institut f\"{u}r extraterrestrische Physik (MPE), Gie{\ss}enbachstra{\ss}e 1, D-85748 Garching bei M\"unchen, Germany
\and
INAF-Osservatorio Astronomico di Brera, Via E. Bianchi 46, I-23807 Merate (LC), Italy 
\and
Como Lake Center for Astrophysics (CLAP), DiSAT, Università degli Studi dell'Insubria, via Valleggio 11, I-22100 Como, Italy
\and
Argelander-Institut für Astronomie (AIfA), Universität Bonn, Auf dem H\"ugel 71, 53121 Bonn, Germany
\and 
Institute for Frontiers in Astronomy and Astrophysics, Beijing Normal University, Beijing 102206, China
\and
Department of Astronomy, University of Geneva, Ch. d'Ecogia 16, CH-1290 Versoix, Switzerland
\and
IRAP, Universit\'e de Toulouse, CNRS, UPS, CNES, F-31028 Toulouse, France
}
    
\date{\today}

\abstract{
This paper presents the construction and validation of complete stellar mass-selected, volume-limited galaxy samples using the Legacy Survey (data release 10) galaxy catalogs, covering $\sim16,800$ deg$^2$ of extra-galactic sky and extending to redshifts of $z<0.35$. We used companion mock catalogs to ensure a controlled galaxy selection.  
We measured the two-point correlation function of these galaxies with tiny statistical uncertainties at the percent level and systematic uncertainties up to 5\%. We fitted a four-parameter halo occupation distribution (HOD) model to retrieve the population of host halos, yielding results on the stellar to halo mass relation that are consistent with the current models of galaxy formation and evolution.  
Using these complete galaxy samples, we measured and analyzed the cross-correlation between galaxies and all soft X-ray photons observed by SRG/eROSITA in the 0.5-2 keV band over $\sim13,000$ deg$^2$. 
The cross-correlation measurements have an unprecedented sub-percent statistical uncertainty and ~5-10\% systematic uncertainty. 

We introduced a novel extension to the halo model to interpret the cross-correlation, decomposing contributions from X-ray point sources, hot gas, satellites, and the two-halo term. 
The model offers a new comprehensive view of the relation between the complete 0.5-2 keV X-ray photon field and complete sets of galaxies at low redshift and their host halos. 
For low stellar mass thresholds ($\log M^*/M_{\odot}>$ 10, 10.25, 10.5), we find that the point source emission dominates the cross-correlation at small separation ($r<80$kpc). Then, in the range of $80<r<2$Mpc, the emission from large halos hosting satellite galaxies dominates. Finally, on scales beyond those considered here ($r>2$Mpc), the two-halo term becomes dominant. 
Interestingly, there is no scale at which the hot gas dominates. In the range ($20<r<200$kpc), the hot gas contributes to more than 10\% of the signal. 
Progressively, with the minimum stellar mass increasing, the hot gas emission increases. 
For the $\log M^*/M_{\odot}>$ 10.75 sample, in the range 50-60kpc, the three components contribute each the same surface brightness.
For the $\log M^*/M_{\odot}>$11 sample, the hot gas is the dominating emission source over the range of 30-200kpc.
Finally, for the $\log M^*/M_{\odot}>$11.25 and (11.5) samples, the hot gas emission dominates over other components until 400 (700) kpc. 
We constrained the slope of the scaling relation between halo mass and X-ray luminosity (over three orders of magnitude in mass) at the 5\% level, using the samples with the lowest mass threshold. We find a slope of $1.629^{+0.091}_{-0.089}$. 

Additional analyses explore the energy dependence of the cross-correlation and differences between red sequence and blue cloud galaxies, revealing sensitivity to galaxy quiescent fractions and opening avenues for a more complex, unified modeling of galaxies, active galactic nuclei (AGNs), and hot gas in the optical and X-rays.}

\maketitle

\section{Introduction} 
The  large-scale structure of the Universe is observable at different wavelengths, each related to different sources, objects, and physical processes. In this work, we measure and interpret the cross-correlation between galaxies selected in the optical in the legacy surveys' tenth data release (LS10 hereafter, \citealt{DeySchlegelLang_2019AJ....157..168D}) and the set of X-ray photons gathered by SRG/eROSITA \citep{PredehlAndritschkeArefiev_2021A&A...647A...1P,SunyaevArefievBabyshkin_2021A&A...656A.132S} over half of the sky. 
These two wavelength ranges provide complementary information about the cohabitation among galaxies, their surrounding circumgalactic medium, and nearby active galactic nuclei (AGNs).

We use the combination of four eROSITA full-sky scans (eRASS:4) accessible to the German eROSITA consortium (Western Galactic hemisphere, eROSITA-DE hereafter). These constitute our Universe's deepest and most homogeneous soft X-ray map. Thanks to its scanning strategy, the X-ray background is well covered over large areas \citep[see e.g.,][]{MerloniLamerLiu_2024A&A...682A..34M, LocatelliPontiZheng_2024A&A...681A..78L, ZhengPontiFreyberg_2024A&A...681A..77Z, YeungPontiFreyberg_2024A&A...690A.399Y}. As demonstrated by the spectral analysis of \citet{PontiZhengLocatelli_2023A&A...674A.195P, PontiSandersLocatelli_2023A&A...670A..99P}, 
this set of events mainly traces (ordered by the total number of detected photons): the cosmic X-ray background \citep{HopkinsHernquistCox_2006ApJS..163....1H},
our Milky-Way's hot circumgalactic medium \citep{LocatelliPontiZheng_2024A&A...681A..78L, ZhengPontiFreyberg_2024A&A...681A..77Z, ZhengPontiLocatelli_2024A&A...689A.328Z}, 
the local hot bubble \citep{YeungPontiFreyberg_2024A&A...690A.399Y}, 
solar wind charge exchange \citep{MerloniLamerLiu_2024A&A...682A..34M},
detected point sources made of AGNs and stars \citep{SchwopeKnauffKurpas_2024A&A...690A.243S,SchwopeKurpasBaecke_2024A&A...686A.110S,SchneiderFreundCzesla_2022A&A...661A...6S,FreundCzeslaPredehl_2024A&A...684A.121F,WaddellNandraBuchner_2024A&A...690A.132W,WaddellBuchnerNandra_2024arXiv240117306W}, 
detected extended source made of hot gaseous halos \citep{LiuBulbulKluge_2024A&A...683A.130L, BaharBulbulGhirardini_2024A&A...691A.188B, SeppiComparatGhirardini_2024A&A...686A.196S, BulbulLiuKluge_2024A&A...685A.106B, KlugeComparatLiu_2024A&A...688A.210K}, and resolved star formation in near-by galaxies \citep{KyritsisZezasHaberl_2025A&A...694A.128K}. 

In the past, the galaxies-X-ray cross-correlation was used to characterize the composition of the X-ray background
\citep[e.g.,][]{SoltanHasingerEgger_1997A&A...320..705S, RefregierHelfandMcMahon_1997ApJ...477...58R,NewsamMcHardyJones_1999MNRAS.310..255N,KaminskyCappellutiHasinger_2025arXiv250209705K}. 
More recently, analyses stacking X-rays (in most cases detected by SRG/eROSITA) at the position of optically selected galaxies managed to grasp the relation between galaxies, their host halos, and their hot gas content \citep{AndersonGaspariWhite_2015MNRAS.449.3806A,ComparatTruongMerloni_2022A&A...666A.156C,ZhangComparatPonti_2024A&A...690A.267Z,ZhangComparatPonti_2024A&A...690A.268Z,PopessoBivianoBulbul_2024MNRAS.527..895P}. 
They leveraged enhanced galaxy catalogs to select central galaxies in their halos \citep{YangMovandenBosch_2005MNRAS.356.1293Y,RobothamNorbergDriver_2011MNRAS.416.2640R,TempelKipperTamm_2016A&A...588A..14T,TempelKruuseKipper_2018A&A...618A..81T,Tinker_2021ApJ...923..154T, Tinker_2022AJ....163..126T}. 
The inference for an individual galaxy of its host halo mass and its position in the halo (central or satellite) is complex. In the Milky-Way mass range (halos of $\sim3\times10^{12}M_\odot$), recent work on simulations \citep{PopessoMariniDolag_2024arXiv241117120P} indicate that the scatter on the determination of the halo mass may be as large as one dex. Thus, when selecting a set of galaxies in a bin of inferred host halo mass, the obtained sample suffers from a high fraction of contaminants. 
Specifically, \citet{PopessoMariniDolag_2024arXiv241117120P} found that contamination fractions may be as large as 40\%. 
To interpret these stacking experiments, we require solid knowledge of the selection function made on the inferred host halo mass, which may limit, or even systematically bias, our understanding of the observations. 
Furthermore, \citet{ComparatTruongMerloni_2022A&A...666A.156C} and \citet{ZhangComparatPonti_2024A&A...690A.267Z} used observations and simulations to model the surface brightness profiles of contaminants: point source emission from AGNs and X-ray binaries (XRB) and satellite contamination (or misclassified centrals). They then subtracted these values to derive a measurement focused on the X-ray surface brightness profile of the hot gas and the scaling relation between, for example, X-ray luminosity and halo mass. 
This fruitful (albeit complex) approach is limited to areas of the sky where nearly complete spectroscopic surveys are available, which currently impedes using the complete eROSITA-DE extra-galactic sky. The study presented here generalizes and extends these earlier works, thanks to larger galaxy catalogs and X-ray data covering most of the eROSITA-DE extra-galactic sky (13,000 deg$^2$). It leverages on the proportionality between the signal-to-noise ratio (S/N) of the cross-correlation (or stacks) and sky coverage. 

\citet{ComparatTruongMerloni_2022A&A...666A.156C} and \citet{ZhangComparatPonti_2024A&A...690A.267Z} showed that both GAMA and SDSS, nearly complete spectroscopic surveys \citep{DriverBellstedtRobotham_2022MNRAS.513..439D, StraussWeinbergLupton_2002AJ....124.1810S}, are well suited to the eROSITA depth for stacking analysis. 
The overlapping area between eROSITA-DE and the GAMA (SDSS) spectroscopic surveys is limited to $\sim$200 ($\sim$4000) square degrees where eRASS exposure is shallowest, thereby limiting the total S/N in the stacks. 
Complete spectroscopic galaxy surveys overlapping with eROSITA-DE over a wider area will only be available in the 2030s after 4MOST completes its first tier of public surveys \citep{deJong_2011Msngr.145...14D, deJongAgertzBerbel_2019Msngr.175....3D, FinoguenovMerloniComparat_2019Msngr.175...39F}. 
In a slightly shorter timescale, DESI \citep{DESICollaborationAghamousaAguilar_2016arXiv161100036D} will provide a complete galaxy spectroscopic survey with its Bright Galaxy Sample \citep[BGS][]{HahnWilsonRuiz-Macias_2023AJ....165..253H} but with a limited overlap with the eROSITA coverage (about 4,000 deg$^2$ at positive Declination). Therefore, since complete spectroscopic surveys do not yet cover the full eROSITA-DE area, here we follow  a different and complementary methodology to understand how the low-redshift large-scale structure appears in X-rays.

Luckily, the quality and depth of the LS10 in the southern hemisphere \citep{DeySchlegelLang_2019AJ....157..168D} enables the inference of precise and unbiased photometric redshifts for galaxies brighter than $r<19.5$ \citep{ZhouNewmanMao_2021MNRAS.501.3309Z}. 
These redshift estimates allow for accurate galaxy clustering measurements, provided that there is access to projected summary statistics integrated up to $100h^{-1}$Mpc. 
To reach our goal of maximal S/N for a galaxy-X-rays cross-correlation experiment, we  constructed galaxy samples following the DESI BGS prescription ($r<19.5$; \citealt{HahnWilsonRuiz-Macias_2023AJ....165..253H}), resulting in a sample similar (but slightly brighter) to the GAMA one \citep{DriverBellstedtRobotham_2022MNRAS.513..439D} and significantly deeper than the SDSS MAIN sample \citep[$r<17.77$;][]{StraussWeinbergLupton_2002AJ....124.1810S}. 
We have covered the entire extra-galactic sky from eROSITA-DE with the LS10' photometric sample. 
We designed nine stellar mass-selected galaxy samples spanning redshifts between 0.05 and 0.35 and stellar masses between 10$^9$ and 10$^{12}$ M$_\odot$. 
We selected complete galaxy samples, close to volume-limited samples, regardless of the position of galaxies in their halo (central or satellite). 
This choice has guaranteed a simple galaxy selection function. 

Here, we present and interpret the measurements of three summary statistics: \textit{(i)}  auto-correlation of stellar-mass selected galaxy samples over $\sim16,000$ deg$^2$ ($w_p(r_p)$); \textit{(ii)}  cross-correlation between these galaxies and soft X-ray events over $\sim13,000$ deg$^2$;  and \textit{(iii)}  stacked X-ray surface brightness profiles around galaxies on the same area. 
We consider the complete photon field in the 0.5-2 keV band without masking detected sources, making the cross-correlation measurement reproducible. 
To interpret the auto-correlation, we use a standard halo occupation distribution model \citep[HOD, e.g.,][for a recent review]{AsgariMeadHeymans_2023OJAp....6E..39A}. With the $w_p(r_p)$, we are able to unambiguously link galaxies to their host halos. 
To enhance our control and understanding of uncertainties related to the galaxy samples, we constructed mock catalogs using the Uchuu simulation augmented by the UniverseMachine empirical model of galaxy evolution \citep{IshiyamaPradaKlypin_2021MNRAS.506.4210I, BehrooziWechslerWu_2013ApJ...762..109B, BehrooziWechslerHearin_2019MNRAS.488.3143B, AungNagaiKlypin_2023MNRAS.519.1648A}. 

By cross-correlating low-redshift galaxies with soft X-ray event maps, we filtered out the signal produced by the Solar System and Milky Way foregrounds and traced these galaxies' nuclear activity (AGNs), X-ray binaries (XRB), and hot gaseous halos instead. 
The complexity of this approach lies in modeling the observed cross-correlation coming from multiple X-ray sources, together (and consistently) with the underlying host dark matter halos. 
To do so, we can augment the HOD model to predict X-ray events (from the hot gas and point sources) to extract the information from the cross-correlation and the stacks. 
This model directly interprets the correlation between the complete galaxy and X-ray fields. 
The strength of this approach is that the uncertainties in the measurement process due to the galaxy selection, the complexities arising from masks on the X-ray map, and the dependency on the stellar-to-halo mass relation will not impede the interpretation in any way \citep[e.g., see discussion in][]{ComparatTruongMerloni_2022A&A...666A.156C,ZhangComparatPonti_2024A&A...690A.268Z,PopessoBivianoBulbul_2024MNRAS.527..895P, MariniPopessoLamer_2024A&A...689A...7M}. 

We structure the paper as follows. 
We describe the observations we used in Sect. \ref{sec:data} (and the creation of galaxy mock catalogues in Appendix \ref{sec:mocks}). We detail the measurement of the galaxy auto-correlation, the galaxy X-ray cross-correlation, and the stacks in Sect. \ref{sec:measurements}. We describe the HOD model in Sect. \ref{sec:method:halo:model} and the results in Sect. \ref{sec:results}. 
We assume a standard cosmology model (flat LCDM) from \citet{PlanckCollaborationAghanimAkrami_2020A&A...641A...6P}. Magnitudes are given in the AB system. 

\section{Observations}
\label{sec:data}

\begin{figure}
\centering
\includegraphics[width=0.95\columnwidth]{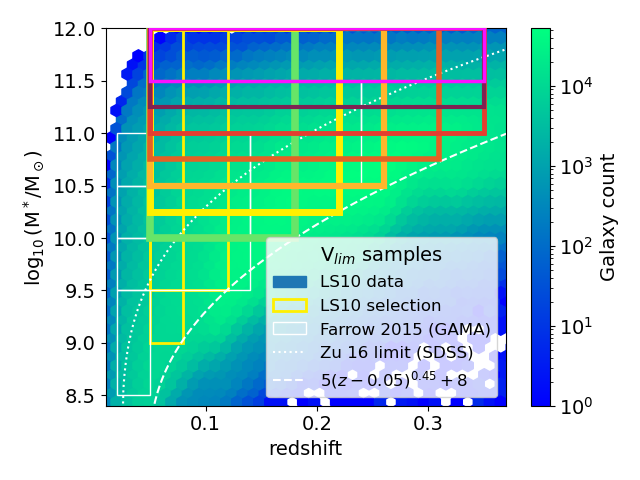}
\includegraphics[width=0.95\columnwidth]{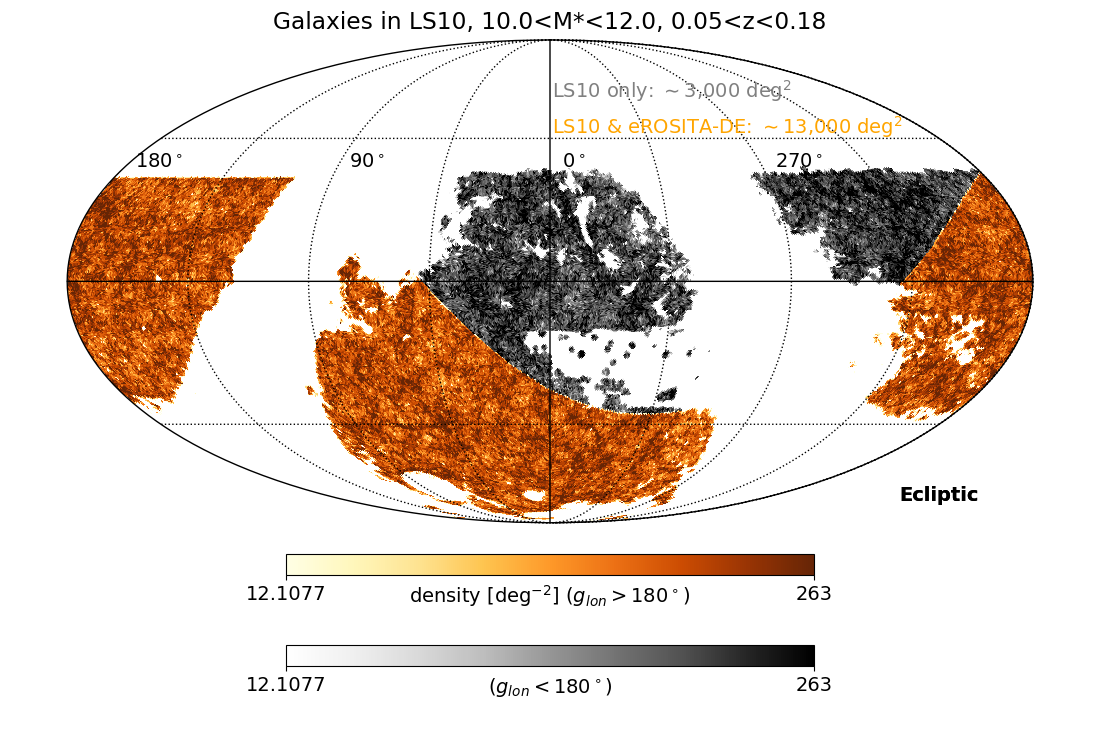}
\caption{\textbf{Top}: Sample definitions in the redshift-stellar mass plane. In the analysis, we considered the samples encompassed by the colored boxes. The white boxes depict previously defined SDSS and GAMA-based galaxy samples \citep{FarrowColeNorberg_2015MNRAS.454.2120F, ZuMandelbaum_2016MNRAS.457.4360Z}. The line $\log10(M*)=5(z-0.05)^{0.45}+8$ corresponds to the volume limit used in this work. The colorbar depicts the number of galaxies in hexagonal (for better visualization) bins. \textbf{Bottom}: Area covered by the sample with $\log_{10}(M^*/M_\odot)>10$ and $0.05<z<0.18$. The filamentary nature of the large-scale structure is visible. The orange (grey) area is (not) included in the eROSITA\_DE footprint.}
\label{fig:Vlim:samples:Mstar}
\end{figure}

\begin{table*}
\begin{center}
\caption{Description of the stellar-mass selected samples. }
\label{tab:Vlim:samples:basic:info}
\begin{tabular}{rrrrrrrrrrrrrrrrr}
\hline \hline
 \multicolumn{3}{c}{redshift} &  \multicolumn{4}{c}{stellar mass} & N & volume & \multicolumn{2}{c}{density} & \multicolumn{2}{c}{completeness} \\
 mean & median & max & min & mean & median & max &  & Gpc$^3$ & deg$^{-2}$ & $10^{-5}$ Mpc$^{-3}$ &\% & $\sigma$[\%] & \\
\hline
 0.067 & 0.068 & 0.08 & 9.0 & 9.89 & 9.81 & 12.0 & 530387 & 0.05 & 31.6 & 988.4 & 99.3 & 2.2 \\
 0.094 & 0.097 & 0.12 & 9.5 & 10.19 & 10.14 & 12.0 & 1383581 & 0.22 & 82.4 & 639.1 & 99.0 & 1.6 \\
 0.136 & 0.142 & 0.18 & 10.0 & 10.49 & 10.43 & 12.0 & 2811951 & 0.74 & 167.4 & 381.0 & 97.9 & 1.1 \\
 0.162 & 0.169 & 0.22 & 10.25 & 10.66 & 10.6 & 12.0 & 3280777 & 1.32 & 195.3 & 248.4 & 97.2 & 0.9 \\
 0.191 & 0.202 & 0.26 & 10.5 & 10.83 & 10.77 & 12.0 & 3287997 & 2.12 & 195.8 & 154.8 & 97.5 & 0.9 \\
 0.226 & 0.238 & 0.31 & 10.75 & 11.02 & 10.97 & 12.0 & 2768066 & 3.47 & 164.8 & 79.7 & 96.4 & 0.9 \\
 0.252 & 0.265 & 0.35 & 11.0 & 11.21 & 11.17 & 12.0 & 1611928 & 4.85 & 96.0 & 33.3 & 97.6 & 1.3 \\
 0.255 & 0.268 & 0.35 & 11.25 & 11.41 & 11.37 & 12.0 & 541919 & 4.85 & 32.3 & 11.2 & 99.9 & 2.7 \\
 0.261 & 0.274 & 0.35 & 11.5 & 11.62 & 11.59 & 12.0 & 121044 & 4.85 & 7.2 & 2.5 & 100.0 & 8.4 \\
  \hline
\end{tabular}
\tablefoot{The minimum redshift is 0.05 for all samples. The values in the stellar mass columns are in $\log_{10}(M_\odot)$. 
N gives the total number of galaxies present in each sample. 
The "volume" column gives the comoving cosmological volume covered by each sample. 
We compute the completeness value (and its uncertainty) with GAMA as a reference that extends 0.3 mag fainter in magnitude than the selection done here.
}
 \end{center}
\end{table*}

We note that this analysis requires X-ray data, which are described in Sect. \ref{subsec:data:xray}. It also draws from the galaxy catalogs described in Sect. \ref{subsec:data:galaxies}.

\subsection{eROSITA soft X-ray events}
\label{subsec:data:xray}

Overall, eROSITA (extended ROentgen Survey with an Imaging Telescope Array) is a wide-field X-ray telescope on board the Russian-German "Spectrum-Roentgen-Gamma" (SRG) observatory \citep{MerloniPredehlBecker_2012arXiv1209.3114M,PredehlAndritschkeArefiev_2021A&A...647A...1P,SunyaevArefievBabyshkin_2021A&A...656A.132S}. With its seven identical Wolter-1 mirror modules, eROSITA serves as a sensitive wide-field X-ray telescope capable of delivering deep, sharp images over vast sky areas in the energy range of 0.2-8 keV, with a maximum sensitivity in the 0.3-2.3 keV range. 
We use the first four all-sky scans of eROSITA (eRASS:4) reduced with the pipeline version c030 \citep{BrunnerLiuLamer_2022A&A...661A...1B, MerloniLamerLiu_2024A&A...682A..34M}; the data 
 are organized and processed in "fields" of approximately 3.6x3.6 square degrees. 
In this version of the observations, the times at which solar flares occurred are masked. Then, we need to select events in the soft X-ray 0.5-2 keV range (no flag selection). 
We did not apply  any further masking throughout the analysis and we considered the complete soft X-ray event field.  

\subsection{LS10 galaxies}
\label{subsec:data:galaxies}

We used the LS10 to select a flux-limited galaxy sample \citep{DeySchlegelLang_2019AJ....157..168D}. LS10 provides source catalogs for Dec$<$32$^\circ$, from which we selected low-redshift galaxies using the DESI BGS algorithm \citep{HahnWilsonRuiz-Macias_2023AJ....165..253H} with a magnitude cut of $r_{AB}<19.5$. Following their criterion, we separated stars from galaxies and selected galaxies with $13<r_{AB}<19.5$ to obtain a set of BGS-like galaxies.

We required the galactic reddening to be small (E(B-V)$<0.1$) and to have at least one observation in each of the $g,r,z$ bands. 
With the mask bits, we removed any secondary detections (\texttt{MASKBIT} 0), as well as sources that touch Tycho sources with \texttt{MAG\_VT} $<$ 13 and Gaia stars with G $<$ 13 (\texttt{MASKBIT} 1), and sources touching a pixel in a globular cluster (\texttt{MASKBIT} 13).
These cuts removed 1.7\% of the LS10 area. 
To further refine the selection, we compared the obtained source list with that of GAMA \citep[fourth data release][]{DriverBellstedtRobotham_2022MNRAS.513..439D} as a benchmark. 
Indeed, GAMA is almost complete to $r$ magnitude of 19.8, deeper than this selection extending to 19.5. 
By looking at the fraction of sources without a GAMA counterpart and with any of the \texttt{FITBITS} on, we found that sources with the \texttt{FITBITS} 1, 6, 10, 11, 12, and 13 are primarily spurious, so we removed them. 
Finally, we removed Gaia duplicates (\texttt{TYPE}="DUP"). We followed the BGS selection to discard artifacts, but with a stricter criterion in color-color space, namely: we kept sources with $-0.2<g-r<2$ and $-0.2<r-z<1.6$. 

We compared the sources selected to GAMA DR4 in the 9hr field and found that 92\% (93\%) of these have a GAMA spectroscopic counterpart within 1 (2) arc seconds.
For the sources with a match, the mean difference in $r$-band magnitude is $\sim$0.1 mag, while the scatter is 0.15 mag, which we might expect given the different filters, images, and pipelines used.
The remaining 7-8\% of target sources closely follow  the distributions of the ones matched to GAMA, simply reflecting the incompleteness of the GAMA survey if it were targeted with LS10.
We carried out this selection on the LS10 data and random data to obtain the samples for auto and cross-correlations. Ultimately, we selected 13,881,761 sources over a footprint of 16,796 square degrees.

The observed number densities of the selected sources (counting sources per square degree), after a correction of the mean r-mag offset, are within a few percent from the GAMA densities for $16.5<r_{AB}<19.5$ (see Table \ref{tab:number:density:galaxies} and Fig. \ref{fig:stellar:mass:density:functions}). This level of disparity is compatible with the cosmic variance of about 2\% for 60 square degrees of the volume subtended by a $r_{AB}=$ 19.8  band flux-limited survey \citep{DriverRobotham_2010MNRAS.407.2131D}.

For this source sample, we measured the density variations across the footprint as a function of EBV, stellar density, depth, and PSF. We found variations within $\sim$10\% of the average density. This is sufficient to obtain an unbiased clustering measurement to feed to HOD models \citep[e.g.,][]{DelubacRaichoorComparat_2017MNRAS.465.1831D, KongBurleighRoss_2020MNRAS.499.3943K}.

\subsubsection{Galaxy properties}

We used the photometric redshifts from \citet{ZhouNewmanMao_2021MNRAS.501.3309Z}. 
The photometric redshifts of the aforementioned selection ($r_{AB}<19.5$) are excellent, with 0.15\% bias and 3\% scatter. 
We computed the stellar mass on the $g, r, i, z, W1, W2$ set of bands using \texttt{LePhare} \citep{IlbertArnoutsMcCracken_2006A&A...457..841I}. 

Compared to previous estimates using deeper observations and more bands \citep[COSMOS, GAMA][]{IlbertMcCrackenLeFevre_2013A&A...556A..55I,DriverBellstedtRobotham_2022MNRAS.513..439D}, our estimated stellar masses are found to be less accurate, with a scatter of 0.18 dex. This accuracy is similar to that obtained by \citet{ZouGaoZhou_2019ApJS..242....8Z} using the ninth data release of legacy surveys. However, the stellar mass values are, on average, almost unbiased ($-0.085$ dex). The derived stellar mass functions agree with the literature (see following sections and Fig. \ref{fig:stellar:mass:functions}). This indicates that it is a fair quantity to use when selecting galaxy samples. 

\subsubsection{Stellar-mass selected samples}

We followed \citet{ZuMandelbaum_2015MNRAS.454.1161Z,FarrowColeNorberg_2015MNRAS.454.2120F} to design the stellar mass selected volume-limited samples.
The most notable difference is the minimum redshift cut. We imposed $z>0.05$ instead of 0.02 in GAMA. Indeed, the cataloging methods used in LS10 come with some uncertainties when handling sources with a large extent on the sky. This is, in part, handled in the Siena Galaxy Atlas \citep{MoustakasLangDey_2023ApJS..269....3M}, but unfortunately, it does not cover the complete footprint of LS10 (only the footprint of the ninth data release). 
This entails incompleteness in the sample at low redshift, which becomes small above $z=0.05$, then galaxies appear smaller in the sky.
The volume loss $0.042$ Gpc$^3$ is small and negligible (at least ten times smaller) for the samples selected with a stellar mass larger than 10$^{10}$M$_\odot$. 
For the low-mass selection, adding it would increase the volume and the significance of the correlation function, but at the cost of completeness and interpretation.
The upcoming Rubin/LSST observations \citep{IvezicKahnTyson_2019ApJ...873..111I} will enable us to push such investigations to to lower redshifts and enhance the future results for  lower mass galaxies.

The sample selections are illustrated in Fig. \ref{fig:Vlim:samples:Mstar}, where the galaxy sample is shown in the redshift-stellar mass plane, where the boxes identify the selections. Table \ref{tab:Vlim:samples:basic:info} summarizes the exact boundaries used and gives the general properties of the selections (redshift and stellar mass ranges, number densities, and completeness). Since the total number of galaxies and the area covered are large, we simplified the definitions compared to the GAMA ones. We chose a set of minimum stellar mass threshold (9, 9.5, 10, 10.25, 10.5, 10.75, 11, 11.25, and 11.5) that we mapped; with $\log_{10}(M^*/M_\odot)=5\times(z-0.05)^{0.45}+8$ (equation form taken from \citealt{ZuMandelbaum_2015MNRAS.454.1161Z} and adapted to our specific case); to a maximum redshift to obtain something close to volume-limited samples. The line used is close to the one used in the GAMA survey (see the GAMA selection in white boxes). 
This line splits the sample where its density starts to decrease. 
We obtained high levels of completeness between 96 and 100\% when using GAMA as a benchmark (which is deeper and extends to 19.8; see values for each sample in Table \ref{tab:Vlim:samples:basic:info}). 
These completeness levels are trustworthy at the high mass end, above the turnover of the stellar mass distribution for a given sample, but certainly not at low mass and moderate redshift (below 9 and abovez$>0.05$); in that case,  the magnitude limit of GAMA (19.8) will, by construction, deliver an incomplete sample.
The stellar mass functions (without weights applied) are in good agreement with similarly estimated (without completeness correction, only volume weighted number count) stellar mass functions from SDSS and GAMA (and from the mock), as shown in Fig. \ref{fig:stellar:mass:functions}.

These selections are beyond the capabilities of the SDSS spectroscopic survey illustrated by the dotted line in Fig. \ref{fig:Vlim:samples:Mstar} (see also Figs. \ref{fig:stellar:mass:density:functions}, \ref{fig:stellar:mass:functions}). Compared to \citet{ZhangComparatPonti_2024A&A...690A.267Z, ZhangWangLuo_2021A&A...650A.155Z}, the selections described here cover (at a given redshift) galaxies with significantly lower stellar masses than that from the SDSS group catalog from \citet{Tinker_2021ApJ...923..154T, Tinker_2022AJ....163..126T}. We obtained a set of GAMA-like galaxy selections over 16,000 deg$^2$. 
The footprint of the $10<\log_{10}(M^*/M_\odot)<12$ samples is shown in Fig. \ref{fig:Vlim:samples:Mstar}. 
Within a narrow redshift range, we can see the largest cosmic structures by eye. The other samples occupy the same sky area.

\section{Measurements}
\label{sec:measurements}

This section describes the method to measure the galaxy and event auto-correlation, the galaxy-event cross-correlation, and the stacks.
We measured three clustering summary statistics: \textit{(i)} the projected galaxy correlation function ($w_p(r_p)$), 
\textit{(ii)} the galaxy-event angular cross-correlation ($w^{X-corr}_{LS93}$), and \textit{(iii)} the stacked surface brightness profiles ($w^{stack}_{DP83}$).

\subsection{Galaxies}
\label{subsec:measurements:galaxy:autocorr}

\begin{figure}
\centering
\includegraphics[width=0.95\columnwidth]{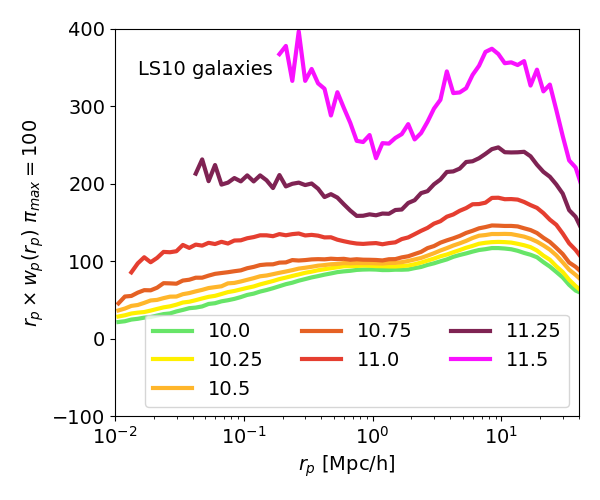}
\caption{Projected galaxy auto-correlation function ($w_p(r_p)$) obtained with the LS10 galaxy samples on $\sim$16,000 deg$^2$. We fit HOD models on these measurements. Differently colored lines represent different samples, identified here by the logarithm of their minimum stellar mass, in solar masses. The color scheme follows that of Fig. \ref{fig:Vlim:samples:Mstar}.}
\label{fig:wprp:final}
\end{figure}

\begin{figure}
    \centering
    \includegraphics[width=0.9\columnwidth]{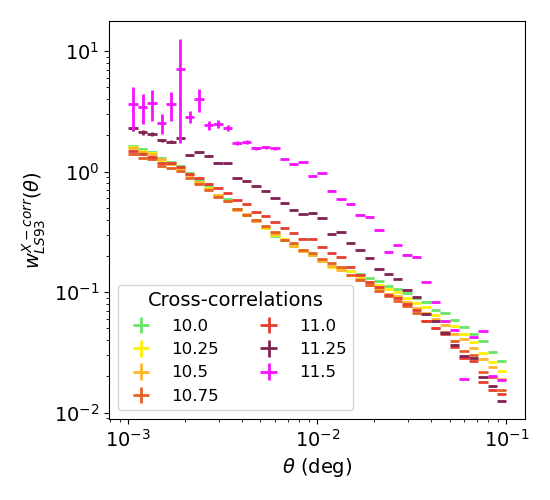}
    \caption{Measured angular cross-correlation between galaxies and X-ray event angular positions. It is estimated in angular space with Eq. \ref{eq:xcorr}. We fit the extension of the HOD model to these measurements. Differently colored lines represent different samples, identified here by the logarithm of their minimum stellar mass, in solar masses. The color scheme follows that of Fig. \ref{fig:Vlim:samples:Mstar}.}
    \label{fig:xcorr:gal:evt:raw}
\end{figure}

\begin{figure}
\centering
\includegraphics[width=0.9\columnwidth]{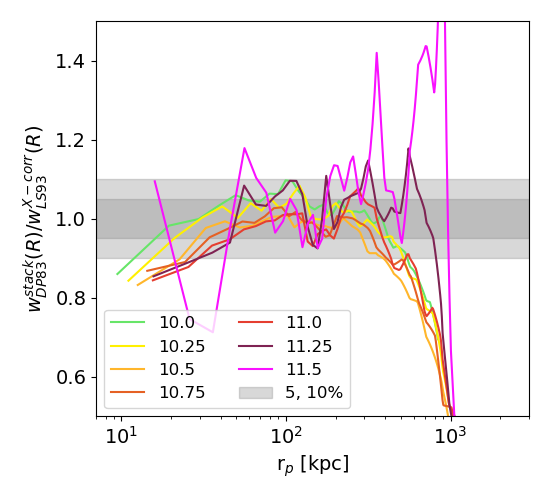}
\caption{\label{fig:xcorr:stacks:estimator} Ratio of galaxy-event cross-correlation function estimators: $w^{stack}_{DP83}/w^{X-corr}_{LS93}$ (see Eq. \ref{eq:xcorr}, \ref{eq:xcorr:DP83}) as a function of separation in proper kpc. 
The discrepancy between estimators is smaller than 5\% in the range $\sim$30-$\sim$300 kpc (exact boundaries depending on the samples). For each sample, we fit models on the measurements where the agreement between estimators is better than 5\%. For the two highest mass selections, we us the range where the agreement is better than 10\%. Differently colored lines represent different samples, identified here by the logarithm of their minimum stellar mass, in solar masses. The color scheme follows that of Fig. \ref{fig:Vlim:samples:Mstar}.}
\end{figure}

\begin{figure}
\centering
\includegraphics[width=0.95\columnwidth]{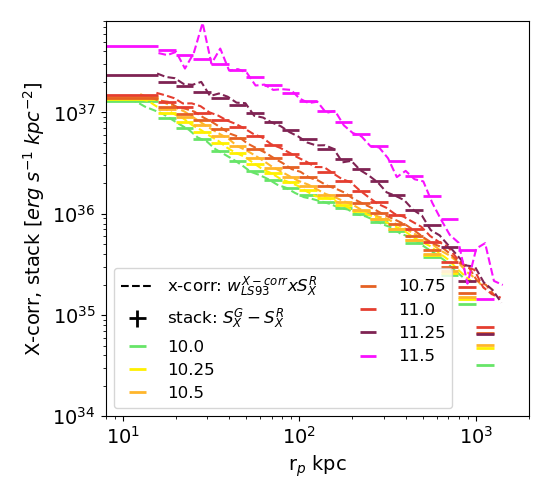}\\
\caption{\label{fig:xcorr:stacks}Comparison of the two cross-correlation estimators in physical units for the stellar mass selected galaxy samples. 
Differently colored lines represent different samples, identified here by the logarithm of their minimum stellar mass, in solar masses. The color scheme follows that of Fig. \ref{fig:Vlim:samples:Mstar}. 
The background subtracted surface brightness profiles in the 0.5-2 keV soft band as a function of proper separation ($r_p$, kpc) is shown in steps ($w^{stack}_{DP83}\times S^R_X$, Eq. \ref{eq:xcorr:DP83}). 
The cross-correlation between galaxies and X-ray events is shown with dashed lines ($w^{X-corr}_{LS93}\times S^R_X$, Eq. \ref{eq:xcorr}). 
The brightness of the emission is correlated with the stellar mass threshold. 
The shape of the profiles changes with the stellar mass threshold. 
The individual best-fit models for these surface brightness profile are presented in Figs. \ref{fig:best:fit:stack:105}, \ref{fig:best:fit:stack:100}, and \ref{fig:best:fit:stack:110}.  
} 
\end{figure}

We used the LS10's random catalog\footnote{\url{https://www.legacysurvey.org/dr10/files}}, created following the procedure from \citet{MyersMoustakasBailey_2023AJ....165...50M}. We applied the same cuts as in the data to create a random catalog with five times the total number in the galaxy samples.
We assigned redshifts to the randoms by shuffling the measured photometric redshifts \citep{RossBautistaTojeiro_2020MNRAS.498.2354R}.

We used the \citet{LandySzalay_1993ApJ...412...64L} estimator and the CorrFunc software \citep{SinhaGarrison_2020MNRAS.491.3022S} to estimate the clustering. We measured the projected clustering (with photometric redshifts and $\pi_{max}=100 h^{-1}$Mpc):\begin{equation}
w_p = \frac{G\times G - 2 G\times R + R\times R}{R\times R}    
,\end{equation}
where $G$ represents the galaxies and $R$ the randoms. 
The measurements are then corrected from systematic effects using the mock catalogs. 
The $w_p(r_p)$ measurements from the observations and the mocks are in good agreement (observations on Fig. \ref{fig:wprp:final}, mocks on Fig. \ref{fig:wprp:final:mocks}). 

We estimated the clustering signal in equal area pixels (of different sizes from tens to hundreds of square degrees). We constructed a distribution of the clustering amplitude around a projected separation of 5 Mpc/h (well in the two-halo term). We excluded all $3\sigma$ outliers and repeat the procedure another time. We excluded a few (on the order of 10) pixels (losing a few hundred square degrees) from the analysis. They reside at the boundary of the survey area either towards the Galactic center or the northern boundary (around a declination of 32$^\circ$). 

The samples defined here do not match those defined in previous analyses \citep{ZehaviZhengWeinberg_2011ApJ...736...59Z,ZuMandelbaum_2015MNRAS.454.1161Z, FarrowColeNorberg_2015MNRAS.454.2120F}, so a face-to-face comparison of the measurements is tedious. 
We find that the measurements are in fair agreement for the same mean stellar mass. 
When we apply the same measurement procedure to the GAMA catalogs, we can obtain clustering measurements in agreement, albeit with large uncertainties (for the GAMA samples). 

The clustering measurements for the lowest stellar mass thresholds are poorly behaved. We would require more accurate photometric redshifts to obtain a robust correlation function in narrow, low-redshift bins. Indeed, in these bins, the photometric redshift uncertainty is comparable to the bin width, and its effects dominate. In the following, we consider only the samples with a stellar mass threshold larger than $10^{10}M_\odot$. We defer the investigation of lower mass samples to spectroscopic surveys.

\subsubsection*{Uncertainties}

To estimate the uncertainties (covariance matrix), we used 1000 re-sampling of the summary statistic, each time leaving out a randomly chosen a 10\% fraction of the sample. Figure \ref{fig:cvcm:10:105:1075:11} shows the correlation coefficient matrix and the diagonal uncertainty. Depending on the sample considered and the scale, the {statistical uncertainty, $\sigma_{stat}$, varies between 5\% (at small separation) and 0.1\% (at large separations).

To ascertain that this measurement is only limited by the sub-percent statistical uncertainties, we would need to control (e.g., with weights) density variations across the footprint to the percent level. 
Reaching this level of precision is difficult and requires complex analysis \citep[e.g.,][]{KongBurleighRoss_2020MNRAS.499.3943K,BerlfeinMandelbaumDodelson_2024MNRAS.531.4954B}. 
Given the variation of densities measured across the footprint (see previous paragraphs), we {assumed a conservative $\sigma_{syst}=5\%$ systematic uncertainty} on all $w_p(r_p)$ measurements. 
We added it in quadrature to the statistical uncertainties derived by resampling {to obtain the total uncertainty, $\sigma^2_{w_p(r_p)}=\sigma^2_{syst}+\sigma^2_{stat}$.} 
An exhaustive study of systematic effects impacting this measurement is beyond the scope of this study. Indeed, massive spectroscopic follow-up is required, which is currently ongoing in DESI.

\subsection{X-rays}
We measured the cross-correlation between galaxies and events for each eROSITA field, where galaxies are available. 
A few eROSITA fields are removed due to contamination from bright galactic (extended) foregrounds (see details in\ Appendix \ref{subsec:app:XRAY:autocorr}). 

The exact area and number of fields removed varies between galaxy samples in this procedure. 
After removing $100-200$ fields, with the remaining $1,400-1,500$ fields, corresponding to approximately 12,000-13,000 deg$^2$, we measured the angular cross-correlation between galaxies and soft X-ray events. We show in Fig. \ref{fig:xcorr:footprint:M10} the footprint of the set of eROSITA fields used for the cross-correlation between X-rays and the sample with $\log_{10}(M^*/M_\odot)>10$ and $0.05<z<0.18$. 

The cross-correlation is obtained with the estimator from  \citet{LandySzalay_1993ApJ...412...64L} with the same set of random points as that used in the galaxy auto-correlation
\begin{equation}
\label{eq:xcorr}
w^{X-corr}_{LS93}(\theta) = \frac{G\times E - G\times R - E\times R + R\times R}{R\times R},    
\end{equation}
Here, $G$ represents the galaxies, $E$ the events, $R$ the randoms, and $G\times E$ the normalized pair counts; all are functions of the angular separation. Figure \ref{fig:xcorr:gal:evt:raw} shows the obtained estimated cross-correlations for each galaxy sample. Given the different mean redshifts of each sample, a direct, quantitative comparison is not straightforward. 
Using the stacking method (as explained below), we obtained the cross-correlation through a different estimator. 
We found an excellent agreement between the two methods (see the next section and Fig. \ref{fig:xcorr:stacks:estimator}). 
The cross-correlation is converted to physical units and shown in Fig. \ref{fig:xcorr:stacks} (dashes). 
Therefore, the ordering of the functions is clear: the higher the stellar mass threshold, the higher the correlation.

\subsubsection{Stacked X-ray surface brightness profiles}

As an alternative estimator of the cross-correlation, to verify and strengthen the results, we stacked the X-rays around the same galaxy sample following the method described in \citet{ComparatTruongMerloni_2022A&A...666A.156C}. By counting pairs of galaxies and events as a function of separation, we obtained a surface brightness profile $S^{G}_X$. It is similar to $G\times E$ in Eq. \ref{eq:xcorr}, however, here we used the redshifts to obtain a result in surface brightness units and proper separation. In addition, we extracted a surface brightness profile around random points by counting pairs of random points and events as a function of separation $S^{R}_X$ (similar to $E\times R$ in Eq. \ref{eq:xcorr}). We took the random points from the same set for the galaxy clustering. With that, we were able to estimate the "background" of the stacking experiment. 
We report the values of $S^{R}_X$ in Table \ref{tab:background:values}. 
These values are similar to that derived by \citet{ZhangComparatPonti_2024A&A...690A.267Z} $\in 0.8-1.2 \times 10^{37}$ erg kpc$^{-2}$ s$^{-1}$, when stacking around different SDSS galaxy samples covering a different redshift range and on a different area of the sky than that used here. 

Then, we form the \citet{DavisPeebles_1983ApJ...267..465D} estimator of the cross-correlation
\begin{equation}
\label{eq:xcorr:DP83}
w^{stack}_{DP83}(R) = \frac{S^{G}_X - S^{R}_X}{S^{R}_X}.    
\end{equation}
From $w^{X-corr}_{LS93}(\theta)$, we can convert the angular separations into proper distances at the mean redshift of each sample to obtain $w^{X-corr}_{LS93}(r_p)$ and compare it to $w^{stack}_{DP83}(r_p)$ (Fig. \ref{fig:xcorr:stacks:estimator}). We find the two estimators in 10\% (5\%) agreement for separations between 20-500 (40-300) kpc. Outside this range, the estimators slowly diverge from one another, with the DP83 lower than the LS93. The large-scale difference may come from excluding the $G\times R$ term as inaccuracies in the photometric redshifts. The difference on small scales may stem from the use of photometric redshifts (in the DP83 estimator) that have a lower accuracy for AGN. To investigate this in detail, one would need event-level simulation with reliable correlation function predictions over Gpc volumes for hot gas, AGN, and XRB events, which is not yet available, but hopefully will be in the future. 

When comparing the two estimators in physical units, the cross-correlation (dashes) and stacked (solid steps) measurement in a log-log representation are in excellent agreement in the ranges mentioned above; see Fig. \ref{fig:xcorr:stacks}. 

\begin{table}[]
    \centering
    \caption{\label{tab:background:values} Background values, $S^{R}_X$, for each stack, their uncertainty, and in parenthesis the relative uncertainty in \%.}
    \begin{tabular}{cc}
\hline
\hline
$M^*_0$ & $S^{R}_X\times(1+\bar{z})^{-2}$ [erg kpc$^{-2}$ s$^{-1}$] (\%) \\ \hline
10.0 & 6.859 $\pm$ 0.032 $\times 10^{36}$ ( 0.47 ) \\
10.25 & 6.799 $\pm$ 0.028 $\times 10^{36}$ ( 0.41 ) \\
10.5 & 6.772 $\pm$ 0.025 $\times 10^{36}$ ( 0.37 ) \\
10.75 & 6.712 $\pm$ 0.027 $\times 10^{36}$ ( 0.39 ) \\
11.0 & 6.693 $\pm$ 0.051 $\times 10^{36}$ ( 0.76 ) \\
11.25 & 6.706 $\pm$ 0.078 $\times 10^{36}$ ( 1.16 ) \\
11.5 & 6.727 $\pm$ 0.109 $\times 10^{36}$ ( 1.61 ) \\
\hline
    \end{tabular}
\end{table}

\subsubsection{Uncertainties}

Thanks to the large area and number of galaxies used, numbers are sufficient to estimate with the Jackknife method the diagonal uncertainties (Fig. \ref{fig:xcorr:gal:evt:raw:uncertainties} top panel) and the correlation coefficient of the stacked profiles (Fig. \ref{fig:xcorr:gal:evt:raw:uncertainties} bottom panel). 
They are below 2\% for the samples with a stellar mass minimum below 11. 
Only for the two highest stellar mass selections are the uncertainties larger, mainly driven by the smaller sample sizes.  
The correlation coefficient matrix is diagonal with some noise, so we will only use the diagonal uncertainties.
They are smaller than the uncertainty on the background values (Table \ref{tab:background:values}; i.e., on the order of a few percent) needed to convert the cross-correlation to a physical value. 

Given the difference obtained with the two estimators of the cross-correlation (Fig. \ref{fig:xcorr:stacks}), until more simulations are available to quantify the accuracy at which each estimator enables us to retrieve the cross-correlation, we deem necessary to add a $\sigma_{syst}=$5\% (10\%) systematic uncertainty to encompass the possible uncertainty on the convergence of the cross-correlation estimator for the samples with threshold in $\log_{10}(M^*/M_\odot)$ of 10, 10.25, 10.5, 11 (11.25, 11.5). 
A complete assessment of systematics uncertainties impacting this measurement (and its estimators) requires extensive work on the models linking the large-scale structure to X-ray emission. 
These new measurements are key to creating new and accurate such models to enhance our understanding of possible systematic uncertainties impacting it. 

In what follows, we explain how we used the measurements obtained with the LS93 estimator of the cross-correlation; here, the total diagonal uncertainties considered are the statistical ones, with the systematic one added in quadrature. 
It is conservative, but it allows us to mine the information contained in the measurements.

When we are working in physical space, the multiplication by the background value incurs an additional statistical uncertainty reported in Table \ref{tab:background:values} (the last value in percent). 
There is also a systematic uncertainty in the estimation of the background. Indeed, the estimates on the different samples should all be consistent with the actual value of the background. 
The seven estimates point to values between 6.7 and 6.85 with a $\sim2\%$ variation, larger than the statistical uncertainties reported in the table. 
To remain conservative, we added a 2\% systematic uncertainty on the background value (when used). 

\section{The halo model}
\label{sec:method:halo:model}

Halo occupation distribution (HOD) models describe how galaxies populate dark matter halos, predicting the number and types of galaxies within a given halo based on its mass. These models are essential for linking observations of galaxy clustering to the underlying dark matter distribution. By incorporating statistical relationships between halo properties and galaxy formation, HOD models help refine cosmological simulations and improve our understanding of large-scale structure formation \citep{CooraySheth_2002PhR...372....1C, AsgariMeadHeymans_2023OJAp....6E..39A}.

We followed \citet{MurrayDiemerChen_2021A&C....3600487M,NishimichiTakadaTakahashi_2019ApJ...884...29N,AsgariMeadHeymans_2023OJAp....6E..39A,MeadAsgari_2023ascl.soft07025M} to implement the HOD model and link the galaxy samples to their host dark matter halo population (Sect. \ref{subsec:hod:autocorr}). 
Using ultranest \citep{Buchner_2016S&C....26..383B,Buchner_2019PASP..131j8005B,Buchner_2021JOSS....6.3001B}, we fit a classic HOD model on the $w_p(r_p)$ measurement, shown on Fig. \ref{fig:wprp:final}.
We added an X-ray dimension to the HOD model (Sect. \ref{subsec:hod:crosscorr}) to fit the cross-correlation measurement, shown in Fig. \ref{fig:xcorr:gal:evt:raw}.

\subsection{Model for the galaxy auto-correlation}
\label{subsec:hod:autocorr}

To model the projected clustering $w_p(r_p)$ of a galaxy sample, we assumed the measurement to be at the mean redshift of the sample.
We relied on the theoretical framework from \citet{vandenBoschMoreCacciato_2013MNRAS.430..725V, MoreMiyatakeMandelbaum_2015ApJ...806....2M}.

\subsubsection{Cosmological setup}

The cosmological parameters are fixed to $\Omega_c = 0.26069$, $\Omega_b = 0.04897$, $\Omega_k = 0.0$, $h = 0.6766$, $As = 2.09\times 10^{-9}$, $n_s = 0.965$, $w = -1.0$, and $w_a = 0.0$, $m_{\nu} = 0.0$ \citep{PlanckCollaborationAghanimAkrami_2020A&A...641A...6P}. The deduced parameters are $\sigma_8 \sim 0.811$, $\omega_b = \Omega_b h^2$, $\omega_c = \Omega_c h^2$,  and $\Omega_m=\Omega_c+\Omega_b=0.30966$.
The linear dark matter power spectrum is then predicted at the redshift of interest using CAMB \citep{LewisChallinorLasenby_2000ApJ...538..473L}.

For the halo model, we used the \citet{TinkerRobertsonKravtsov_2010ApJ...724..878T} halo mass function with masses expressed in units of 200 times the mean density of the Universe ($M_{200}$ in $M_{\odot}/h$).
Halo quantities are computed for halos with a mass starting at 10$^9$ $M_{\odot}/h$.
Given the halo mass, we computed the virial radius, $r_{v}$, based on the halo mass and overdensity condition and the average concentration of the halo $c_s(M)$ using the \citep{DuffySchayeKay_2008MNRAS.390L..64D} relation.
We performed the Fourier transform of all corresponding halo profiles (assuming \citet{NavarroFrenkWhite_1997ApJ...490..493N}, NFW hereafter, for the set of virial radii) on the wavenumber grid used for the power spectrum to obtain the halo window function $U_k(k, r_v, c_s)$.
With these, we predict the dark matter halo profile.

\subsubsection{Parameter grid and occupation function}
{Conceptually, we followed  the HOD parametrization of \citet[][Eq. 2, 5]{ZhengCoilZehavi_2007ApJ...667..760Z}. Regarding the complete galaxy samples considered here, the satellite galaxies start populating halos at higher halo masses than central ones. We can replace the $\langle N_c \rangle (M)$ factor in front of the satellite occupation with a Heaviside step function for computational speed. Indeed, in that case, the two formulations are numerically very close. 
We parametrize the occupation number of central (satellite) galaxies in distinct halos, as follows,}
\begin{equation}
\label{eq:Ncen}
\langle N_c \rangle (M) = 0.5 \left[ 1 + erf \left( \frac{ log_{10}(M)- \log_{10}(M_{\rm min}) } { \sigma_{\log(M)} } \right) \right]\,,
\end{equation}
\begin{equation}
\label{eq:Nsat}
\langle N_s \rangle (M) = \mathcal{H}(M-M_0, 1.) \left( \frac{M-M_0}{M'_1} \right)^\alpha_{\rm sat}\,.
\end{equation}
$erf$ is the error function and $\mathcal{H}$ is the Heaviside function. 

We tie the two parameters $M'_1$ and $M_0$ with:
$\log_{10}(M'_1) = \log_{10}(M_0) + 1$. Indeed, theoretical \citep{ZhengBerlindWeinberg_2005ApJ...633..791Z} and observational \citep{ZhengCoilZehavi_2007ApJ...667..760Z,ZehaviZhengWeinberg_2011ApJ...736...59Z} application of this model showed that $M_1$ and $M_0$ typically vary by a factor of 10. 
{To avoid sampling un-physical values of $M_0$, the parameter
passed to the fitting routine is $\Delta_{M_0}=\log_{10}(M_0) - \log_{10}(M_{\rm min})$}
{This parametrization can capture the clustering of the (almost complete) samples designed here.} We adopted a uniform prior with ranges:
$\log_{10}(M_{\rm min})\in[10,15]$,
$\alpha_{\rm sat}\in[0.01,1.5]$,
$\sigma_{\log(M)}\in[0.01, 1.5]$,
$\Delta_{M_0}\in[-1,2]$. 
We limited$M_{\rm min}$ to a smaller interval for the higher mass samples.

\subsubsection{Fitting function}

Given a set of parameters and at each fit iteration, the following is evaluated and compared to the measured signal.
We computed the expected (mean) number of central and satellite galaxies as a function of halo mass: $N_{\rm cen}$, $N_{\rm sat}$.
We computed the expected variance (and covariance) in the numbers of central and satellite galaxies assuming Bernoulli statistics for centrals and Poisson statistics for satellites: $V_{\rm cen}$, $V_{\rm sat}$.
Then we integrated the occupation function with the halo mass function to retrieve the mean density of galaxies: $\rho_g[(Mpc/h)^{-3}] = \int \langle N\rangle (M) n(M)dM$.

For central galaxies, we used a delta function profile. Satellite galaxies follow the NFW profile.
The central profile depends on $N_{\rm cen}$, $\rho_g$, $V_{\rm cen}$, and the satellite profile on the corresponding quantities.
Finally, with all the ingredients mentioned above, we predict the galaxy power spectrum contributed by central satellites and their cross terms: $P_{gg}(k) = P_{cc}(k)  + 2 P_{cs}(k) + P_{ss}(k)$. 
Each comprises two components: a one-halo term and a two-halo term. The implementation used to compute the power spectrum is that of \citet[][Eqs. 20-29]{AsgariMeadHeymans_2023OJAp....6E..39A}. 
With the Hankel transform, we obtained the configuration space real space correlation function $\xi_{gg}$, which we project into a projected correlation function $w_p(r_p)$ by doing the Abel transform up to $\pi_{max}=100 h^{-1}$Mpc.
We interpolated the obtained function on the same grid as the measurement and compute a $\chi^2$ summary statistic. 

\subsubsection{Angular clustering prediction}

The angular clustering was predicted using the \citet{Limber_1953ApJ...117..134L} approximation.
We used the implementation of \citet{MurrayDiemerChen_2021A&C....3600487M}. 

\subsubsection{Validation with the mock catalogs}

We tested the HOD implementation with the mock catalogs (see Appendix \ref{sec:mocks}). We retrieved the input HOD within a few per cent after fitting the predicted correlation function.
The range of scales where the algorithm works seamlessly is 0.01-30 Mpc/h and at the few percent precision level. 
To reach beyond that precision more detailed HOD models are required \citep[e.g.,][]{ZuMandelbaum_2015MNRAS.454.1161Z, ZuMandelbaum_2016MNRAS.457.4360Z, ContrerasAnguloZennaro_2021MNRAS.504.5205C, YuanGarrisonHadzhiyska_2022MNRAS.510.3301Y}. In our case, an accuracy of a few percent  is sufficient.

\subsection{A model for the galaxy-event cross-correlation}
\label{subsec:hod:crosscorr}

This section aims to obtain a proof-of-concept model that accounts for the cross-correlation measurement to grasp the information content to the first order.
To predict the cross-correlation between soft X-ray photons and galaxies, we fix the parameters of the HOD models to the ones obtained when fitting the $w_p(r_p)$ (see Sect. \ref{subsec:results:wprp}). 

\subsubsection{Hot gas component}

In this model, the hot gas emission comprises the circumgalactic, intragroup, and intracluster medium. It corresponds to any X-ray emission coming from a hot gas component. In the following, for simplicity, we designate the hot gas filling the dark matter halo volume, regardless of halo mass (or temperature), by "hot gas" or the circumgalactic medium (CGM).
Recent advances in the studies of cosmic rays indicate that a fraction of the X-ray flux may find its origin there \citep[e.g.,][]{HopkinsQuataertPonnada_2025arXiv250118696H}. We opt here for a classic hot gas model with only Bremsstrahlung. We leave the implementation of cosmic rays in the HOD model to future works. 

We converted (at  mean redshift) the halo mass distribution from the HOD from a 200m definition to a 500c definition using the \textsc{colossus} software \citep{Diemer_2018ApJS..239...35D}. 
Instead of using the \citet{DuffySchayeKay_2008MNRAS.390L..64D} relation (as in the halo model), we use that of \citet{IshiyamaPradaKlypin_2021MNRAS.506.4210I} that has a native mass definition on both M500c and M200c and thus provides a slightly more precise conversion. 
Then, we predict the average X-ray luminosity for each halo with the following scaling relation with 
\begin{equation}
\log_{10}(L^{0.5-2\; keV}_X) = L_0 + \alpha_{SR} (\log_{10}\left(\frac{M_{500c}}{10^{15}}\right) + 2\log_{10}(E(\bar{z})),
\end{equation}
where $\alpha_{SR}$ and $L_0$ are the slope and amplitude of the scaling relation and $E(\bar{z})=H(\bar{z})/H(0)$ is the normalized Hubble parameter. 

We convert the mass-luminosity relation into a mass-count relation to predict the average number of counts received as a function of halo mass.
\begin{equation}
\log_{10}(N_{ct}) = \log_{10}\left(\frac{ L^{0.5-2\; keV}_X  t_{exp} \; ARF}{4\pi C } \right) -  2 \log_{10}(d_L) .
\end{equation}
There, we denote $d_L$ as the luminosity distance to the mean redshift of the sample. We convert the individual photon energy in keV to erg with a constant factor $C=1.602177\times10^{-9}$ [erg/keV] without assuming a spectral model and virtually giving all the counts an energy of one keV.
Using the mock catalogs from \citet{ComparatEckertFinoguenov_2020OJAp....3E..13C, SeppiComparatBulbul_2022A&A...665A..78S}, where detailed spectra are used to convert the rest-frame 0.5-2keV luminosity into an observed frame flux, we find that a more accurate $C$ has minimal effect on the predicted average count. The scatter increases by $\sim10\%$, while the mean and the slope remain unchanged.
We take the vignetting corrected Field-of-view-averaged Auxiliary Response File ($ARF$) for all seven telescope modules combined at 1keV of 1,000 cm$^2$ to obtain at a given mass the average number of counts expected in the band.
We assume the median unvignetted exposure time of eRASS:4 of $t_{exp}=550$s. 
The cross-correlation is sensitive to the slope of the scaling relation, not to its amplitude, so the exact values of $t_{exp}$, $ARF$ matter little in the prediction. 
If we were also to model the luminosity functions, one would need to fold in the complete distribution of exposure times. 
As the samples studied are at low redshift, we ignore the impact of K-correction (and temperature) as a function of mass. As stated above, this assumption has a minimal impact.
The sky area considered has low foreground absorption; we ignore its effect. 
In Sect. \ref{subsec:energy:dependence} we show the energy dependence of the cross-correlation. In the future, we will use models with energy dependence (and spectral models) upon the complete set of cross-correlations as a function of energy. 

Using the best-fit parameters of the auto-correlation function, the halo mass function, and the volume covered by the sample, we deduce the number of distinct halos occupied in each halo mass bin. 
For each mass bin and each distinct halo we draw an observed luminosity from a Gaussian distribution centered on $\log_{10}(L^{0.5-2\; keV}_X, M_{500c})$ with a fixed intrinsic scatter $\sigma_{LX}=0.3$ compatible with the value inferred by \citet{LovisariReiprichSchellenberger_2015A&A...573A.118L,BulbulChiuMohr_2019ApJ...871...50B,ComparatEckertFinoguenov_2020OJAp....3E..13C,SeppiComparatGhirardini_2024A&A...686A.196S}. 
For all masses, we deduce the average number of counts emitted and its standard deviation, which we use as the amplitude and variance of the X-ray profile. 

We then need a functional form for the 3D profile of the events. 
For an isothermal optically thin plasma, the soft X-ray flux (event rate) is proportional to the square of the electron density. 
The generic profile formulated by \citet{VikhlininKravtsovForman_2006ApJ...640..691V} (or the simple beta profile) informs on the general shape of the electron density. 
Here, we need a 3D profile for the events, not the electron density, so previous profile parameters are not directly applicable.
The most direct way to obtain a simple analytic formula for such a profile is to use the eROSITA digital twin and events emitted by clusters and groups \citep{SeppiComparatBulbul_2022A&A...665A..78S} to find a suitable functional form in 3D that will, by construction, in 2D, follow the surface brightness profile expected. 
We find that the event profiles of distinct halos in the simulation are well-described by
\begin{equation}
p_e(r, r_{200m}, c_{200m} )
        \propto
                \frac{ 1}{ \left(\frac{r}{ \frac{r_{200m}}{ c_{200m}} }\right)^{\alpha_{prof}} }
                \times
                \frac{1}{\left(1+\left(\frac{ r}{ \frac{r_{200m}}{ c_{200m}}}\right)^{2}\right)^{p2}},\end{equation}
where $c_{200m}$ is the halo concentration and $r_{200m}$ its radius. 
Their ratio constitutes the scale radius ($\frac{r_{200m}}{ c_{200m}}$) of the halo.
It is a characteristic radius that defines the transition between the inner, dense core and the outer, more diffuse region of the halo.
We use ($\alpha_{prof}$, $p2$) = (0.9, 1.6), close to the analytical formulae from  \citet{VikhlininKravtsovForman_2006ApJ...640..691V}. 
Converting to a surface brightness, this corresponds on large scales to a beta profile with $\beta=0.53(=1.6/3)$, consistent with measurements in clusters \citep[e.g.,][albeit with different profile functional forms, the slope on large scales is broadly consistent]{NeumannArnaud_1999A&A...348..711N,MantzAllenMorris_2016MNRAS.456.4020M,SandersFabianRussell_2018MNRAS.474.1065S,EckertGhirardiniEttori_2019A&A...621A..40E, EttoriGhirardiniEckert_2019A&A...621A..39E, LiuBulbulGhirardini_2022A&A...661A...2L, BulbulLiuKluge_2024A&A...685A.106B}, but steeper than values measured for lower mass halos by \citet[][that find 0.4]{ZhangComparatPonti_2024A&A...690A.267Z}. 
Current observations indicate the need a slope varying with mass to guarantee consistency.
Here, we use a value consistent with clusters where most of the events come from, but inconsistent with the measurement from \citet{ZhangComparatPonti_2024A&A...690A.267Z}. We leave to future models the exploration on how to free the profile slope as a function of mass.

In the observation, we cross-correlate all the events with a complete set of galaxies. This is reflected in the profile's normalization, which gives the cross-correlation amplitude.
From simulations \citep{SeppiComparatBulbul_2022A&A...665A..78S}, we estimate the fraction of hot gas events correlated to galaxies to be $w_0=$1-2\% (depending on the redshift range). 
Knowing the exact value of $w_0$ is complex and depends on models of the X-ray foregrounds and backgrounds. 

The power spectrum is the sum of the cross components: events with central and satellite galaxies. 
It contains contributions from the one-halo and the two-halo approach in each term.
The one-halo term accounts for galaxy correlations within the same halo, which is important on small scales.
The two-halo contains galaxy correlations between separate halos, it dominates on large scales.  
We compute the Hankel transform to obtain the real-space correlation function and the Limber equation to obtain an angular correlation function, denoted $ w_{\rm Gas \times Gal}$. 
Finally, we convolve the cross-correlation function with the eROSITA PSF \citep{MerloniLamerLiu_2024A&A...682A..34M}. 
This prediction works when the set of events solely comes from hot gas around galaxies following that profile and scaling relation. 

In future model incarnations, the scaling relation and profile parametrizations (and their scatter) should vary with halo mass \citep{SchayeKugelSchaller_2023MNRAS.526.4978S}. Using the scaling relation parametrization from \citet[][Eq. 67]{ChiuGhirardiniLiu_2022A&A...661A..11C}, the profile parametrization from \citet[][Eq. 3]{VikhlininKravtsovForman_2006ApJ...640..691V}, we can have up to 5 + 10 = 15 parameters (with degeneracies). The X-ray spectral aspect should be taken into account, possibly by parametrizing the mass-temperature scaling relation to deduce the mass-luminosity relation in different energy bands \citep{LovisariEttoriGaspari_2021Univ....7..139L}.

\subsubsection{Point sources}

In observations, there are X-ray point sources (XRB and AGN) in the galaxies that contribute to the signal. 
Using \citet{AirdCoilGeorgakakis_2017MNRAS.465.3390A} and \citet{ComparatMerloniSalvato_2019MNRAS.487.2005C}, we can estimate that the XRB contribution amounts to $\lesssim10\%$ of the total point source luminosity. We refer to Fig. \ref{fig:model:ps:comparison}, where the blue circles corresponding to XRB are $\sim$10 times lower than the total model emission in red stars. 
We opted for a simple formulation of the X-ray point sources occupation in galaxies driven by the AGN HOD occupation. 
In \citet{ComparatLuoMerloni_2023A&A...673A.122C}, we measured the halo occupation distribution \citep[using the formalism of][]{MoreMiyatakeMandelbaum_2015ApJ...806....2M} of detected AGN. 
Here, the galaxy sample also contains undetected AGNs and XRBs. We used the HOD parameters given in the Table 2 (top set of values) from \citet{ComparatLuoMerloni_2023A&A...673A.122C} except for $M_{\rm min}$ that is lowered by $\Delta_{M_{\rm min}}$ to include undetected AGN and XRB.
For simplicity, we assumed only AGNs in central galaxies, since the way in which AGNs populate satellite galaxies remains an open question \citep{LeauthaudJ.BensonCivano_2015MNRAS.446.1874L,ComparatLuoMerloni_2023A&A...673A.122C,PowellKrumpeCoil_2024A&A...686A..57P}. 
Possible signals from the cross-correlation with satellite galaxies hosting AGN are degenerate (and thus absorbed) with that of hot gas events correlated with satellite galaxies.
We obtained a power spectrum for the point sources that we convert to the angular correlation function. 
After the fitting procedure, we extracted the average point source luminosity in central galaxies in the sample that contains contributions from AGNs and XRBs. Similarly to the hot gas, a possible X-ray point source model may be (in the future) formulated with physical links to the halos, galaxies, and AGNs as in \citet[][with more than 20 parameters]{ZhangBehrooziVolonteri_2023MNRAS.518.2123Z}, and XRB as in \citet[][with three parameters]{AirdCoilGeorgakakis_2017MNRAS.465.3390A}.

\subsubsection{Complete model}

The model for the measured cross-correlation ($w^{X-corr}_{LS93}(\theta)$) is expressed as
\begin{equation}
w^{model}(\theta) =  w_{PSF}(\theta, \Delta_{M{min}}, w_{0}) +  w_{Gas\; \times\; Gal}(\theta, \alpha_{SR}, w_{0}) ,
\end{equation}
with three free parameters $\Delta_{M{min}}$ (X-ray point sources), $\alpha_{SR}$ (hot gas), and $w_0$ (the fraction of correlated events in the complete soft X-ray field). 
With this model, we can decompose the signal as a function of scale into its main components. 
The normalization parameter ($w_{0}$) would not be necessary if we could cross-correlate exactly the photons coming from the large-scale structure in which the galaxies reside. 

Further efforts in adding physics in each component and freeing parameters in the X-ray model interfaced with the HOD model is a complex task left for future investigation. 
Indeed, we do  need to account for the physics of the hot gas (AGN feedback, metallicity, density profiles) across five orders of magnitude in halo mass \citep[e.g.,][expCGM\footnote{\url{https://gmvoit.github.io/ExpCGM}}]{OppenheimerSchaye_2013MNRAS.434.1043O,FlenderNagaiMcDonald_2017ApJ...837..124F,BiffiDolagMerloni_2018MNRAS.481.2213B, ComparatEckertFinoguenov_2020OJAp....3E..13C, PopHernquistNagai_2022arXiv220511528P,OrenSternbergMcKee_2024ApJ...974..291O, SinghLauFaerman_2024MNRAS.532.3222S, LauNagaiBogdan_2024arXiv241204559L, LauNagaiFarahi_2025ApJ...980..122L}, the incidence and brightness of AGNs as a function of stellar mass and star formation rate \citep[e.g.,][]{AirdCoilGeorgakakis_2015MNRAS.451.1892A,GeorgakakisAirdSchulze_2017MNRAS.471.1976G,BiffiDolagMerloni_2018MNRAS.481.2213B, ComparatMerloniSalvato_2019MNRAS.487.2005C, ComparatLuoMerloni_2023A&A...673A.122C, GeorgakakisComparatMerloni_2019MNRAS.487..275G, AllevatoShankarMarsden_2021ApJ...916...34A}, and of the emission of X-ray binaries as a function of stellar mass and star formation rate \citep[e.g.,][]{AirdCoilGeorgakakis_2017MNRAS.465.3390A,LehmerBasu-ZychMineo_2016ApJ...825....7L,Vladutescu-ZoppBiffiDolag_2023A&A...669A..34V} to construct a comprehensive model. 

This model serves as a proof-of-concept to show the complexity of the data vector considered. 
Unifying the halo model to predict summary statistics related to galaxies, AGNs, and hot gas both in the optical and X-ray will take significant effort and time. This should serve as a springboard for the community to engage in this complex modeling task. 

\section{Results}
\label{sec:results}

\begin{table*}[ht]
\begin{center}
\caption{Best-fit parameters obtained by fitting the HOD model on the galaxy auto-correlation functions.}
\label{tab:Vlim:samples:HODbestfitParams}
\begin{tabular}{|r|rrrrr|rrrrrr|}
\hline \hline
sample & \multicolumn{5}{|c|}{Halo Occupation distribution model} & \multicolumn{6}{|c|}{Deduced quantities} \\

$M^*_0$ & \multicolumn{4}{c}{Best-fit parameters} & $\frac{\chi^2}{ndof}$ & \multicolumn{2}{c}{Densities} & $f_{sat}$ & b & \multicolumn{2}{c|}{$\log_{10}(\bar{M}_{200m})$ } \\
& $M_{\rm min}$ & $\alpha_{\rm sat}$ & $\sigma_{\log(M)}$ & $\Delta M_{0}$ &  & deg$^{-2}$ & $10^{-5}$ Mpc$^{-3}$ &\% & & $\bar{M}_{\rm cen}$ & $\bar{M}_{\rm sat}$ \\
\hline
10 &
12.113$_{-0.257}^{+0.409}$ &
1.184$_{-0.029}^{+0.059}$ &
0.666$_{-0.236}^{+0.335}$ &
0.052$_{-0.067}^{+0.182}$ &
0.49 &
103.9 &
235.8 &
23.0 &
1.22 &
12.66 &
14.14 
\\
10.25 & 
12.26$_{-0.276}^{+0.468}$ &
1.178$_{-0.031}^{+0.062}$ &
0.619$_{-0.303}^{+0.437}$ &
0.014$_{-0.083}^{+0.204}$ &
0.43 &
128.4 &
162.9 &
23.0 &
1.29 &
12.79 &
14.13 
\\
10.5 & 
12.362$_{-0.23}^{+0.483}$ &
1.163$_{-0.03}^{+0.063}$ &
0.538$_{-0.312}^{+0.506}$ &
0.016$_{-0.113}^{+0.208}$ &
0.57 &
148.5 &
117.1 &
22.1 &
1.34 &
12.9 &
14.12 
\\
10.75 &
12.327$_{-0.047}^{+0.132}$ &
1.091$_{-0.011}^{+0.023}$ &
0.228$_{-0.09}^{+0.231}$ &
0.031$_{-0.035}^{+0.056}$ &
0.84 &
208.6 &
100.7 &
25.3 &
1.43 &
12.98 &
14.04 
\\
11 & 
12.674$_{-0.039}^{+0.105}$ &
1.131$_{-0.013}^{+0.026}$ &
0.202$_{-0.072}^{+0.194}$ &
0.018$_{-0.03}^{+0.046}$ &
0.8 &
123.0 &
42.5 &
20.5 &
1.6 &
13.24 &
14.12 
\\
11.25 &
13.096$_{-0.03}^{+0.077}$ &
1.159$_{-0.028}^{+0.053}$ &
0.123$_{-0.082}^{+0.199}$ &
0.036$_{-0.019}^{+0.031}$ &
1.02 &
39.2 &
13.6 &
13.9 &
1.89 &
13.57 &
14.26 
\\
11.5 &
13.483$_{-0.021}^{+0.057}$ &
1.261$_{-0.064}^{+0.119}$ &
0.1$_{-0.062}^{+0.167}$ &
0.002$_{-0.016}^{+0.031}$ &
1.93 &
12.7 &
4.4 &
9.9 &
2.32 &
13.87 &
14.42 
\\

  \hline
\end{tabular}
\tablefoot{Best-fit parameters (with 1$\sigma$ uncertainties) and deduced parameters obtained by fitting the HOD model on the galaxy auto-correlation functions (Fig. \ref{fig:wprp:final}). The parameters are described in Sect. \ref{sec:method:halo:model}. The deduced quantities model densities of tracers, satellite fraction, large-scale halo bias, and mean halo mass hosting central and satellite galaxies are computed for the best-fit model.}

 \end{center}
\end{table*}

\begin{figure*}
\centering
\includegraphics[width=1.8\columnwidth]{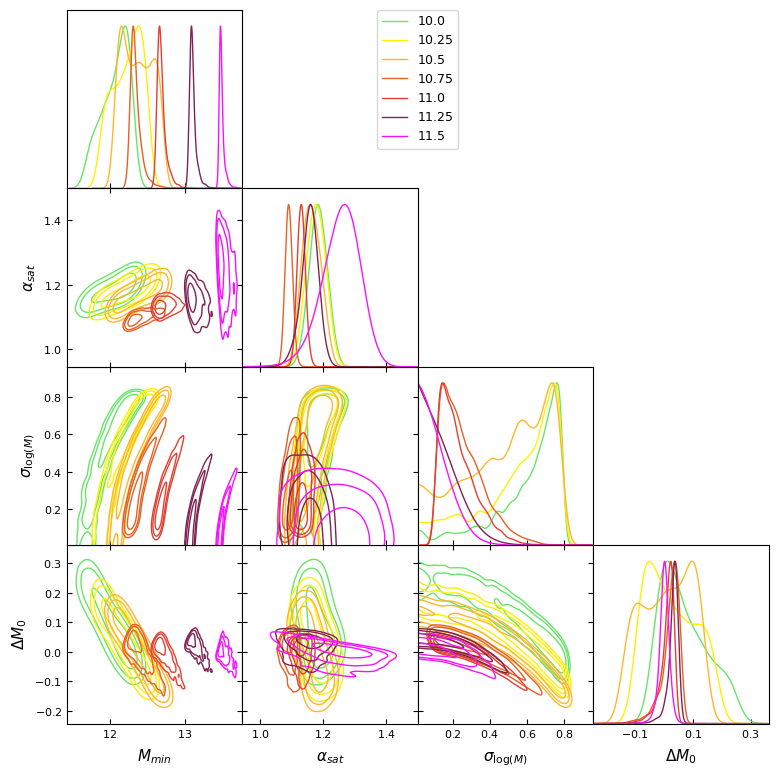}
\includegraphics[width=0.9\columnwidth]{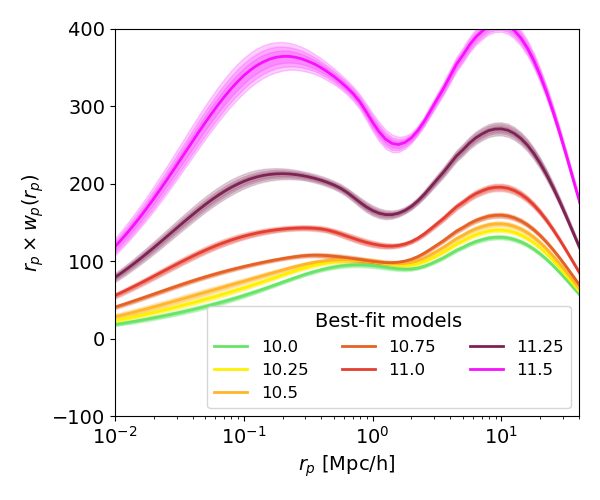}
\includegraphics[width=0.9\columnwidth]{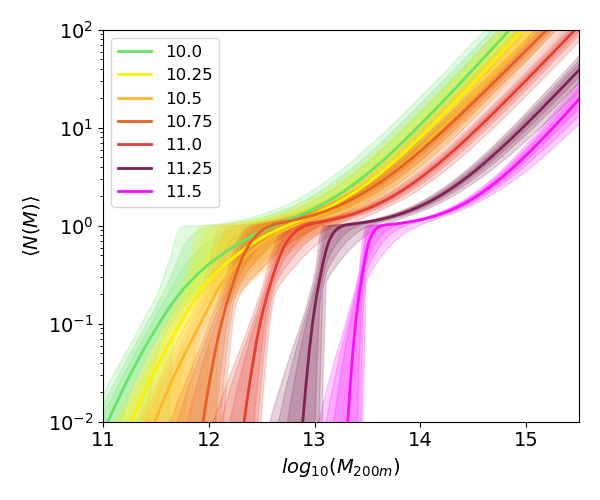}
\caption{\label{fig:HOD:best:fit} HOD best-fit model parameters obtained by adjusting the model from Sect. \ref{sec:method:halo:model} on the $w_p(r_p)$ measurements obtained in Sect. \ref{sec:measurements} and shown in Fig. \ref{fig:wprp:final}. 
Differently colored lines represent different samples, identified here by the logarithm of their minimum stellar mass, in solar masses. The color scheme follows that of Fig. \ref{fig:Vlim:samples:Mstar}.
We show the corner plot of the parameters and their 1,2,3$\sigma$ contours in the top panel. 
The trends of the parameters obtained and their correlation with the stellar mass threshold of the different galaxy samples is sensible: $M_{min}$ is correlated with the minimum stellar mass, while $\sigma_{\log(M)}$ is anti-correlated.  
Model $w_p(r_p)$ obtained and their 1,2,3$\sigma$ contours (bottom left). Posterior HOD with 1,2,3 $\sigma$ contours (bottom right). 
The ordering of the HOD curves follows the stellar mass selection and follows expectations.
An individual comparison of the models and the measurements are shown in Fig. \ref{fig:wprpBFmodels:10:105:1075:11}.
}

\end{figure*}

\begin{figure}
    \centering
    \includegraphics[width=.9\columnwidth]{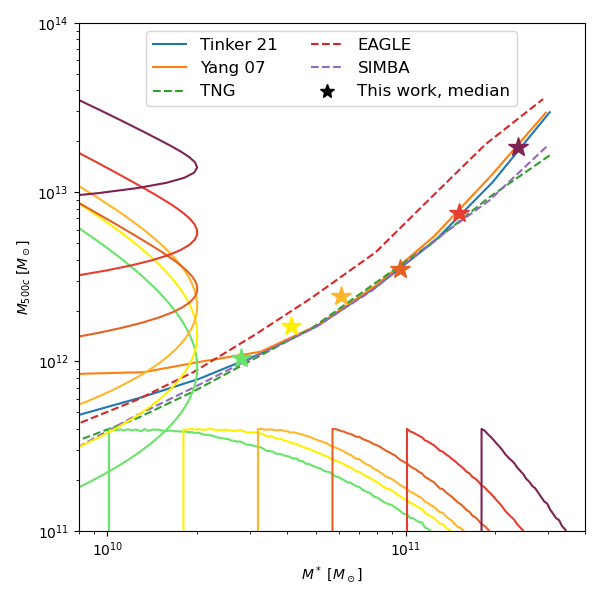}
    \caption{Inferred stellar-to-halo-mass relation for each sample observed. 
    The stars are located at the median values. 
    The distributions of the halo and stellar masses of the sample studied are shown with colored solid lines.
    There is a fair agreement with previous models \citet{YangMovandenBosch_2005MNRAS.356.1293Y,Tinker_2021ApJ...923..154T} and with the prediction from the TNG and SIMBA hydrodynamical simulations \citep{PillepichNelsonHernquist_2018MNRAS.475..648P, DaveAngles-AlcazarNarayanan_2019MNRAS.486.2827D,SchayeCrainBower_2015MNRAS.446..521S}.}
    \label{fig:SMHMR:hod}
\end{figure}

We present here the best-fit models on the galaxy auto-correlation (Sect. \ref{subsec:results:wprp}) and on the galaxy-event cross-correlation (Sect. \ref{subsec:results:xcorr}). 

\subsection{Best-fit HOD models of the galaxy projected correlation function ($w_p(r_p)$)}
\label{subsec:results:wprp}
The best-fit models account for the observed correlation function within a few percent, with acceptable reduced $\chi^2$. 
We show the best-fit model, together with the measurements (see Fig. \ref{fig:wprpBFmodels:10:105:1075:11}).
A few $\chi^2$ values are quite significantly below 1, hinting that the systematic uncertainty budget of 5\% may be too generous in these cases (the ones with stellar mass threshold of 10, 10.25, and 10.5).
The best-fit parameters and their 1$\sigma$ uncertainties are listed in Table \ref{tab:Vlim:samples:HODbestfitParams}.  The value of
$M_{min}$ ($\sigma_{logM}$) is correlated (anti-correlated) with the stellar mass threshold. The satellite slope ($\Delta M_0$) is consistent with 1.2 (0) for all samples. 
The corner plot with the 1,2,3 $\sigma$ contours, shown in Fig. \ref{fig:HOD:best:fit} (top panel), illustrates the trends. The bottom panels show the best-fit models for $w_p(r_p)$ and the halo occupation $\langle N(M) \rangle$. 
The evolution of the parameters with the different stellar mass thresholds is sensible. 

The highest mass bin's summary statistic suggests the model does not perfectly account for the observations (see Table \ref{tab:Vlim:samples:HODbestfitParams}). The reduced $chi^2$ is too large at 1.93. In Fig. \ref{fig:wprpBFmodels:10:105:1075:11}, bottom right panel, we see deviation between the data and the model at the transition between the one-halo and the two-halo term (1-3 Mpc/h range).
The small stellar mass range and the extended redshift range likely play a role in this disagreement. 
This sample requires further model refinements to fully capture the information present in the measurement. 

From the best-fit models, we deduced the fraction of satellites and the large-scale halo bias (Table \ref{tab:Vlim:samples:HODbestfitParams}). 
The deduced parameters' trends are sensible: $f_{\rm sat}$ decreases from 23\% to 10\% throughout the samples. The large-scale halo bias (mean halo mass hosting central galaxies $\log_{10}(M_{200m}/M_\odot)$) increases from 1.22 to 2.32 (12.66 to 13.87). 
The samples very well probe the MW-mass galaxies and M31-mass galaxies up to galaxy groups. 
Throughout the samples, satellite galaxies are hosted by clusters with a mean mass above $\log_{10}(M_{200m}/M_\odot)>14$.
Qualitatively, these parameters and trends are consistent with previous studies \citep[e.g.,][]{ZehaviZhengWeinberg_2005ApJ...630....1Z, ZehaviZhengWeinberg_2011ApJ...736...59Z, FarrowColeNorberg_2015MNRAS.454.2120F, ZuMandelbaum_2015MNRAS.454.1161Z, ZuMandelbaum_2016MNRAS.457.4360Z}. Quantitative comparisons with these studies are technically challenging due to the differences in the models used.

The deduced densities (per square degree and cubic Mpc) are in broad agreement with observations. The differences between the two may stem from, in order of importance: \textit{(i)} redshift evolution of the sample and \textit{(ii)} uncertainties on photometric redshifts and stellar masses. It indicates a direction for future improvement of the HOD model for these samples.

Each sample, with its model, finds its location on the stellar to halo mass relation (Fig. \ref{fig:SMHMR:hod}). We get consistent values with previous models \citep{YangMovandenBosch_2005MNRAS.356.1293Y, Tinker_2021ApJ...923..154T} or predictions \citep{PillepichNelsonHernquist_2018MNRAS.475..648P, DaveAngles-AlcazarNarayanan_2019MNRAS.486.2827D,SchayeCrainBower_2015MNRAS.446..521S,McAlpineHellySchaller_2016A&C....15...72M}. Overall, the HOD parameters represent the observed auto-correlation functions well. 
We fixed the HOD model parameters to their best-fit values to model the cross-correlation. 

\subsection{Models of the cross-correlation and stacks}
\label{subsec:results:xcorr}

\begin{figure}
\centering
\includegraphics[width=0.9\columnwidth]{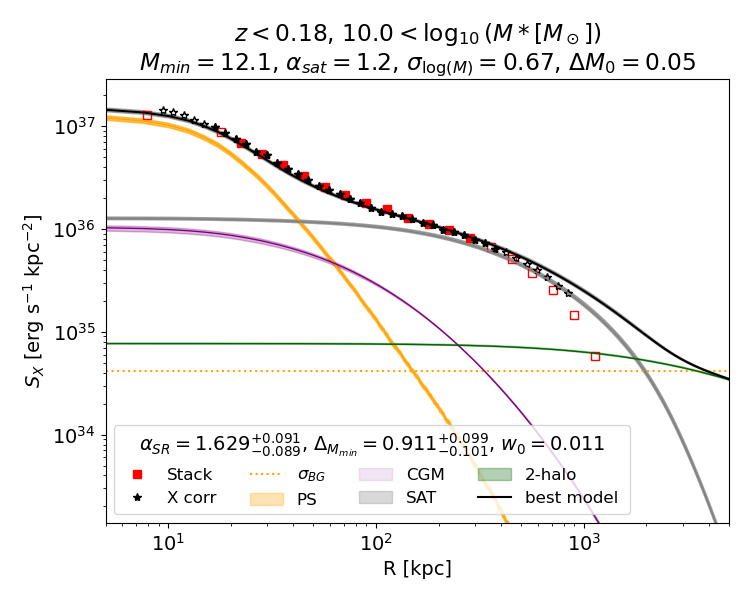}
\vspace*{-0.5cm}
\includegraphics[width=0.9\columnwidth]{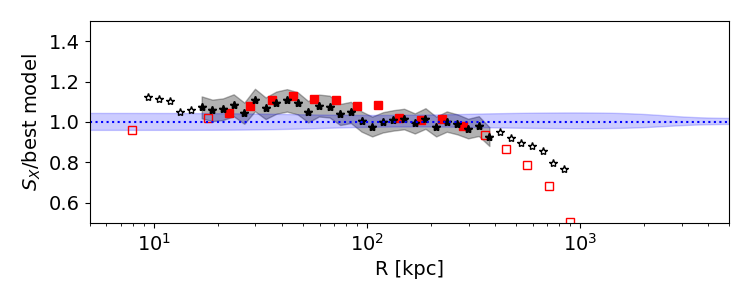}
\caption{\label{fig:best:fit:stack:105} Best-fit model for the galaxy X-ray cross-correlation obtained with the stellar mass-selected sample with mass $M^*>10^{10} M_\odot$ (10.0). The black star ( red square) symbols indicate the data points obtained with the estimator from the LS93 estimator, see Eq. \ref{eq:xcorr} (DP83 estimator, Eq. \ref{eq:xcorr:DP83}). 
The symbols are filled in the range where good agreement is obtained between estimators. 
These values are used for fitting the model. 
Empty symbols are measurements outside of that range. They are not used for fitting. 
We compare the observed cross-correlation (black stars) and stacks (red squares) to the best-fit model (black line) and its components: circumgalactic medium (purple), point sources (yellow), satellites (grey), and two-halo term (green). The uncertainties are 1$\sigma$. The HOD parameters used are fixed at the median of the best-fit model. We indicate the values in the title. The median best-fit cross-correlation parameters are indicated in the legend title (and in Table \ref{tab:xcorr:parameters}). The reduced $\chi^2$ is 0.918, indicating that the model accounts for the data. 
The horizontal dotted line shows the uncertainty on the background determination for this sample. 
The observed cross-correlation is dominated at small separation by the point source emission, then at larger separation by the emission of the hot gas hosted by large halos ($M_{200m}>10^{14}M_\odot$) seen from the positions of satellite galaxies. 
Beyond separations of 3 Mpc, the two-halo term dominates. 
The contribution to the total emission of average emission from the hot gas in distinct halos hosting central galaxies in the inner 200 kpc varies between 5 and 15\%. 
We show the ratio of the cross-correlation (stack) and the best-fit model in the bottom panel. In the range of 20-400 kpc, the residuals are consistent with the model within 1$\sigma$ uncertainties. Outside of the fitting range, on small scales, the cross-correlation is up to 20\% larger than the best-fit model. On large scales, at 1Mpc, the measurements are smaller than the best-fit model by 10-20\%. 
The result for the other samples are shown in Figs. \ref{fig:best:fit:stack:100} and \ref{fig:best:fit:stack:110}.}
\end{figure}

\begin{figure*}
\centering
\includegraphics[width=0.9\columnwidth]{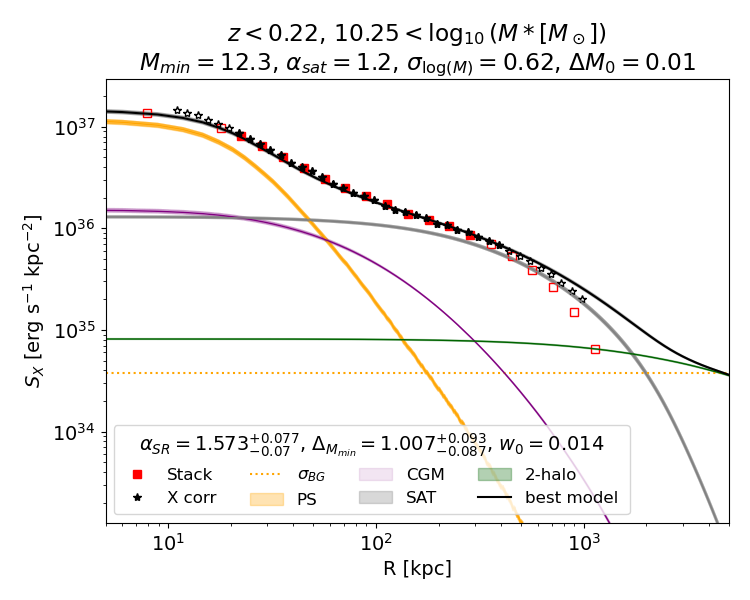}
\includegraphics[width=0.9\columnwidth]{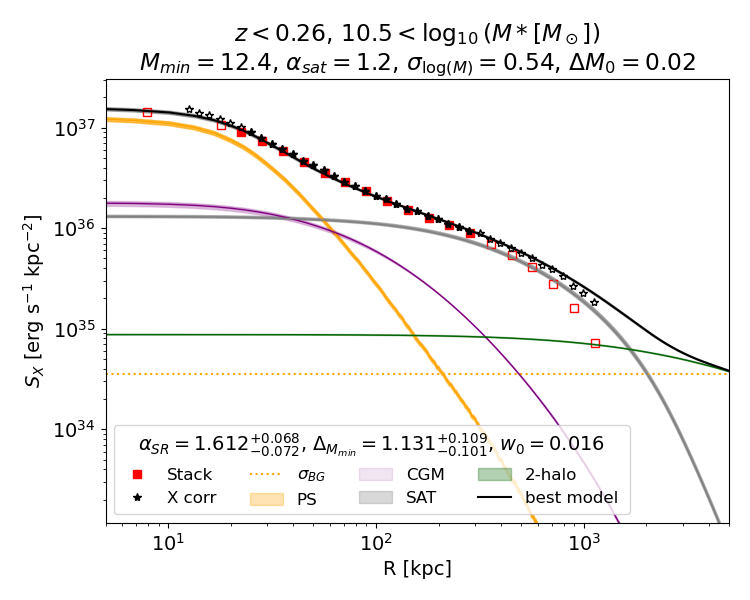}
\\
\vspace*{-0.5cm}
\includegraphics[width=0.9\columnwidth]{xcorr3/LS10_VLIM_ANY_10.0_Mstar_12.0_0.05_z_0.18_N_27592380.011-xcorr-grid-best-model-residual.png}
\includegraphics[width=0.9\columnwidth]{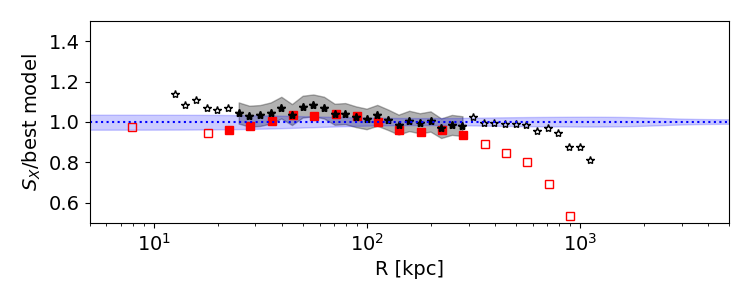}\\
\includegraphics[width=0.9\columnwidth]{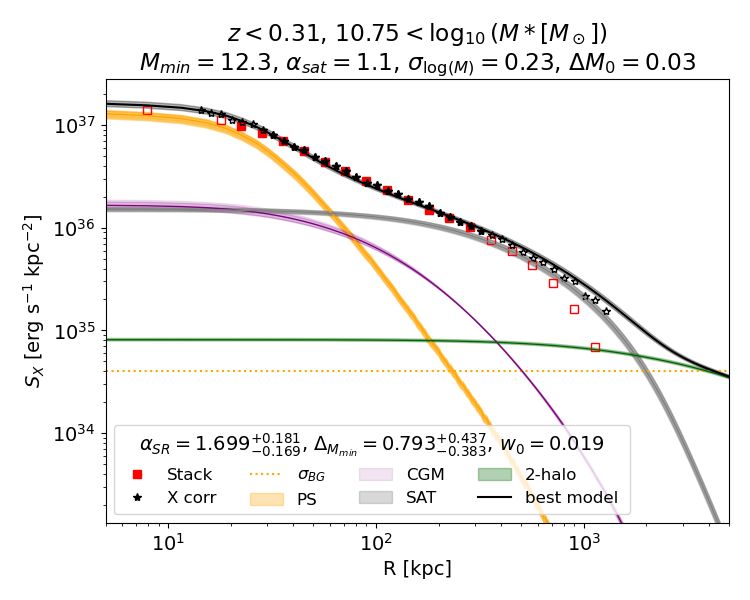}
\includegraphics[width=0.9\columnwidth]{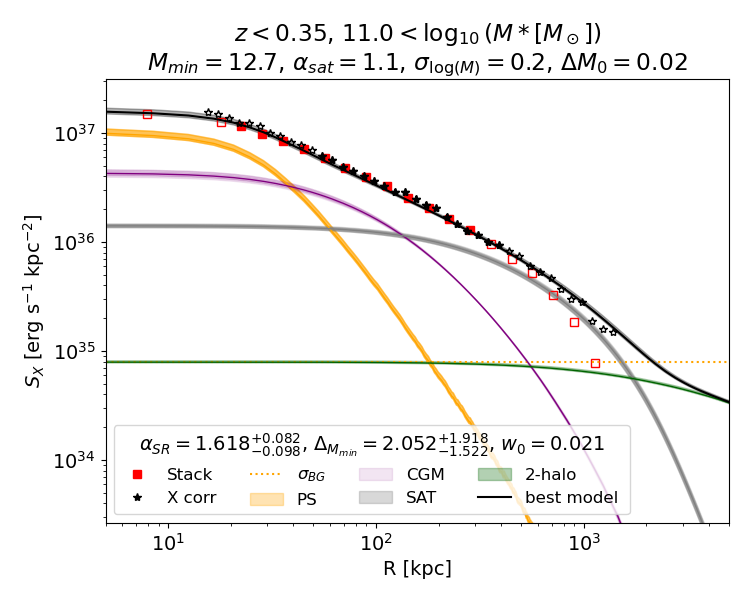}\\
\vspace*{-0.5cm}
\includegraphics[width=0.9\columnwidth]{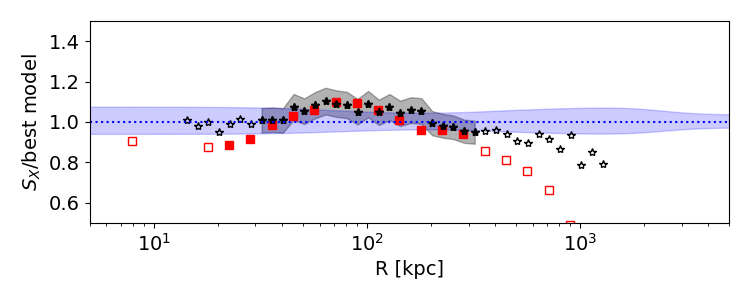}
\includegraphics[width=0.9\columnwidth]{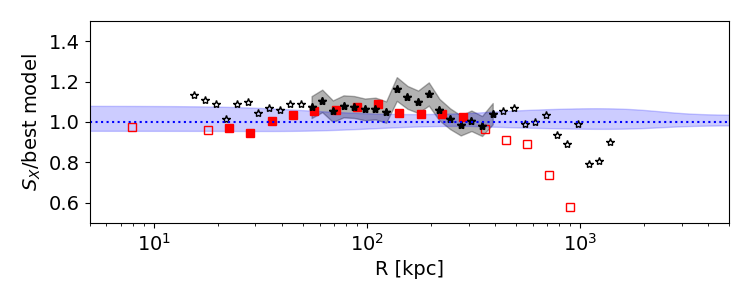}
\caption{\label{fig:best:fit:stack:100}  Fig. \ref{fig:best:fit:stack:105} continued. Models of the soft X-ray surface brightness profiles for the stellar mass-selected sample with mass 
$M^*>1.7\times10^{10} M_\odot$ (top left, 10.25), $M^*>3.1\times10^{10} M_\odot$ (top right, 10.5),
$M^*>5.6\times10^{10} M_\odot$ (bottom left, 10.75), $M^*>10^{11} M_\odot$ (bottom right, 11.0). 
The model accounts well for the measurements.
The point source, satellite and two-halo terms show little evolution. The hot gas model brightness increases significantly with the stellar mass threshold.
}
\end{figure*}

\begin{figure*}
\centering
\includegraphics[width=0.9\columnwidth]{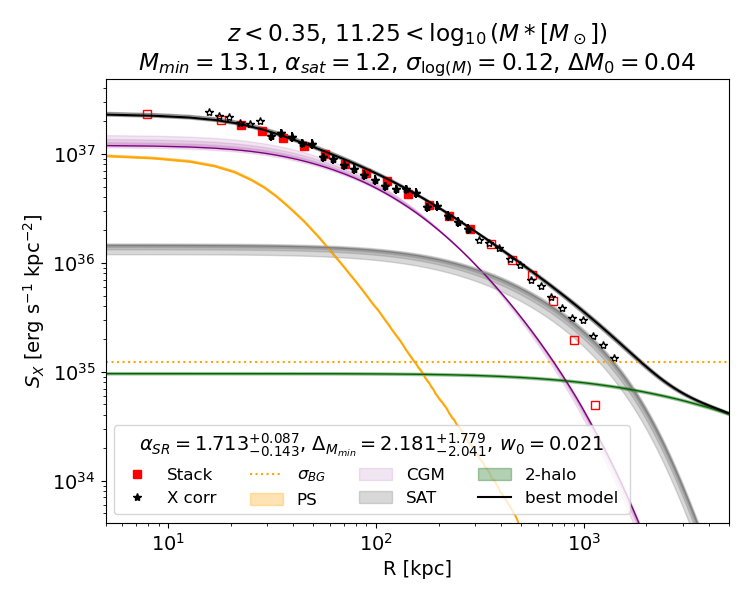}
\includegraphics[width=0.9\columnwidth]{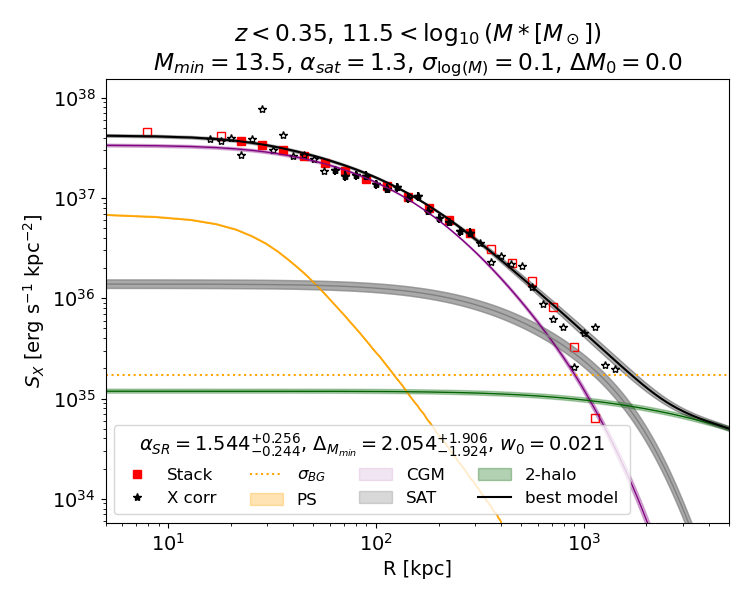}\\
\vspace*{-0.5cm}
\includegraphics[width=0.9\columnwidth]{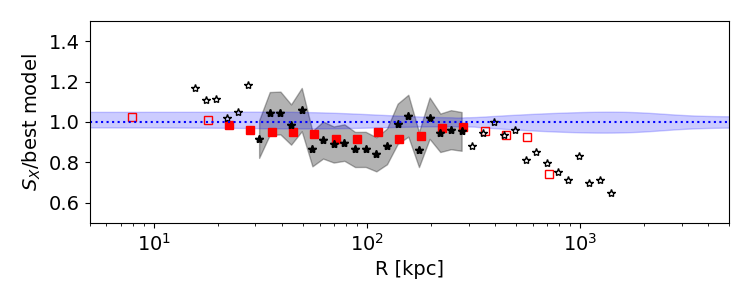}
\includegraphics[width=0.9\columnwidth]{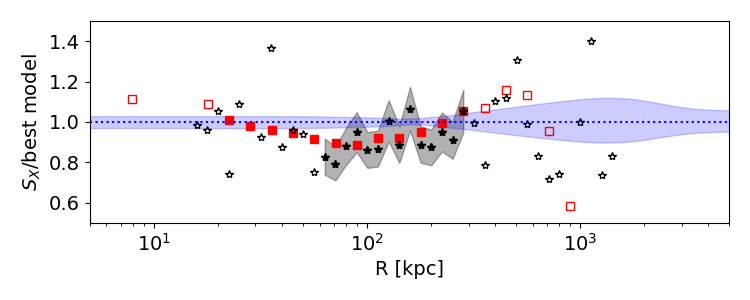}
\caption{\label{fig:best:fit:stack:110}  Fig. \ref{fig:best:fit:stack:105} continued. Models of the soft X-ray surface brightness profiles for the stellar mass-selected sample with mass 
$M^*>1.7\times10^{11} M_\odot$ (left, 11.25) and $M^*>3.1\times10^{11} M_\odot$ (11.5, right). 
The deviations between the model and the data indicate either that the model needs more complexity to account for the data or that the estimator suffers from additional systematic uncertainties. These data sets will constitute an important benchmark for future models. 
The models indicate that the hot gas emission becomes dominant for these samples. 
}
\end{figure*}

\begin{table*}
\begin{center}
    \caption{Best-fit parameters obtained by fitting the model on the galaxy-event cross-correlation.}
    \label{tab:xcorr:parameters}
    \begin{tabular}{|c|cccc|cc|}
    \hline \hline
    sample &  \multicolumn{4}{|c|}{Surface brightness profile model} & \multicolumn{2}{|c|}{Deduced L$_X$ [$10^{40}$erg s$^{-1}$]} \\
$M^*_0$ 
& $\alpha_{SR}$
& $\Delta_{M_{min}}$
& $w_{0}$
& $\frac{\chi^2}{ndof}$ 
& point source
& hot gas  
\\          
      \hline
10.0 & $ 1.629^{+0.091}_{-0.089}$ & $ 0.911^{+0.099}_{-0.101}$ &  0.011 & $0.918^{+0.5114}_{-0.5501}$  & $3.59^{+0.16}_{-0.15}$ & $7.81^{+0.16}_{-0.16}$ \\
10.25 & $ 1.573^{+0.077}_{-0.07}$ & $ 1.007^{+0.093}_{-0.087}$ &  0.014 & $0.83^{+0.5237}_{-0.5423}$   & $4.54^{+0.19}_{-0.18}$ & $11.6^{+0.18}_{-0.18}$ \\
10.5 & $ 1.612^{+0.068}_{-0.072}$ & $ 1.131^{+0.109}_{-0.101}$ &  0.016 & $0.752^{+0.5395}_{-0.534}$   & $6.2^{+0.22}_{-0.28}$ & $15.64^{+0.18}_{-0.18}$ \\
10.75 & $ 1.654^{+0.166}_{-0.104}$ & $ 0.851^{+0.358}_{-0.331}$ &  0.019 & $1.083^{+0.448}_{-0.6232}$  & $8.63^{+0.66}_{-0.54}$ & $18.27^{+0.56}_{-0.55}$ \\
11.0 & $ 1.634^{+0.066}_{-0.094}$  & $ 1.221^{+0.279}_{-0.421}$ &  0.021 & $1.103^{+0.7386}_{-0.3429}$ & $7.6^{+0.75}_{-0.28}$ & $42.11^{+0.93}_{-1.08}$ \\
11.25 & $ 1.713^{+0.087}_{-0.143}$ & $ 2.181^{+1.779}_{-2.041}$ &  0.021 & $0.865^{+0.7073}_{-0.3646}$ & $7.43^{+0.21}_{-0.05}$ & $137.91^{+3.07}_{-3.09}$ \\
11.5 & $ 1.544^{+0.256}_{-0.244}$ & $ 2.054^{+1.906}_{-1.924}$ & 0.021 & $0.788^{+0.2152}_{-0.0819}$   & $5.53^{+0.07}_{-0.01}$ & $401.2^{+12.32}_{-12.99}$ \\

\hline
\end{tabular}
\tablefoot{Best-fit parameters ( $\alpha_{SR}$, $\Delta_{M_{min}}$ and their $1\sigma$ uncertainties) obtained by fitting the model on the galaxy-event cross-correlation (Fig. \ref{fig:xcorr:gal:evt:raw}). The $w_{0}$ is parameter fixed.  The parameters are described in Sect. \ref{sec:method:halo:model}. Deduced quantities are obtained by integrating the model profiles.}
\end{center}    
\end{table*}

We considered three parameters ($\alpha_{SR}$, $\Delta_{M_{min}}$, $w_0$) in this step. 
We manually adjusted $w_0$ to match the cross-correlation on large scales (beyond the point sources' influence), giving values between 1 and 2 percent as simulations indicate. We report the values found by hand in Table \ref{tab:xcorr:parameters}.
We computed a set of models on a grid of the other two parameters: $\alpha_{SR}\in$[1.4,1.8], and $\Delta_{M_{min}}\in[0,4]$.
We fixed the scaling relation amplitude parameter to $\log_{10}(L_0)=44.7$ to force the compatibility with scaling relations obtained at high mass for galaxy clusters \citep{SeppiComparatGhirardini_2024A&A...686A.196S, ChiuGhirardiniLiu_2022A&A...661A..11C, MantzAllenMorris_2016MNRAS.456.4020M, LovisariReiprichSchellenberger_2015A&A...573A.118L, BulbulChiuMohr_2019ApJ...871...50B, PrattCrostonArnaud_2009A&A...498..361P}. 

We obtained a reasonable reduced $\chi^2$ (see 
Table \ref{tab:xcorr:parameters}). 
We show the best-fit models and the cross-correlation measurements in Figs. 
\ref{fig:best:fit:stack:105},
\ref{fig:best:fit:stack:100}, and
\ref{fig:best:fit:stack:110}. 
For the lower stellar mass thresholds (10, 10.25, 10.5), we find that the point source (PS, yellow) emission dominates the cross-correlation at small separation ($r<80$kpc). In the range ($80<r<2$Mpc), the emission from large halos hosting satellite galaxies dominates (SAT, grey). Finally, on scales beyond that measured here ($r>2$Mpc), the two-halo term (green) becomes dominant.
Interestingly, there is no scale at which the circumgalactic medium (CGM, purple) dominates. In the range ($20<r<200$kpc), the CGM contributes to more than 10\% of the signal. 
Progressively, with the minimum stellar mass increasing, the CGM emission increases. 
For the 10.75 sample, the trend is essentially the same, except in the range 50-60kpc, where the three components (CGM, PS, SAT) contribute each the same surface brightness.
For the mass threshold $M^*>11$, the CGM is the dominating emission source over the range of 30-200kpc.
Finally, for the mass thresholds 11.25 and (11.5), Fig. \ref{fig:best:fit:stack:110}, the CGM emission dominates over other components until 400 (700) kpc. 

\subsubsection{Point source emission}

The point source emission has a total average luminosity in the soft band (0.5-2 keV) between 3 and 9 $\times10^{40}$ erg s$^{-1}$. The luminosity increases with mean stellar mass until $\log_{10}(M^*/M_\odot)=$11, then stalls and declines (Fig. \ref{fig:model:ps:comparison}). 
This result is in disagreement with AGN models where the luminosity correlates with the stellar mass \citep[e.g.,][]{ComparatMerloniSalvato_2019MNRAS.487.2005C}. Indeed, they predict that such a relationship will increase monotonically. 
This result will constitute a strong benchmark for creating future X-ray AGN models realistically embedded in the galaxy population. 
It suggests that models sampling the specific accretion rate distribution may perform better in this dimension \citep[e.g.,][]{GeorgakakisComparatMerloni_2019MNRAS.487..275G,PowellAllenCaglar_2022ApJ...938...77P} and that X-ray AGN activity needs modulation with the environment \citep{MartiniMillerBrodwin_2013ApJ...768....1M, SabaterBestArgudo-Fernandez_2013MNRAS.430..638S, SabaterBestHardcastle_2019A&A...622A..17S, PoggiantiJaffeMoretti_2017Natur.548..304P, KoulouridisGkiniDrigga_2024A&A...684A.111K}.
It opens a new route of exploration for models to jointly predict the entire galaxy population and the average emissions of AGNs in the soft X-rays \citep[e.g., the combination of the UniverseMachine and the Trinity models][]{BehrooziWechslerHearin_2019MNRAS.488.3143B, ZhangBehrooziVolonteri_2023MNRAS.518.2123Z}. 
We further discuss this finding in Sect. \ref{sec:discussion}.

\begin{figure}
    \centering
    \includegraphics[width=.9\columnwidth]{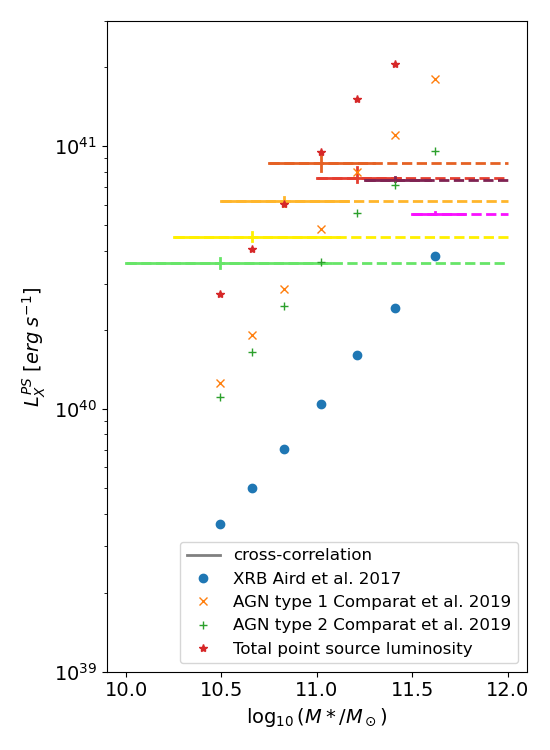}
\caption{\label{fig:model:ps:comparison}Comparison of the average point source emission components obtained for each sample as a function of the stellar mass. The horizontal solid line represents the mean and standard deviation of the stellar masses in the samples. The horizontal dashes show the extent of the selection. 
Different colored crosses represent different samples, identified here by the logarithm of their minimum stellar mass, in solar masses following the color scheme of Fig. \ref{fig:Vlim:samples:Mstar}.
The inferred average point source luminosity increases with mean stellar mass from $\log_{10}(M*/M_\odot)=$10.6 until 11 and stalls, then decreases slightly. 
The point source luminosities in each sample are consistently more luminous than the predicted XRB luminosity for each sample using \citet{AirdCoilGeorgakakis_2017MNRAS.465.3390A}. 
The predicted AGN luminosities from the \citet{ComparatMerloniSalvato_2019MNRAS.487.2005C} model for type 1 (orange crosses) and type 2 (green pluses) are lower than the observations for the thresholds 10, 10.25, 10.5, 10.75, but they then are higher for the thresholds 11, 11.25 and 11.5. 
The total predicted luminosity in point sources (red stars) with these models agrees with the observations only for three samples (10.25, 10.5, 10.75 selections) out of seven. 
These inferred quantities will be used as a benchmark to test upcoming AGN models in the large-scale structure, for example to test mechanisms of X-ray AGN suppression in high halo mass \citep[e.g.,][]{MunozRodriguezGeorgakakisShankar_2024MNRAS.532..336M}.}
\end{figure}

\subsubsection{CGM emission}

The brightness of the model CGM emission correlates with the stellar mass threshold; see Fig. \ref{fig:model:comparison:CGM}. 
It follows the expectation based on previous measurements indicating that the scaling relation between soft X-ray luminosity and stellar (or halo) mass continues below the group regime \citep{ZhangComparatPonti_2024A&A...690A.268Z, ZhangComparatPonti_2025A&A...693A.197Z, PopessoMariniDolag_2024arXiv241117120P}.
Since the scaling relation between halo mass and X-ray luminosity is steep, one must interpret these values cautiously. 
In Fig. \ref{fig:CGM:profFRAC:SR} (top row of panels), we show the fraction of emission in the profile as a function of its originating halo mass. 
In the bottom row of panels, we show the posterior scaling relation obtained from the fit (blue shaded area) and, for each halo mass, its contribution to the luminosity obtained when integrating the average model profile. 
For the $\log_{10}(M^*/M_\odot)>$10 sample (left column of panels), halos from $10^{11}M_\odot$ to $10^{14}M_\odot$ contribute to the average profile. 
The most common halo hosting these galaxies ($10^{12}M_\odot$) contributes very little to the average hot gas emission.
For the $\log_{10}(M^*/M_\odot)>$11 sample (right column of panels), halos from $10^{12}M_\odot$ to $10^{14}M_\odot$ contribute to the average profile. 
The most common halo hosting these galaxies ($10^{13}M_\odot$) contributes more to the average hot gas emission.
The model identifies the individual contribution of each dark matter halo to the total luminosity of a cross-correlation measurement. 
This stresses the importance of considering the complete distribution of halos hosting galaxies when measuring the average luminosity in the presence of a steep slope. 
The cross-correlation measurements are compatible with a steep scaling relation (slope $\sim1.63$).
They are compatible with the results of \citet{ZhangComparatPonti_2024A&A...690A.268Z, ZhangComparatPonti_2025A&A...693A.197Z, PopessoMariniDolag_2024arXiv241117120P} as well as literature measurements of galaxy clusters as collected in \citet{ComparatEckertFinoguenov_2020OJAp....3E..13C}.

The statements above depend on the formulation of the hot gas profile, the structure of the scaling relation, and their possible variations or evolution with mass and redshift. It shows the importance of embedding detailed models of the hot GCM in the large-scale structure \citep[see e.g.,][]{FlenderNagaiMcDonald_2017ApJ...837..124F,ComparatEckertFinoguenov_2020OJAp....3E..13C,LauNagaiFarahi_2025ApJ...980..122L,LaPostaAlonsoChisari_2024arXiv241212081L,KeruzoreBleemFrontiere_2024OJAp....7E.116K}.

\begin{figure}
    \centering
    \includegraphics[width=.9\columnwidth]{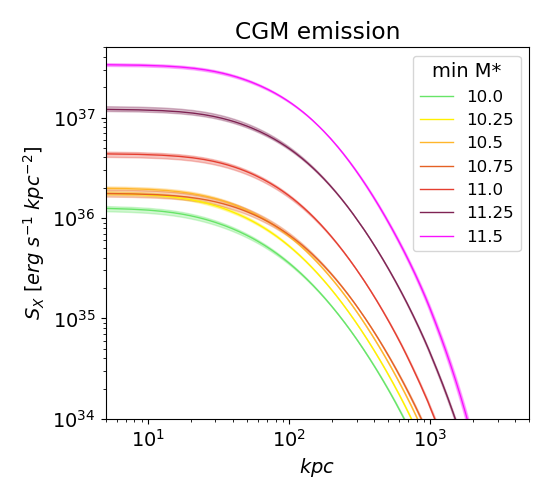}
\caption{\label{fig:model:comparison:CGM}Comparison of average hot gas surface brightness profiles (cross-correlation between events and central galaxies) inferred with the model for each sample. The differently colored, shaded areas represent different samples, identified here by the logarithm of their minimum stellar mass, in solar masses following the color scheme of Fig. \ref{fig:Vlim:samples:Mstar}. 
This emission average is made over a large range of halo masses. 
The average emission increases with the stellar mass threshold. 
We decompose the average emission as a function of halo mass in Fig. \ref{fig:CGM:profFRAC:SR} (see top row). 
}
\end{figure}

\begin{figure*}
    \centering
    \includegraphics[width=.9\columnwidth]{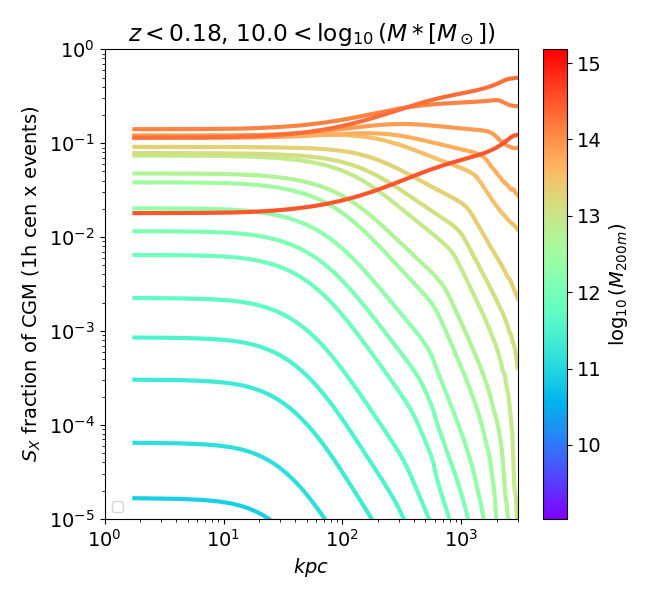}
    \includegraphics[width=.9\columnwidth]{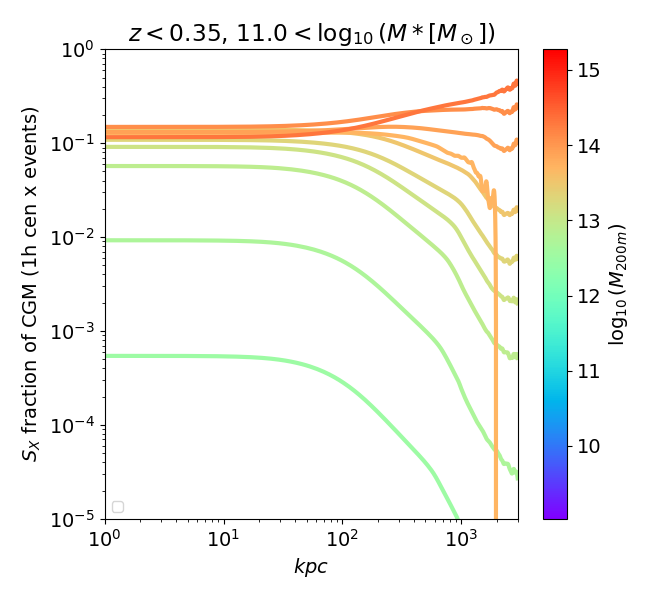}
    \includegraphics[width=.9\columnwidth]{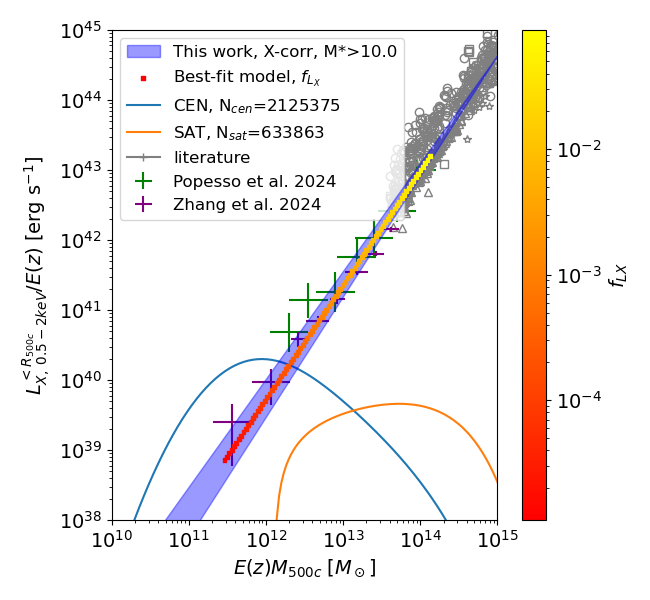}
    \includegraphics[width=.9\columnwidth]{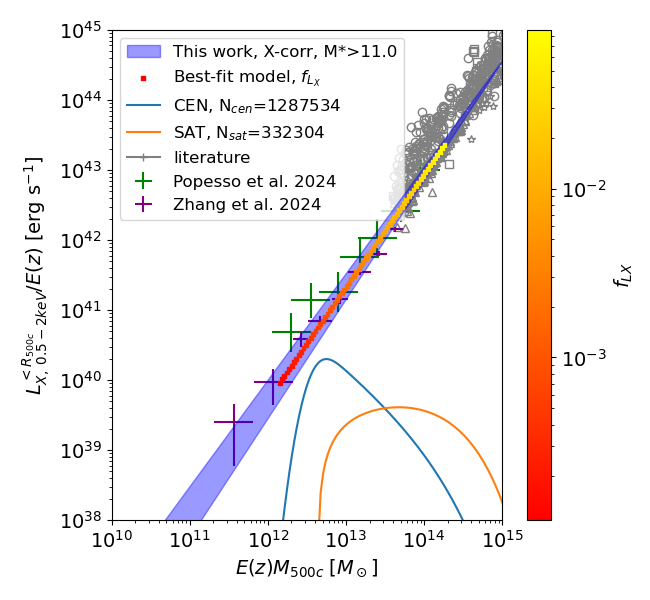} \caption{\label{fig:CGM:profFRAC:SR}\textbf{Top}: Decomposition in fractions of the average hot gas model surface brightness profile (see Fig. \ref{fig:model:comparison:CGM}) as a function of radius (separation in proper kpc) and halo mass (colorbar) for the samples with $\log_{10}(M^*/M_\odot)>10$ (left column) and with $\log_{10}(M^*/M_\odot)>11$ (right column). 
    It shows the different origin of the average emission as a function of separation. 
    The emission at large separation is clearly dominated by higher mass distinct halos. At small separations, smaller distinct halos also contribute to the average signal. 
    \textbf{Bottom}: Scaling relation between mass and X-ray luminosity. The posterior on the scaling relation obtained from fitting the cross-correlation (mainly sensitive to its slope) is depicted by the blue shaded area. On the left (right) panel, the width of the contour corresponds to the uncertainty on the scaling relation slope: $\alpha_{SR}= 1.629^{+0.091}_{-0.089}$ ($1.634^{+0.066}_{-0.094}$). 
    The shaded area overlaps with literature measurements shown with crosses and gray symbols \citep{LovisariReiprichSchellenberger_2015A&A...573A.118L, LovisariSchellenbergerSereno_2020ApJ...892..102L, MantzAllenMorris_2016MNRAS.456.4020M,AdamiGilesKoulouridis_2018A&A...620A...5A,SchellenbergerReiprich_2017MNRAS.469.3738S, BulbulChiuMohr_2019ApJ...871...50B, LiuBulbulGhirardini_2022A&A...661A...2L,ZhangComparatPonti_2024A&A...690A.268Z, PopessoMariniDolag_2024arXiv241117120P}. 
    Each galaxy sample is hosted by different populations of dark matter halos. The posterior normalized distribution (multiplied by $2\times10^{40}$ so that it appears in the panel) of halo masses, obtained with the HOD model, hosting central (satellite) galaxies is shown at the bottom of the panel with a blue (orange) solid line. 
    It shows that the constraint on the slope comes mostly from the central (satellite) galaxies at low (high) halo mass. 
    Fitting for a complete sample (including satellite) gives stronger constraints than using central galaxies alone, that are hampered by small statistics at high mass. 
    The integrated X-ray luminosity (one-halo central, see decomposition on the top row of panels) at a given halo mass contributes a fraction ($f_{L_{X}}$) to the average luminosity inferred (Table \ref{tab:xcorr:parameters}, last column) in the cross-correlation fit. 
    This fraction is shown using the red-yellow color bar. 
    On the bottom left panel, the color is limited to $10^{-5}$. The galaxies with halo masses smaller than 10$^{11}$M$_\odot$ have almost no contribution to the average luminosity. 
    }
    
\end{figure*}

\subsubsection{Satellites and two-halo term emission}

The emission from the two-halo term dominates on large scales, $R>3$Mpc. 
For the sample $\log_{10}(M^*/M_\odot)>10.5$, the model's two-halo term brightness at 5Mpc is 4$\times10^{34}$ erg kpc$^{-2}$ s$^{-1}$ and constitutes $\sim$100\% of the model emission. 
At 1Mpc, the model two-halo term brightness is less dominant with 8$\times10^{34}$ erg kpc$^{-2}$ s$^{-1}$, compared that of the one-halo satellite term at 2$\times10^{35}$ erg kpc$^{-2}$ s$^{-1}$ (the one-halo central term is much smaller: 5$\times10^{33}$ erg kpc$^{-2}$ s$^{-1}$). 
The two-halo term emission is directly related to the large-scale halo bias, defined as the clustering strength of halos compared to that of the underlying dark matter distribution, see Table \ref{tab:Vlim:samples:HODbestfitParams}.
Naturally, as stellar mass traces halo mass, it also correlates with the stellar mass threshold, see Fig. \ref{fig:model:components:2halo:sat} top panel. 

The emission from the cross-correlation between satellite galaxies and hot gas events is nearly identical in every sample (Fig. \ref{fig:model:components:2halo:sat} bottom panel), reflecting the fact that only the satellite galaxies living in clusters contribute to this correlation. 
Lower mass halos' satellite contributes much less due to the steepness of the mass-luminosity scaling relation. 
In that regard, masking clusters is a valid approach to discard most of this component. 

Importantly, this illustrates how subtracting a constant background value from a surface brightness profile depends on the large-scale structure in which galaxies are embedded \citep[see discussions in][]{ZhangComparatPonti_2024A&A...690A.268Z, ShreeramComparatMerloni_2024arXiv240910397S}. 

\begin{figure}
    \centering
    \includegraphics[width=.9\columnwidth]{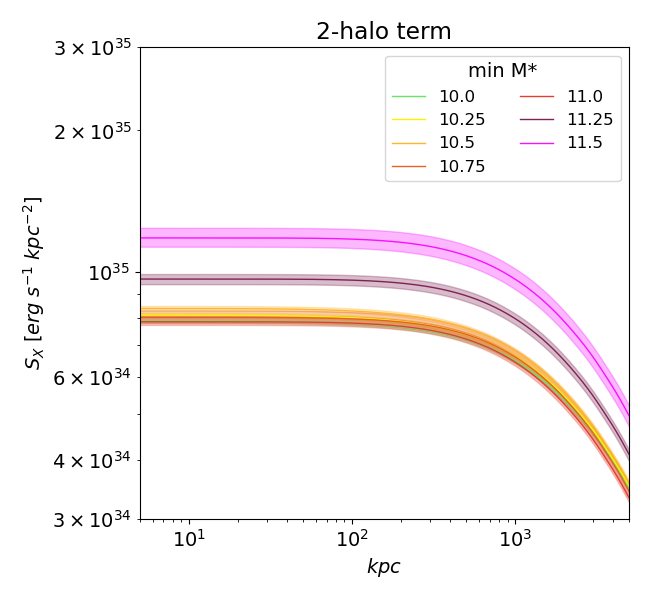}
    \includegraphics[width=.9\columnwidth]{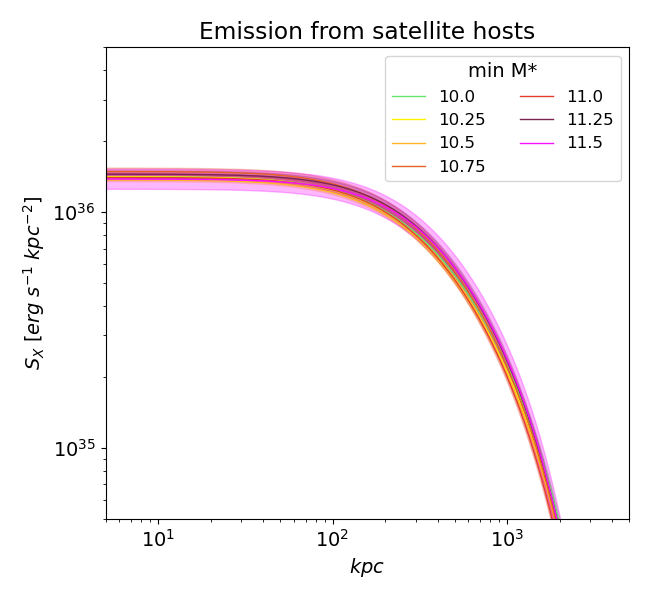}
    \caption{\label{fig:model:components:2halo:sat}Comparison of components obtained for each sample: the two-halo term (\textbf{top}) and satellite hosts (\textbf{bottom}). Differently colored shaded areas represent different samples, identified here by the logarithm of their minimum stellar mass, in solar masses following the color scheme of Fig. \ref{fig:Vlim:samples:Mstar}. 
    The evolution of the two-halo term (satellite) is linked to the evolution of the large-scale halo bias (mean halo mass hosting satellite galaxies), as shown in Table \ref{tab:Vlim:samples:HODbestfitParams}, column "b" ($\bar{M}_{sat}$). 
    }
    
\end{figure}

\section{Discussion and outlook}
\label{sec:discussion}

The measured galaxy auto-correlation and the galaxy-X-rays cross-correlation are rich in information. 
With this current combination of surveys, eROSITA + LS10, the statistical uncertainties on these summary statistics are at the percent level (or smaller).
The measurement may still suffer from systematic uncertainties of the order of 5\% (or even 10\% for the higher mass samples). 
To fully exploit this rich information source, understand its uncertainties and covariance, and gain insights on the connection between the large-scale structure and its dark matter halos, galaxies, hot gas, AGN, and XRB, future efforts must be put into the modeling. 

We discuss here two additional sources of information present in this data set, namely: its dependence on the galaxy classification (Sect. \ref{subsec:RSBC:dependence}) and on the energy (Sect. \ref{subsec:energy:dependence}). Both of these aspects may help in devising relevant future models.

\subsection{Dependence on galaxy classification: red-sequence, blue cloud and green valley galaxies}
\label{subsec:RSBC:dependence}

As pointed out by \citet{TruongPillepichNelson_2021MNRAS.508.1563T} the CGM and AGN properties may interestingly be correlated to the galaxy's specific star formation rate. 
We used a red sequence (RS) model calibrated on clusters of galaxies using redmapper \citep{RykoffRozoBusha_2014ApJ...785..104R, KlugeComparatLiu_2024A&A...688A.210K} and obtained its mean \textit{g-z} color (DECam) as a function of redshift, given in Table \ref{tab:rs:model}. 
Then we selected galaxies as being in the red sequence when they have a \textit{g-z} color above the mean of the RS minus $c_{RS}=0.15$ mag with $g_{AB}-z_{AB}>RS(z)-c_{RS}$. 
By making a comparison to GAMA \citep{DriverBellstedtRobotham_2022MNRAS.513..439D}, we found that the selection has an 80\% completeness and 66\% purity. A completeness of 90\% could be reached with $c_{RS}=0.23,$ but would correspond to a lower purity of 62\%. Maximizing purity is at the cost of completeness. With this method, the maximum attainable purity is 69\% for $c_{RS}=0.06,$ but this corresponds to a relatively low completeness 55\%. 
We selected blue cloud (BC) galaxies, as they are far away from the red sequence, with a \textit{g-z} color below the mean of the RS minus $c_{BC}=0.22$ mag with $g_{AB}-z_{AB}<RS(z)-c_{BC}$. This cut entails a completeness (purity) of 80\% (94\%). 
We classify the remaining galaxies (in between) as green valley galaxies. 
Appendix \ref{sec:app:SF:QU:selection} details the trade-off between completeness and purity. 

Figure \ref{fig:stacks:BC:RS} shows the fraction of emission in the stacks that originate from red-sequence, blue cloud, and green valley galaxies. 
We see how the profiles differ when selecting red-sequence and blue-cloud galaxies. 
For the high-mass samples, the emission is entirely dominated by that around red sequence galaxies that reside in groups and clusters. Indeed the quenched fraction steadily increases with stellar mass \citep[e.g.,][]{IlbertMcCrackenLeFevre_2013A&A...556A..55I, MuzzinMarchesiniStefanon_2013ApJ...777...18M}. 
For the lower mass sample, the decomposition shows interesting trends. The central surface brightness is equally split between the red sequence and the blue cloud. That vastly dominates the outer surface brightness (dominated by satellite emission) around red sequence galaxies. 
The red sequence galaxies that are not the brightest cluster galaxies (BCG), namely, the low-mass red galaxies, are not in the center of groups and clusters \citep[e.g.,][]{BensonDzanovicFrenk_2007MNRAS.379..841B,MoustakasCoilAird_2013ApJ...767...50M}. Therefore, the X-ray emission at their position is lower because the X-ray emission of clusters decreases with radius. Also, the X-ray emission further away from their emission is strong because it (also) includes the region close to the cluster center where the intracluster medium is the brightest.
This decomposition can inform a HOD model where the fraction of central and satellite quiescent galaxies is parametrized as a function of stellar and halo mass \citep[e.g.,][]{ZuMandelbaum_2016MNRAS.457.4360Z,BehrooziWechslerHearin_2019MNRAS.488.3143B, Tinker_2022AJ....163..126T}. 

\begin{figure*}
    \centering
\includegraphics[width=0.65\columnwidth]{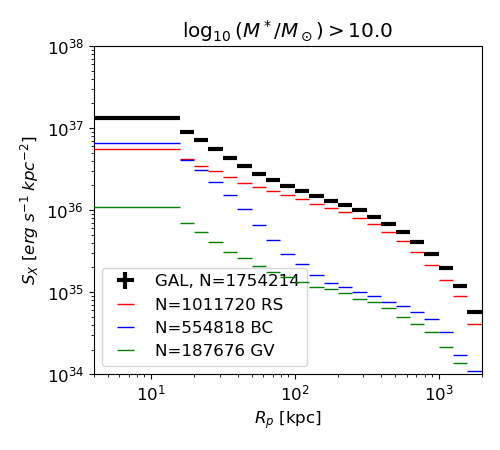}
\includegraphics[width=0.65\columnwidth]{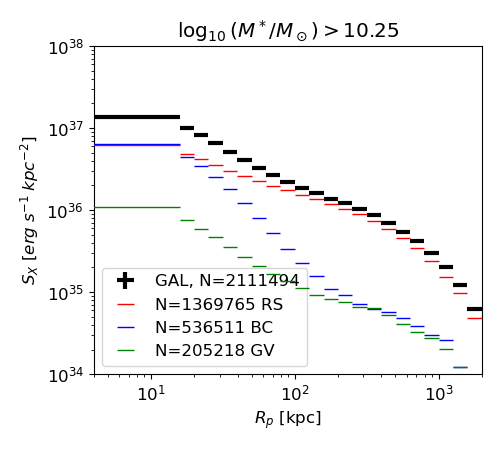}
\includegraphics[width=0.65\columnwidth]{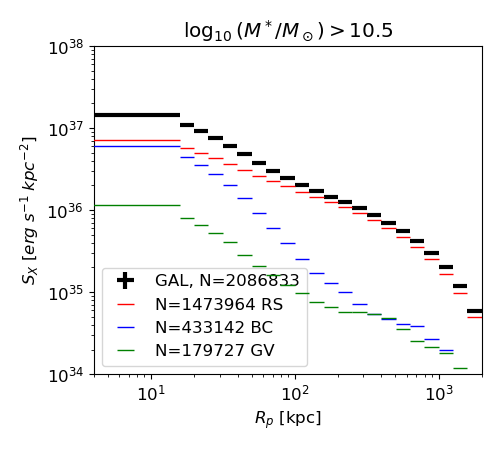}
\includegraphics[width=0.65\columnwidth]{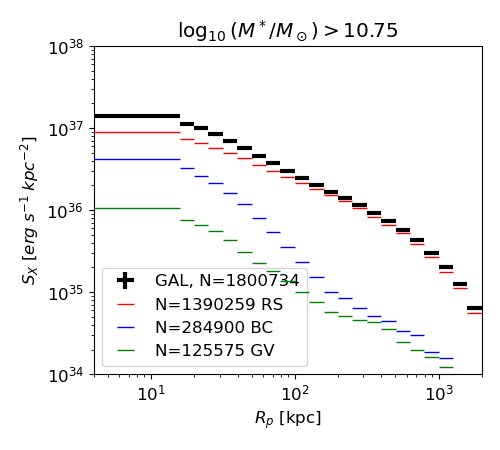}
\includegraphics[width=0.65\columnwidth]{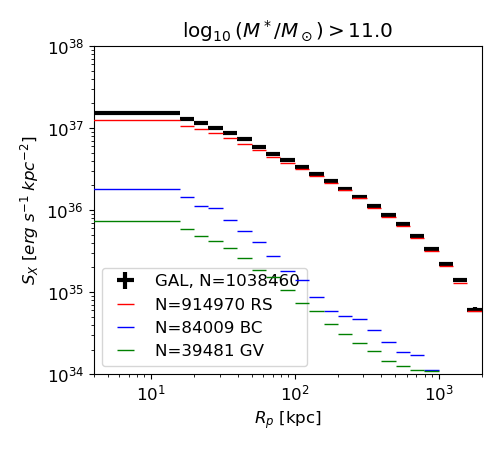}
\includegraphics[width=0.65\columnwidth]{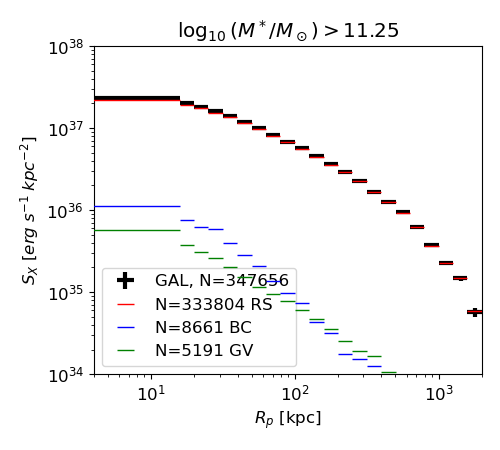}
    \caption{Stacked surface brightness profiles as a function of radius (proper kpc). The total emission (black lines, identical to that reported in colors in Fig. \ref{fig:xcorr:stacks}) is decomposed into the fractional contribution of the blue cloud (BC, blue), green valley (GV, green), and red sequence (RS, red) galaxies. The classification into these categories is detailed in Appendix \ref{sec:app:SF:QU:selection}. 
    This decomposition carries information about quenched fraction as a function host halo center and halo mass. 
    It may constrain future models of the galaxy-gas-AGN-halo connection as a function of specific star formation rate and in dense environment \citep[e.g.,][]{NishizawaOguriOogi_2018PASJ...70S..24N, AungNagaiKlypin_2023MNRAS.519.1648A, ShreeramComparatMerloni_2024arXiv240910397S}.
}
    \label{fig:stacks:BC:RS}
\end{figure*}

\subsection{Energy dependence of the cross-correlation}
\label{subsec:energy:dependence}

The eROSITA telescope measures photons and their energy to about 80 eV precision. 
For the sample with threshold $\log_{10}(M^*/M_\odot)>11$, we measured the cross-correlation in bins on 100 eV in the range 0.5-2 keV and find genuine differences as a function of scale (Fig. \ref{fig:Edep:110}). 
We found that the shape of the relative change of the cross-correlation varies with angular separation. 
This neatly connects with the physical components inferred by the model (point sources on small scales, hot gas on intermediate scales, and a two-halo term on large scales).  
The relative change does not follow the ARF, indicating its physical origin. 
This should be a composite effect. For example, low-energy photons may be dominated by CGM and low-mass halos, high-energy photons from AGNs, and massive halos. The redshift distribution of the sample and k-correction may also play a role here. 
The cross-correlation at 1keV is $\sim20\%$ larger than the broad band, while the one at 0.6 keV is 50\% lower.
This illustrates the extent of information available in the energy-dependent cross-correlation for interpretation with future models. 

\begin{figure}
    \centering
    \includegraphics[width=0.9\columnwidth]{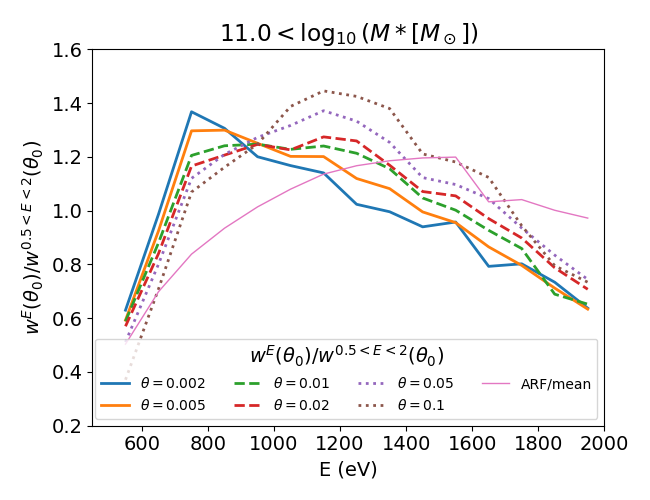}
    \includegraphics[width=0.95\columnwidth]{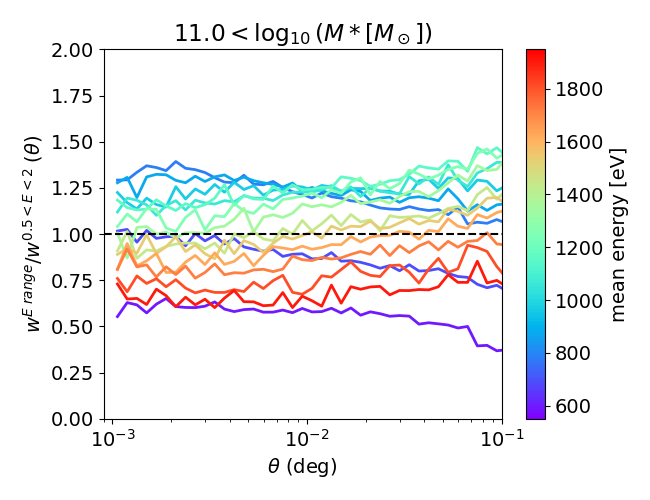}
    \caption{\label{fig:Edep:110} Illustration of the energy dependence of the galaxy-X-ray cross-correlation with the galaxy sample selected with stellar mass larger than $10^{11}M_\odot$ (see best-fit model of the broad band cross-correlation in Fig. \ref{fig:best:fit:stack:105} bottom right panel). 
    \textbf{Top}: Ratio of the cross-correlations at a fixed set of separations (meaning different sources of emission) as a function of energy.  
    We find that the spectral shape obtained depends on the separation. 
    At small angular separations of 0.002 and 0.005 degree, 30-75 kpc, (blue and orange solid lines), where the point source emission dominates, the ratio to the broad band increases between 600 and 800 eV and decreases thereafter. 
    At large separations (0.05 and 0.1 degree, 700-1400 kpc), where emission coming from satellite and the two-halo dominate (hot gas for large halos), the ratio increases until 1200 eV and decreases thereafter. 
    At intermediate separation (0.01 and 0.02 degree, 150-300 kpc) where the emission comes from half half from the CGM and the satellites, the evolution with energy is in-between the two. 
    Leveraging the energy dependence is key in future models to further ascertain the physical origin of the emitting components. 
    \textbf{Bottom}: Same ratio between the cross-correlations in different energy bins as a function of angular separation.}
\end{figure}

\subsection{On future observations}

In the future and as suggested by \citet{LaPostaAlonsoChisari_2024arXiv241212081L}, we anticipate that interesting synergies with lensing to constrain further the link between X-rays, galaxies, and their host dark matter halos \citep[e.g., with the Euclid and LSST experiments][]{Euclid2024arXiv240513491E, IvezicKahnTyson_2019ApJ...873..111I}.  In addition, synergies with the cross-correlation between the same galaxies and thermal SZ maps will further inform the hot gas profiles \citep[e.g., from ACT, CMB-4, SO, ][]{DasChiangMathur_2023ApJ...951..125D,CMB-S4_AbazajianAdsheadAhmed_2016arXiv161002743A, SO_AdeAguirreAhmed_2019JCAP...02..056A}.
Over time, with NewAthena \citep{NandraBarretBarcons_2013arXiv1306.2307N,CruiseGuainazziAird_2025NatAs...9...36C} AXIS \citep{mushotzky2019advanced}, and HUBS \citep{cui2020hubs}, these measurements will be extended to a redshift of $z\sim1$.

\section{Summary}
In this paper, we present the construction and validation of stellar mass selected (volume-limited) samples using the Legacy Survey (data release 10) catalogs limited to $r<19.5$ covering $\sim16,800$ deg$^2$ of the extra-galactic sky (Sect. \ref{sec:data}, Fig. \ref{fig:Vlim:samples:Mstar}) and companion mock catalogs (Appendix \ref{sec:mocks}). The selection of these galaxies was amply controlled (see Appendix \ref{sec:app:galaxy:density}). 
We  measured their two-point correlation function (Sect. \ref{subsec:measurements:galaxy:autocorr}, Fig. \ref{fig:wprp:final}). The agreement between the simulations and the observations is excellent. The statistical uncertainties are at the percent level, but systematical uncertainties may be as high as 5\%.
Quantifying the exact systematic uncertainty budget is left for future studies. 

We fit a HOD model (Sect. \ref{sec:method:halo:model}) to the galaxy auto-correlation function to retrieve the population of halos hosting them (Sect. \ref{sec:results}). We find that the four-parameter HOD model is sufficient to capture the information of interest and reproduce observations (Fig. \ref{fig:HOD:best:fit}). The deduced stellar-mass-to-halo-mass relation retrieved agrees with literature estimates (Fig. \ref{fig:SMHMR:hod}). 
The best-fit parameters obtained (Table \ref{tab:Vlim:samples:HODbestfitParams}) are sensible and in agreement with previous studies.

Using this controlled galaxy sample and its HOD model, we then measured and interpreted their cross-correlation, as well as their stacked surface brightness profiles, using the complete field of 0.5-2 keV X-ray photons observed by eROSITA over 2 years of scanning (Sects. \ref{sec:data} and \ref{sec:measurements}). 
The cross-correlation is measured to unprecedented accuracy thanks to the large area used. 
Its covariance matrix is mostly diagonal (Fig. \ref{fig:xcorr:gal:evt:raw}). 
With the stacked surface brightness profile, we constructed an alternative estimator of the cross-correlation function and obtained an excellent (better than 5\%) agreement for separations between 20-300 kpc (Fig. \ref{fig:xcorr:stacks:estimator}).
In this way, we extended the 9,000 deg$^2$ "FULL photometric" measurement from \citet{ZhangComparatPonti_2024A&A...690A.267Z} on the complete eROSITA extra-galactic sky of 13,000 deg$^2$. We increased the area by more than 40\% to include areas where the eROSITA's exposure times are longest. 
We cover  the theoretical link between stacked profile and the cross-correlation via a background measured at the 2\% percent level precision in detail (see Table \ref{tab:background:values}). 
The statistical uncertainty on the cross-correlation is unprecedented at the sub-percent level. The systematic uncertainties are estimated to be at the 5\% level, except for the higher mass samples, where it reaches the 10\% level. 
We used the complete set of events and a complete set of galaxies, making this measurement extremely stable and robust. 
These measurements constitute a key benchmark for galaxy evolution models linking galaxies, hot gas, and AGNs to their host dark matter halos.

We wrote a novel extension to the halo model to extract information from the cross-correlation between galaxies and X-rays (Sect. \ref{sec:method:halo:model}). 
The best-fit models (parameters in Table \ref{tab:xcorr:parameters}) are presented in Sect. \ref{sec:results} and Figs. \ref{fig:best:fit:stack:105}, \ref{fig:best:fit:stack:100}, and \ref{fig:best:fit:stack:110}. 
The models account for the observations and permit the decomposition of the observed signal. 
We obtained interesting posterior on the model components for the X-ray point sources (Fig. \ref{fig:model:ps:comparison}) and for the hot gas (Figs. \ref{fig:model:comparison:CGM}, \ref{fig:CGM:profFRAC:SR}), the satellite, and two-halo term (Fig. \ref{fig:model:components:2halo:sat}). 
The model posteriors are compatible with previous measurements from \citet{ZhangComparatPonti_2024A&A...690A.267Z, ZhangComparatPonti_2024A&A...690A.268Z, PopessoMariniDolag_2024arXiv241117120P}. 
We obtained a tight 5\% constraint on the slope of the mass-luminosity scaling relation over three orders of magnitude. 
The essential novelty is that we have grasped the complete picture of the hot gas emission within the complete large-scale structure, without any selection biases. 
This study opens a new venue for writing models that unify galaxies, AGNs, and their hot gas (in optical and X-rays), which can then be constrained by the observations presented here. 
Finally, we elaborate on the richness of the data presented by stacking red sequence and blue cloud galaxies (Fig. \ref{fig:stacks:BC:RS}),  showing the energy dependence of the cross-correlation (Fig. \ref{fig:Edep:110}). We demonstrate that these additional data are sensitive to the galaxy quiescent fraction. 

\section*{Data availability} 

The measurements presented are available on \href{https://zenodo.org/records/15111974}{Zenodo}. 
We hand out the relevant data arrays to support the development of future models: \textit{(i)} galaxy samples \textit{(ii)} auto and cross correlation functions, and stacked surface brightness profiles.\\

\begin{acknowledgements}
\label{sec:app:benchmark:products}

J.C. thanks J. Schaye, B. Oppenheimer, D. Nagai for insightful discussions on this analysis. \\

G.P. acknowledges financial support from the European Research Council (ERC) under the European Union's Horizon 2020 research and innovation program "Hot Milk" (grant agreement No. 865637) and support from the Framework per l'Attrazione e il Rafforzamento delle Eccellenze (FARE) per la ricerca in Italia (R20L5S39T9).

This work is based on data from eROSITA, the soft X-ray instrument aboard SRG, a joint Russian-German science mission supported by the Russian Space Agency (Roskosmos), in the interests of the Russian Academy of Sciences represented by its Space Research Institute (IKI), and the Deutsches Zentrum f\"ur Luft- und Raumfahrt (DLR). The SRG spacecraft was built by Lavochkin Association (NPOL) and its subcontractors, and is operated by NPOL with support from the Max Planck Institute for Extraterrestrial Physics (MPE).

The development and construction of the eROSITA X-ray instrument was led by MPE, with contributions from the Dr. Karl Remeis Observatory Bamberg \& ECAP (FAU Erlangen-N\"urnberg), the University of Hamburg Observatory, the Leibniz Institute for Astrophysics Potsdam (AIP), and the Institute for Astronomy and Astrophysics of the University of T\"ubingen, with the support of DLR and the Max Planck Society. The Argelander Institute for Astronomy of the University of Bonn and the Ludwig Maximilians Universit\"at Munich also participated in the science preparation for eROSITA. The eROSITA data shown here were processed using the eSASS software system developed by the German eROSITA consortium.

The Legacy Surveys consist of three individual and complementary projects: the Dark Energy Camera Legacy Survey (DECaLS; Proposal ID \#2014B-0404; PIs: David Schlegel and Arjun Dey), the Beijing-Arizona Sky Survey (BASS; NOAO Prop. ID \#2015A-0801; PIs: Zhou Xu and Xiaohui Fan), and the Mayall z-band Legacy Survey (MzLS; Prop. ID \#2016A-0453; PI: Arjun Dey). DECaLS, BASS and MzLS together include data obtained, respectively, at the Blanco telescope, Cerro Tololo Inter-American Observatory, NSF’s NOIRLab; the Bok telescope, Steward Observatory, University of Arizona; and the Mayall telescope, Kitt Peak National Observatory, NOIRLab. Pipeline processing and analyses of the data were supported by NOIRLab and the Lawrence Berkeley National Laboratory (LBNL). The Legacy Surveys project is honored to be permitted to conduct astronomical research on Iolkam Du’ag (Kitt Peak), a mountain with particular significance to the Tohono O’odham Nation.

NOIRLab is operated by the Association of Universities for Research in Astronomy (AURA) under a cooperative agreement with the National Science Foundation. LBNL is managed by the Regents of the University of California under contract to the U.S. Department of Energy.

This project used data obtained with the Dark Energy Camera (DECam), which was constructed by the Dark Energy Survey (DES) collaboration. Funding for the DES Projects has been provided by the U.S. Department of Energy, the U.S. National Science Foundation, the Ministry of Science and Education of Spain, the Science and Technology Facilities Council of the United Kingdom, the Higher Education Funding Council for England, the National Center for Supercomputing Applications at the University of Illinois at Urbana-Champaign, the Kavli Institute of Cosmological Physics at the University of Chicago, Center for Cosmology and Astro-Particle Physics at the Ohio State University, the Mitchell Institute for Fundamental Physics and Astronomy at Texas A\&M University, Financiadora de Estudos e Projetos, Fundacao Carlos Chagas Filho de Amparo, Financiadora de Estudos e Projetos, Fundacao Carlos Chagas Filho de Amparo a Pesquisa do Estado do Rio de Janeiro, Conselho Nacional de Desenvolvimento Cientifico e Tecnologico and the Ministerio da Ciencia, Tecnologia e Inovacao, the Deutsche Forschungsgemeinschaft and the Collaborating Institutions in the Dark Energy Survey. The Collaborating Institutions are Argonne National Laboratory, the University of California at Santa Cruz, the University of Cambridge, Centro de Investigaciones Energeticas, Medioambientales y Tecnologicas-Madrid, the University of Chicago, University College London, the DES-Brazil Consortium, the University of Edinburgh, the Eidgenossische Technische Hochschule (ETH) Zurich, Fermi National Accelerator Laboratory, the University of Illinois at Urbana-Champaign, the Institut de Ciencies de l’Espai (IEEC/CSIC), the Institut de Fisica d’Altes Energies, Lawrence Berkeley National Laboratory, the Ludwig Maximilians Universitat Munchen and the associated Excellence Cluster Universe, the University of Michigan, NSF’s NOIRLab, the University of Nottingham, the Ohio State University, the University of Pennsylvania, the University of Portsmouth, SLAC National Accelerator Laboratory, Stanford University, the University of Sussex, and Texas A\&M University.

BASS is a key project of the Telescope Access Program (TAP), which has been funded by the National Astronomical Observatories of China, the Chinese Academy of Sciences (the Strategic Priority Research Program “The Emergence of Cosmological Structures” Grant \# XDB09000000), and the Special Fund for Astronomy from the Ministry of Finance. The BASS is also supported by the External Cooperation Program of Chinese Academy of Sciences (Grant \# 114A11KYSB20160057), and Chinese National Natural Science Foundation (Grant \# 12120101003, \# 11433005).

The Legacy Survey team makes use of data products from the Near-Earth Object Wide-field Infrared Survey Explorer (NEOWISE), which is a project of the Jet Propulsion Laboratory/California Institute of Technology. NEOWISE is funded by the National Aeronautics and Space Administration.

The Legacy Surveys imaging of the DESI footprint is supported by the Director, Office of Science, Office of High Energy Physics of the U.S. Department of Energy under Contract No. DE-AC02-05CH1123, by the National Energy Research Scientific Computing Center, a DOE Office of Science User Facility under the same contract; and by the U.S. National Science Foundation, Division of Astronomical Sciences under Contract No. AST-0950945 to NOAO.

The Photometric Redshifts for the Legacy Surveys (PRLS) catalog used in this paper was produced thanks to funding from the U.S. Department of Energy Office of Science, Office of High Energy Physics via grant DE-SC0007914.

\end{acknowledgements}

\bibliographystyle{aa}
\bibliography{references}

\begin{thebibliography}{166}
\expandafter\ifx\csname natexlab\endcsname\relax\def\natexlab#1{#1}\fi

\bibitem[{{Abazajian} {et~al.}(2016){Abazajian}, {Adshead}, {Ahmed}, {Allen},
  {Alonso}, {Arnold}, {Baccigalupi}, {Bartlett}, {Battaglia}, {Benson}, \&
  et~al.}]{CMB-S4_AbazajianAdsheadAhmed_2016arXiv161002743A}
{Abazajian}, K.~N., {Adshead}, P., {Ahmed}, Z., {et~al.} 2016, arXiv e-prints,
  arXiv:1610.02743

\bibitem[{{Adami} {et~al.}(2018){Adami}, {Giles}, {Koulouridis}, {Pacaud},
  {Caretta}, {Pierre}, {Eckert}, {Ramos-Ceja}, {Gastaldello}, {Fotopoulou}, \&
  et~al.}]{AdamiGilesKoulouridis_2018A&A...620A...5A}
{Adami}, C., {Giles}, P., {Koulouridis}, E., {et~al.} 2018, \aap, 620, A5

\bibitem[{{Ade} {et~al.}(2019){Ade}, {Aguirre}, {Ahmed}, {Aiola}, {Ali},
  {Alonso}, {Alvarez}, {Arnold}, {Ashton}, {Austermann}, \&
  et~al.}]{SO_AdeAguirreAhmed_2019JCAP...02..056A}
{Ade}, P., {Aguirre}, J., {Ahmed}, Z., {et~al.} 2019, \jcap, 2019, 056

\bibitem[{{Aird} {et~al.}(2017){Aird}, {Coil}, \&
  {Georgakakis}}]{AirdCoilGeorgakakis_2017MNRAS.465.3390A}
{Aird}, J., {Coil}, A.~L., \& {Georgakakis}, A. 2017, \mnras, 465, 3390

\bibitem[{{Aird} {et~al.}(2015){Aird}, {Coil}, {Georgakakis}, {Nandra},
  {Barro}, \&
  {P{\'e}rez-Gonz{\'a}lez}}]{AirdCoilGeorgakakis_2015MNRAS.451.1892A}
{Aird}, J., {Coil}, A.~L., {Georgakakis}, A., {et~al.} 2015, \mnras, 451, 1892

\bibitem[{{Allevato} {et~al.}(2021){Allevato}, {Shankar}, {Marsden}, {Rasulov},
  {Viitanen}, {Georgakakis}, {Ferrara}, \&
  {Finoguenov}}]{AllevatoShankarMarsden_2021ApJ...916...34A}
{Allevato}, V., {Shankar}, F., {Marsden}, C., {et~al.} 2021, \apj, 916, 34

\bibitem[{{Anderson} {et~al.}(2015){Anderson}, {Gaspari}, {White}, {Wang}, \&
  {Dai}}]{AndersonGaspariWhite_2015MNRAS.449.3806A}
{Anderson}, M.~E., {Gaspari}, M., {White}, S. D.~M., {Wang}, W., \& {Dai}, X.
  2015, \mnras, 449, 3806

\bibitem[{{Asgari} {et~al.}(2023){Asgari}, {Mead}, \&
  {Heymans}}]{AsgariMeadHeymans_2023OJAp....6E..39A}
{Asgari}, M., {Mead}, A.~J., \& {Heymans}, C. 2023, The Open Journal of
  Astrophysics, 6, 39

\bibitem[{{Aung} {et~al.}(2023){Aung}, {Nagai}, {Klypin}, {Behroozi},
  {Abdullah}, {Ishiyama}, {Prada}, {P{\'e}rez}, {L{\'o}pez Cacheiro}, \&
  {Ruedas}}]{AungNagaiKlypin_2023MNRAS.519.1648A}
{Aung}, H., {Nagai}, D., {Klypin}, A., {et~al.} 2023, \mnras, 519, 1648

\bibitem[{{Bahar} {et~al.}(2024){Bahar}, {Bulbul}, {Ghirardini}, {Sanders},
  {Zhang}, {Liu}, {Clerc}, {Artis}, {Balzer}, {Biffi}, \&
  et~al.}]{BaharBulbulGhirardini_2024A&A...691A.188B}
{Bahar}, Y.~E., {Bulbul}, E., {Ghirardini}, V., {et~al.} 2024, \aap, 691, A188

\bibitem[{{Behroozi} {et~al.}(2019){Behroozi}, {Wechsler}, {Hearin}, \&
  {Conroy}}]{BehrooziWechslerHearin_2019MNRAS.488.3143B}
{Behroozi}, P., {Wechsler}, R.~H., {Hearin}, A.~P., \& {Conroy}, C. 2019,
  \mnras, 488, 3143

\bibitem[{{Behroozi} {et~al.}(2013){Behroozi}, {Wechsler}, \&
  {Wu}}]{BehrooziWechslerWu_2013ApJ...762..109B}
{Behroozi}, P.~S., {Wechsler}, R.~H., \& {Wu}, H.-Y. 2013, \apj, 762, 109

\bibitem[{{Benson} {et~al.}(2007){Benson}, {D{\v{z}}anovi{\'c}}, {Frenk}, \&
  {Sharples}}]{BensonDzanovicFrenk_2007MNRAS.379..841B}
{Benson}, A.~J., {D{\v{z}}anovi{\'c}}, D., {Frenk}, C.~S., \& {Sharples}, R.
  2007, \mnras, 379, 841

\bibitem[{{Berlfein} {et~al.}(2024){Berlfein}, {Mandelbaum}, {Dodelson}, \&
  {Schafer}}]{BerlfeinMandelbaumDodelson_2024MNRAS.531.4954B}
{Berlfein}, F., {Mandelbaum}, R., {Dodelson}, S., \& {Schafer}, C. 2024,
  \mnras, 531, 4954

\bibitem[{{Biffi} {et~al.}(2018){Biffi}, {Dolag}, \&
  {Merloni}}]{BiffiDolagMerloni_2018MNRAS.481.2213B}
{Biffi}, V., {Dolag}, K., \& {Merloni}, A. 2018, \mnras, 481, 2213

\bibitem[{{Brunner} {et~al.}(2022){Brunner}, {Liu}, {Lamer}, {Georgakakis},
  {Merloni}, {Brusa}, {Bulbul}, {Dennerl}, {Friedrich}, {Liu}, \&
  et~al.}]{BrunnerLiuLamer_2022A&A...661A...1B}
{Brunner}, H., {Liu}, T., {Lamer}, G., {et~al.} 2022, \aap, 661, A1

\bibitem[{{Buchner}(2016)}]{Buchner_2016S&C....26..383B}
{Buchner}, J. 2016, Statistics and Computing, 26, 383

\bibitem[{{Buchner}(2019)}]{Buchner_2019PASP..131j8005B}
{Buchner}, J. 2019, \pasp, 131, 108005

\bibitem[{{Buchner}(2021)}]{Buchner_2021JOSS....6.3001B}
{Buchner}, J. 2021, The Journal of Open Source Software, 6, 3001

\bibitem[{{Bulbul} {et~al.}(2019){Bulbul}, {Chiu}, {Mohr}, {McDonald},
  {Benson}, {Bautz}, {Bayliss}, {Bleem}, {Brodwin}, {Bocquet}, \&
  et~al.}]{BulbulChiuMohr_2019ApJ...871...50B}
{Bulbul}, E., {Chiu}, I.~N., {Mohr}, J.~J., {et~al.} 2019, \apj, 871, 50

\bibitem[{{Bulbul} {et~al.}(2024){Bulbul}, {Liu}, {Kluge}, {Zhang}, {Sanders},
  {Bahar}, {Ghirardini}, {Artis}, {Seppi}, {Garrel}, \&
  et~al.}]{BulbulLiuKluge_2024A&A...685A.106B}
{Bulbul}, E., {Liu}, A., {Kluge}, M., {et~al.} 2024, \aap, 685, A106

\bibitem[{{Chiu} {et~al.}(2022){Chiu}, {Ghirardini}, {Liu}, {Grandis},
  {Bulbul}, {Bahar}, {Comparat}, {Bocquet}, {Clerc}, {Klein}, \&
  et~al.}]{ChiuGhirardiniLiu_2022A&A...661A..11C}
{Chiu}, I.~N., {Ghirardini}, V., {Liu}, A., {et~al.} 2022, \aap, 661, A11

\bibitem[{{Comparat} {et~al.}(2020){Comparat}, {Eckert}, {Finoguenov},
  {Schmidt}, {Sanders}, {Nagai}, {Lau}, {K{\"a}}, {fer}, {Pacaud}, \&
  et~al.}]{ComparatEckertFinoguenov_2020OJAp....3E..13C}
{Comparat}, J., {Eckert}, D., {Finoguenov}, A., {et~al.} 2020, The Open Journal
  of Astrophysics, 3, 13

\bibitem[{{Comparat} {et~al.}(2023){Comparat}, {Luo}, {Merloni}, {More},
  {Salvato}, {Krumpe}, {Miyaji}, {Brandt}, {Georgakakis}, {Akiyama}, \&
  et~al.}]{ComparatLuoMerloni_2023A&A...673A.122C}
{Comparat}, J., {Luo}, W., {Merloni}, A., {et~al.} 2023, \aap, 673, A122

\bibitem[{{Comparat} {et~al.}(2019){Comparat}, {Merloni}, {Salvato}, {Nandra},
  {Boller}, {Georgakakis}, {Finoguenov}, {Dwelly}, {Buchner}, {Del Moro}, \&
  et~al.}]{ComparatMerloniSalvato_2019MNRAS.487.2005C}
{Comparat}, J., {Merloni}, A., {Salvato}, M., {et~al.} 2019, \mnras, 487, 2005

\bibitem[{{Comparat} {et~al.}(2022){Comparat}, {Truong}, {Merloni},
  {Pillepich}, {Ponti}, {Driver}, {Bellstedt}, {Liske}, {Aird}, {Br{\"u}ggen},
  \& et~al.}]{ComparatTruongMerloni_2022A&A...666A.156C}
{Comparat}, J., {Truong}, N., {Merloni}, A., {et~al.} 2022, \aap, 666, A156

\bibitem[{{Contreras} {et~al.}(2021){Contreras}, {Angulo}, \&
  {Zennaro}}]{ContrerasAnguloZennaro_2021MNRAS.504.5205C}
{Contreras}, S., {Angulo}, R.~E., \& {Zennaro}, M. 2021, \mnras, 504, 5205

\bibitem[{{Cooray} \& {Sheth}(2002)}]{CooraySheth_2002PhR...372....1C}
{Cooray}, A. \& {Sheth}, R. 2002, \physrep, 372, 1

\bibitem[{{Cruise} {et~al.}(2025){Cruise}, {Guainazzi}, {Aird}, {Carrera},
  {Costantini}, {Corrales}, {Dauser}, {Eckert}, {Gastaldello}, {Matsumoto}, \&
  et~al.}]{CruiseGuainazziAird_2025NatAs...9...36C}
{Cruise}, M., {Guainazzi}, M., {Aird}, J., {et~al.} 2025, Nature Astronomy, 9,
  36

\bibitem[{Cui {et~al.}(2020)Cui, Chen, Gao, Guo, Jin, Wang, Wang, Wang, Wang,
  Wang, {et~al.}}]{cui2020hubs}
Cui, W., Chen, L.-B., Gao, B., {et~al.} 2020, Journal of Low Temperature
  Physics, 199, 502

\bibitem[{{Das} {et~al.}(2023){Das}, {Chiang}, \&
  {Mathur}}]{DasChiangMathur_2023ApJ...951..125D}
{Das}, S., {Chiang}, Y.-K., \& {Mathur}, S. 2023, \apj, 951, 125

\bibitem[{{Dav{\'e}} {et~al.}(2019){Dav{\'e}}, {Angl{\'e}s-Alc{\'a}zar},
  {Narayanan}, {Li}, {Rafieferantsoa}, \&
  {Appleby}}]{DaveAngles-AlcazarNarayanan_2019MNRAS.486.2827D}
{Dav{\'e}}, R., {Angl{\'e}s-Alc{\'a}zar}, D., {Narayanan}, D., {et~al.} 2019,
  \mnras, 486, 2827

\bibitem[{{Davis} \& {Peebles}(1983)}]{DavisPeebles_1983ApJ...267..465D}
{Davis}, M. \& {Peebles}, P.~J.~E. 1983, \apj, 267, 465

\bibitem[{{de Jong}(2011)}]{deJong_2011Msngr.145...14D}
{de Jong}, R. 2011, The Messenger, 145, 14

\bibitem[{{de Jong} {et~al.}(2019){de Jong}, {Agertz}, {Berbel}, {Aird},
  {Alexander}, {Amarsi}, {Anders}, {Andrae}, {Ansarinejad}, {Ansorge}, \&
  et~al.}]{deJongAgertzBerbel_2019Msngr.175....3D}
{de Jong}, R.~S., {Agertz}, O., {Berbel}, A.~A., {et~al.} 2019, The Messenger,
  175, 3

\bibitem[{{Delubac} {et~al.}(2017){Delubac}, {Raichoor}, {Comparat}, {Jouvel},
  {Kneib}, {Y{\`e}che}, {Zou}, {Brownstein}, {Abdalla}, {Dawson}, \&
  et~al.}]{DelubacRaichoorComparat_2017MNRAS.465.1831D}
{Delubac}, T., {Raichoor}, A., {Comparat}, J., {et~al.} 2017, \mnras, 465, 1831

\bibitem[{{DESI Collaboration} {et~al.}(2016){DESI Collaboration}, {Aghamousa},
  {Aguilar}, {Ahlen}, {Alam}, {Allen}, {Allende Prieto}, {Annis}, {Bailey},
  {Balland}, \& et~al.}]{DESICollaborationAghamousaAguilar_2016arXiv161100036D}
{DESI Collaboration}, {Aghamousa}, A., {Aguilar}, J., {et~al.} 2016, arXiv
  e-prints, arXiv:1611.00036

\bibitem[{{Dey} {et~al.}(2019){Dey}, {Schlegel}, {Lang}, {Blum}, {Burleigh},
  {Fan}, {Findlay}, {Finkbeiner}, {Herrera}, {Juneau}, \&
  et~al.}]{DeySchlegelLang_2019AJ....157..168D}
{Dey}, A., {Schlegel}, D.~J., {Lang}, D., {et~al.} 2019, \aj, 157, 168

\bibitem[{{Diemer}(2018)}]{Diemer_2018ApJS..239...35D}
{Diemer}, B. 2018, \apjs, 239, 35

\bibitem[{{Driver} {et~al.}(2022){Driver}, {Bellstedt}, {Robotham}, {Baldry},
  {Davies}, {Liske}, {Obreschkow}, {Taylor}, {Wright}, {Alpaslan}, \&
  et~al.}]{DriverBellstedtRobotham_2022MNRAS.513..439D}
{Driver}, S.~P., {Bellstedt}, S., {Robotham}, A. S.~G., {et~al.} 2022, \mnras,
  513, 439

\bibitem[{{Driver} \& {Robotham}(2010)}]{DriverRobotham_2010MNRAS.407.2131D}
{Driver}, S.~P. \& {Robotham}, A. S.~G. 2010, \mnras, 407, 2131

\bibitem[{{Duffy} {et~al.}(2008){Duffy}, {Schaye}, {Kay}, \& {Dalla
  Vecchia}}]{DuffySchayeKay_2008MNRAS.390L..64D}
{Duffy}, A.~R., {Schaye}, J., {Kay}, S.~T., \& {Dalla Vecchia}, C. 2008,
  \mnras, 390, L64

\bibitem[{{Eckert} {et~al.}(2019){Eckert}, {Ghirardini}, {Ettori}, {Rasia},
  {Biffi}, {Pointecouteau}, {Rossetti}, {Molendi}, {Vazza}, {Gastaldello}, \&
  et~al.}]{EckertGhirardiniEttori_2019A&A...621A..40E}
{Eckert}, D., {Ghirardini}, V., {Ettori}, S., {et~al.} 2019, \aap, 621, A40

\bibitem[{{Ettori} {et~al.}(2019){Ettori}, {Ghirardini}, {Eckert},
  {Pointecouteau}, {Gastaldello}, {Sereno}, {Gaspari}, {Ghizzardi},
  {Roncarelli}, \& {Rossetti}}]{EttoriGhirardiniEckert_2019A&A...621A..39E}
{Ettori}, S., {Ghirardini}, V., {Eckert}, D., {et~al.} 2019, \aap, 621, A39

\bibitem[{{Euclid Collaboration} {et~al.}(2024){Euclid Collaboration},
  {Mellier}, {Abdurro'uf}, {Acevedo Barroso}, {Ach{\'u}carro}, {Adamek},
  {Adam}, {Addison}, {Aghanim}, {Aguena}, {Ajani}, {Akrami}, {Al-Bahlawan},
  {Alavi}, {Albuquerque}, {Alestas}, {Alguero}, {Allaoui}, {Allen}, {Allevato},
  {Alonso-Tetilla}, {Altieri}, {Alvarez-Candal}, {Alvi}, {Amara}, {Amendola},
  {Amiaux}, {Andika}, {Andreon}, {Andrews}, {Angora}, {Angulo}, {Annibali},
  {Anselmi}, {Anselmi}, {Arcari}, {Archidiacono}, {Aric{\`o}}, {Arnaud},
  {Arnouts}, {Asgari}, {Asorey}, {Atayde}, {Atek}, {Atrio-Barandela}, {Aubert},
  {Aubourg}, {Auphan}, {Auricchio}, {Aussel}, {Aussel}, {Avelino},
  {Avgoustidis}, {Avila}, {Awan}, {Azzollini}, {Baccigalupi}, {Bachelet},
  {Bacon}, {Baes}, {Bagley}, {Bahr-Kalus}, {Balaguera-Antolinez}, {Balbinot},
  {Balcells}, {Baldi}, {Baldry}, {Balestra}, {Ballardini}, {Ballester},
  {Balogh}, {Ba{\~n}ados}, {Barbier}, {Bardelli}, {Baron}, {Barreiro},
  {Barrena}, {Barriere}, {Barros}, {Barthelemy}, {Bartolo}, {Basset},
  {Battaglia}, {Battisti}, {Baugh}, {Baumont}, {Bazzanini}, {Beaulieu},
  {Beckmann}, {Belikov}, {Bel}, {Bellagamba}, {Bella}, {Bellini}, {Benabed},
  {Bender}, {Benevento}, {Bennett}, {Benson}, {Bergamini}, {Bermejo-Climent},
  {Bernardeau}, {Bertacca}, {Berthe}, {Berthier}, {Bethermin}, {Beutler},
  {Bevillon}, {Bhargava}, {Bhatawdekar}, {Bianchi}, {Bisigello}, {Biviano},
  {Blake}, {Blanchard}, {Blazek}, {Blot}, {Bosco}, {Bodendorf}, {Boenke},
  {B{\"o}hringer}, {Boldrini}, {Bolzonella}, {Bonchi}, {Bonici}, {Bonino},
  {Bonino}, {Bonvin}, {Bon}, {Booth}, {Borgani}, {Borlaff}, {Borsato}, {Bosco},
  {Bose}, {Botticella}, {Boucaud}, {Bouche}, {Boucher}, {Boutigny}, {Bouvard},
  {Bouwens}, {Bouy}, {Bowler}, {Bozza}, {Bozzo}, {Branchini}, {Brando},
  {Brau-Nogue}, {Brekke}, {Bremer}, {Brescia}, {Breton}, {Brinchmann},
  {Brinckmann}, {Brockley-Blatt}, {Brodwin}, {Brouard}, {Brown}, {Bruton},
  {Bucko}, {Buddelmeijer}, {Buenadicha}, {Buitrago}, {Burger}, {Burigana},
  {Busillo}, {Busonero}, {Cabanac}, {Cabayol-Garcia}, {Cagliari}, {Caillat},
  {Caillat}, {Calabrese}, {Calabro}, {Calderone}, {Calura}, {Camacho Quevedo},
  {Camera}, {Campos}, {Canas-Herrera}, {Candini}, {Cantiello}, {Capobianco},
  {Cappellaro}, {Cappelluti}, {Cappi}, {Caputi}, {Cara}, {Carbone}, {Cardone},
  {Carella}, {Carlberg}, {Carle}, {Carminati}, {Caro}, {Carrasco}, {Carretero},
  {Carrilho}, \& {Carron Duque}}]{Euclid2024arXiv240513491E}
{Euclid Collaboration}, {Mellier}, Y., {Abdurro'uf}, {et~al.} 2024, arXiv
  e-prints, arXiv:2405.13491

\bibitem[{{Farrow} {et~al.}(2015){Farrow}, {Cole}, {Norberg}, {Metcalfe},
  {Baldry}, {Bland-Hawthorn}, {Brown}, {Hopkins}, {Lacey}, {Liske}, \&
  et~al.}]{FarrowColeNorberg_2015MNRAS.454.2120F}
{Farrow}, D.~J., {Cole}, S., {Norberg}, P., {et~al.} 2015, \mnras, 454, 2120

\bibitem[{{Finoguenov} {et~al.}(2019){Finoguenov}, {Merloni}, {Comparat},
  {Nandra}, {Salvato}, {Tempel}, {Raichoor}, {Richard}, {Kneib}, {Pillepich},
  \& et~al.}]{FinoguenovMerloniComparat_2019Msngr.175...39F}
{Finoguenov}, A., {Merloni}, A., {Comparat}, J., {et~al.} 2019, The Messenger,
  175, 39

\bibitem[{{Flender} {et~al.}(2017){Flender}, {Nagai}, \&
  {McDonald}}]{FlenderNagaiMcDonald_2017ApJ...837..124F}
{Flender}, S., {Nagai}, D., \& {McDonald}, M. 2017, \apj, 837, 124

\bibitem[{{Freund} {et~al.}(2024){Freund}, {Czesla}, {Predehl}, {Robrade},
  {Salvato}, {Schneider}, {Starck}, {Wolf}, \&
  {Schmitt}}]{FreundCzeslaPredehl_2024A&A...684A.121F}
{Freund}, S., {Czesla}, S., {Predehl}, P., {et~al.} 2024, \aap, 684, A121

\bibitem[{{Georgakakis} {et~al.}(2017){Georgakakis}, {Aird}, {Schulze},
  {Dwelly}, {Salvato}, {Nandra}, {Merloni}, \&
  {Schneider}}]{GeorgakakisAirdSchulze_2017MNRAS.471.1976G}
{Georgakakis}, A., {Aird}, J., {Schulze}, A., {et~al.} 2017, \mnras, 471, 1976

\bibitem[{{Georgakakis} {et~al.}(2019){Georgakakis}, {Comparat}, {Merloni},
  {Ciesla}, {Aird}, \&
  {Finoguenov}}]{GeorgakakisComparatMerloni_2019MNRAS.487..275G}
{Georgakakis}, A., {Comparat}, J., {Merloni}, A., {et~al.} 2019, \mnras, 487,
  275

\bibitem[{{Hahn} {et~al.}(2023){Hahn}, {Wilson}, {Ruiz-Macias}, {Cole},
  {Weinberg}, {Moustakas}, {Kremin}, {Tinker}, {Smith}, {Wechsler}, \&
  et~al.}]{HahnWilsonRuiz-Macias_2023AJ....165..253H}
{Hahn}, C., {Wilson}, M.~J., {Ruiz-Macias}, O., {et~al.} 2023, \aj, 165, 253

\bibitem[{{Hopkins} {et~al.}(2006){Hopkins}, {Hernquist}, {Cox}, {Di Matteo},
  {Robertson}, \& {Springel}}]{HopkinsHernquistCox_2006ApJS..163....1H}
{Hopkins}, P.~F., {Hernquist}, L., {Cox}, T.~J., {et~al.} 2006, \apjs, 163, 1

\bibitem[{{Hopkins} {et~al.}(2025){Hopkins}, {Quataert}, {Ponnada}, \&
  {Silich}}]{HopkinsQuataertPonnada_2025arXiv250118696H}
{Hopkins}, P.~F., {Quataert}, E., {Ponnada}, S.~B., \& {Silich}, E. 2025, arXiv
  e-prints, arXiv:2501.18696

\bibitem[{{Ider Chitham} {et~al.}(2020){Ider Chitham}, {Comparat},
  {Finoguenov}, {Clerc}, {Kirkpatrick}, {Damsted}, {Kukkola}, {Capasso},
  {Nandra}, {Merloni}, \&
  et~al.}]{IderChithamComparatFinoguenov_2020MNRAS.499.4768I}
{Ider Chitham}, J., {Comparat}, J., {Finoguenov}, A., {et~al.} 2020, \mnras,
  499, 4768

\bibitem[{{Ilbert} {et~al.}(2006){Ilbert}, {Arnouts}, {McCracken},
  {Bolzonella}, {Bertin}, {Le F{\`e}vre}, {Mellier}, {Zamorani}, {Pell{\`o}},
  {Iovino}, \& et~al.}]{IlbertArnoutsMcCracken_2006A&A...457..841I}
{Ilbert}, O., {Arnouts}, S., {McCracken}, H.~J., {et~al.} 2006, \aap, 457, 841

\bibitem[{{Ilbert} {et~al.}(2013){Ilbert}, {McCracken}, {Le F{\`e}vre},
  {Capak}, {Dunlop}, {Karim}, {Renzini}, {Caputi}, {Boissier}, {Arnouts}, \&
  et~al.}]{IlbertMcCrackenLeFevre_2013A&A...556A..55I}
{Ilbert}, O., {McCracken}, H.~J., {Le F{\`e}vre}, O., {et~al.} 2013, \aap, 556,
  A55

\bibitem[{{Ishiyama} {et~al.}(2021){Ishiyama}, {Prada}, {Klypin}, {Sinha},
  {Metcalf}, {Jullo}, {Altieri}, {Cora}, {Croton}, {de la Torre}, \&
  et~al.}]{IshiyamaPradaKlypin_2021MNRAS.506.4210I}
{Ishiyama}, T., {Prada}, F., {Klypin}, A.~A., {et~al.} 2021, \mnras, 506, 4210

\bibitem[{{Ivezi{\'c}} {et~al.}(2019){Ivezi{\'c}}, {Kahn}, {Tyson}, {Abel},
  {Acosta}, {Allsman}, {Alonso}, {AlSayyad}, {Anderson}, {Andrew}, \&
  et~al.}]{IvezicKahnTyson_2019ApJ...873..111I}
{Ivezi{\'c}}, {\v{Z}}., {Kahn}, S.~M., {Tyson}, J.~A., {et~al.} 2019, \apj,
  873, 111

\bibitem[{{Kaminsky} {et~al.}(2025){Kaminsky}, {Cappelluti}, {Hasinger},
  {Peca}, {Casey}, {Drakos}, {Faisst}, {Gozaliasl}, {Ilbert}, {Kartaltepe}, \&
  et~al.}]{KaminskyCappellutiHasinger_2025arXiv250209705K}
{Kaminsky}, A., {Cappelluti}, N., {Hasinger}, G., {et~al.} 2025, arXiv
  e-prints, arXiv:2502.09705

\bibitem[{{K{\'e}ruzor{\'e}} {et~al.}(2024){K{\'e}ruzor{\'e}}, {Bleem},
  {Frontiere}, {Krishnan}, {Buehlmann}, {Emberson}, {Habib}, \&
  {Larsen}}]{KeruzoreBleemFrontiere_2024OJAp....7E.116K}
{K{\'e}ruzor{\'e}}, F., {Bleem}, L.~E., {Frontiere}, N., {et~al.} 2024, The
  Open Journal of Astrophysics, 7, 116

\bibitem[{{Kluge} {et~al.}(2024){Kluge}, {Comparat}, {Liu}, {Balzer}, {Bulbul},
  {Ider Chitham}, {Ghirardini}, {Garrel}, {Bahar}, {Artis}, \&
  et~al.}]{KlugeComparatLiu_2024A&A...688A.210K}
{Kluge}, M., {Comparat}, J., {Liu}, A., {et~al.} 2024, \aap, 688, A210

\bibitem[{{Kong} {et~al.}(2020){Kong}, {Burleigh}, {Ross}, {Moustakas},
  {Chuang}, {Comparat}, {de Mattia}, {du Mas des Bourboux}, {Honscheid}, {Lin},
  \& et~al.}]{KongBurleighRoss_2020MNRAS.499.3943K}
{Kong}, H., {Burleigh}, K.~J., {Ross}, A., {et~al.} 2020, \mnras, 499, 3943

\bibitem[{{Koulouridis} {et~al.}(2024){Koulouridis}, {Gkini}, \&
  {Drigga}}]{KoulouridisGkiniDrigga_2024A&A...684A.111K}
{Koulouridis}, E., {Gkini}, A., \& {Drigga}, E. 2024, \aap, 684, A111

\bibitem[{{Kyritsis} {et~al.}(2025){Kyritsis}, {Zezas}, {Haberl}, {Weber},
  {Basu-Zych}, {Vulic}, {Maitra}, {H{\"a}mmerich}, {Laktionov}, {Wilms}, \&
  et~al.}]{KyritsisZezasHaberl_2025A&A...694A.128K}
{Kyritsis}, E., {Zezas}, A., {Haberl}, F., {et~al.} 2025, \aap, 694, A128

\bibitem[{{La Posta} {et~al.}(2024){La Posta}, {Alonso}, {Chisari}, {Ferreira},
  \& {Garc{\'\i}a-Garc{\'\i}a}}]{LaPostaAlonsoChisari_2024arXiv241212081L}
{La Posta}, A., {Alonso}, D., {Chisari}, N.~E., {Ferreira}, T., \&
  {Garc{\'\i}a-Garc{\'\i}a}, C. 2024, arXiv e-prints, arXiv:2412.12081

\bibitem[{{Landy} \& {Szalay}(1993)}]{LandySzalay_1993ApJ...412...64L}
{Landy}, S.~D. \& {Szalay}, A.~S. 1993, \apj, 412, 64

\bibitem[{{Lau} {et~al.}(2024){Lau}, {Nagai}, {Bogd{\'a}n}, {Medlock},
  {Oppenheimer}, {Battaglia}, {Angl{\'e}s-Alc{\'a}zar}, {Genel}, {Ni}, \&
  {Villaescusa-Navarro}}]{LauNagaiBogdan_2024arXiv241204559L}
{Lau}, E.~T., {Nagai}, D., {Bogd{\'a}n}, {\'A}., {et~al.} 2024, arXiv e-prints,
  arXiv:2412.04559

\bibitem[{{Lau} {et~al.}(2025){Lau}, {Nagai}, {Farahi}, {Ishiyama}, {Miyatake},
  {Osato}, \& {Shirasaki}}]{LauNagaiFarahi_2025ApJ...980..122L}
{Lau}, E.~T., {Nagai}, D., {Farahi}, A., {et~al.} 2025, \apj, 980, 122

\bibitem[{{Leauthaud} {et~al.}(2015){Leauthaud}, {J. Benson}, {Civano}, {L.
  Coil}, {Bundy}, {Massey}, {Schramm}, {Schulze}, {Capak}, {Elvis}, \&
  et~al.}]{LeauthaudJ.BensonCivano_2015MNRAS.446.1874L}
{Leauthaud}, A., {J. Benson}, A., {Civano}, F., {et~al.} 2015, \mnras, 446,
  1874

\bibitem[{{Lehmer} {et~al.}(2016){Lehmer}, {Basu-Zych}, {Mineo}, {Brandt},
  {Eufrasio}, {Fragos}, {Hornschemeier}, {Luo}, {Xue}, {Bauer}, \&
  et~al.}]{LehmerBasu-ZychMineo_2016ApJ...825....7L}
{Lehmer}, B.~D., {Basu-Zych}, A.~R., {Mineo}, S., {et~al.} 2016, \apj, 825, 7

\bibitem[{{Lewis} {et~al.}(2000){Lewis}, {Challinor}, \&
  {Lasenby}}]{LewisChallinorLasenby_2000ApJ...538..473L}
{Lewis}, A., {Challinor}, A., \& {Lasenby}, A. 2000, \apj, 538, 473

\bibitem[{{Limber}(1953)}]{Limber_1953ApJ...117..134L}
{Limber}, D.~N. 1953, \apj, 117, 134

\bibitem[{{Liu} {et~al.}(2022){Liu}, {Bulbul}, {Ghirardini}, {Liu}, {Klein},
  {Clerc}, {{\"O}zsoy}, {Ramos-Ceja}, {Pacaud}, {Comparat}, \&
  et~al.}]{LiuBulbulGhirardini_2022A&A...661A...2L}
{Liu}, A., {Bulbul}, E., {Ghirardini}, V., {et~al.} 2022, \aap, 661, A2

\bibitem[{{Liu} {et~al.}(2024){Liu}, {Bulbul}, {Kluge}, {Ghirardini}, {Zhang},
  {Sanders}, {Artis}, {Bahar}, {Balzer}, {Br{\"u}ggen}, \&
  et~al.}]{LiuBulbulKluge_2024A&A...683A.130L}
{Liu}, A., {Bulbul}, E., {Kluge}, M., {et~al.} 2024, \aap, 683, A130

\bibitem[{{Locatelli} {et~al.}(2024){Locatelli}, {Ponti}, {Zheng}, {Merloni},
  {Becker}, {Comparat}, {Dennerl}, {Freyberg}, {Sasaki}, \&
  {Yeung}}]{LocatelliPontiZheng_2024A&A...681A..78L}
{Locatelli}, N., {Ponti}, G., {Zheng}, X., {et~al.} 2024, \aap, 681, A78

\bibitem[{{Lovisari} {et~al.}(2021){Lovisari}, {Ettori}, {Gaspari}, \&
  {Giles}}]{LovisariEttoriGaspari_2021Univ....7..139L}
{Lovisari}, L., {Ettori}, S., {Gaspari}, M., \& {Giles}, P.~A. 2021, Universe,
  7, 139

\bibitem[{{Lovisari} {et~al.}(2015){Lovisari}, {Reiprich}, \&
  {Schellenberger}}]{LovisariReiprichSchellenberger_2015A&A...573A.118L}
{Lovisari}, L., {Reiprich}, T.~H., \& {Schellenberger}, G. 2015, \aap, 573,
  A118

\bibitem[{{Lovisari} {et~al.}(2020){Lovisari}, {Schellenberger}, {Sereno},
  {Ettori}, {Pratt}, {Forman}, {Jones}, {Andrade-Santos}, {Randall}, \&
  {Kraft}}]{LovisariSchellenbergerSereno_2020ApJ...892..102L}
{Lovisari}, L., {Schellenberger}, G., {Sereno}, M., {et~al.} 2020, \apj, 892,
  102

\bibitem[{{Mantz} {et~al.}(2016){Mantz}, {Allen}, {Morris}, \&
  {Schmidt}}]{MantzAllenMorris_2016MNRAS.456.4020M}
{Mantz}, A.~B., {Allen}, S.~W., {Morris}, R.~G., \& {Schmidt}, R.~W. 2016,
  \mnras, 456, 4020

\bibitem[{{Marini} {et~al.}(2024){Marini}, {Popesso}, {Lamer}, {Dolag},
  {Biffi}, {Vladutescu-Zopp}, {Dev}, {Toptun}, {Bulbul}, {Comparat}, \&
  et~al.}]{MariniPopessoLamer_2024A&A...689A...7M}
{Marini}, I., {Popesso}, P., {Lamer}, G., {et~al.} 2024, \aap, 689, A7

\bibitem[{{Martini} {et~al.}(2013){Martini}, {Miller}, {Brodwin}, {Stanford},
  {Gonzalez}, {Bautz}, {Hickox}, {Stern}, {Eisenhardt}, {Galametz}, \&
  et~al.}]{MartiniMillerBrodwin_2013ApJ...768....1M}
{Martini}, P., {Miller}, E.~D., {Brodwin}, M., {et~al.} 2013, \apj, 768, 1

\bibitem[{{McAlpine} {et~al.}(2016){McAlpine}, {Helly}, {Schaller}, {Trayford},
  {Qu}, {Furlong}, {Bower}, {Crain}, {Schaye}, {Theuns}, \&
  et~al.}]{McAlpineHellySchaller_2016A&C....15...72M}
{McAlpine}, S., {Helly}, J.~C., {Schaller}, M., {et~al.} 2016, Astronomy and
  Computing, 15, 72

\bibitem[{{Mead} \& {Asgari}(2023)}]{MeadAsgari_2023ascl.soft07025M}
{Mead}, A.~J. \& {Asgari}, M. 2023, {pyhalomodel: Halo-model implementation for
  power spectra}, Astrophysics Source Code Library, record ascl:2307.025

\bibitem[{{Merloni} {et~al.}(2024){Merloni}, {Lamer}, {Liu}, {Ramos-Ceja},
  {Brunner}, {Bulbul}, {Dennerl}, {Doroshenko}, {Freyberg}, {Friedrich}, \&
  et~al.}]{MerloniLamerLiu_2024A&A...682A..34M}
{Merloni}, A., {Lamer}, G., {Liu}, T., {et~al.} 2024, \aap, 682, A34

\bibitem[{{Merloni} {et~al.}(2012){Merloni}, {Predehl}, {Becker},
  {B{\"o}hringer}, {Boller}, {Brunner}, {Brusa}, {Dennerl}, {Freyberg},
  {Friedrich}, \& et~al.}]{MerloniPredehlBecker_2012arXiv1209.3114M}
{Merloni}, A., {Predehl}, P., {Becker}, W., {et~al.} 2012, arXiv e-prints,
  arXiv:1209.3114

\bibitem[{{More} {et~al.}(2015){More}, {Miyatake}, {Mandelbaum}, {Takada},
  {Spergel}, {Brownstein}, \&
  {Schneider}}]{MoreMiyatakeMandelbaum_2015ApJ...806....2M}
{More}, S., {Miyatake}, H., {Mandelbaum}, R., {et~al.} 2015, \apj, 806, 2

\bibitem[{{Moustakas} {et~al.}(2013){Moustakas}, {Coil}, {Aird}, {Blanton},
  {Cool}, {Eisenstein}, {Mendez}, {Wong}, {Zhu}, \&
  {Arnouts}}]{MoustakasCoilAird_2013ApJ...767...50M}
{Moustakas}, J., {Coil}, A.~L., {Aird}, J., {et~al.} 2013, \apj, 767, 50

\bibitem[{{Moustakas} {et~al.}(2023){Moustakas}, {Lang}, {Dey}, {Juneau},
  {Meisner}, {Myers}, {Schlafly}, {Schlegel}, {Valdes}, {Weaver}, \&
  et~al.}]{MoustakasLangDey_2023ApJS..269....3M}
{Moustakas}, J., {Lang}, D., {Dey}, A., {et~al.} 2023, \apjs, 269, 3

\bibitem[{{Mu{\~n}oz Rodr{\'\i}guez} {et~al.}(2024){Mu{\~n}oz Rodr{\'\i}guez},
  {Georgakakis}, {Shankar}, {Ruiz}, {Bonoli}, {Comparat}, {Fu}, {Koulouridis},
  {Lapi}, \& {Almeida}}]{MunozRodriguezGeorgakakisShankar_2024MNRAS.532..336M}
{Mu{\~n}oz Rodr{\'\i}guez}, I., {Georgakakis}, A., {Shankar}, F., {et~al.}
  2024, \mnras, 532, 336

\bibitem[{{Murray} {et~al.}(2021){Murray}, {Diemer}, {Chen}, {Neuhold},
  {Schnapp}, {Peruzzi}, {Blevins}, \&
  {Engelman}}]{MurrayDiemerChen_2021A&C....3600487M}
{Murray}, S.~G., {Diemer}, B., {Chen}, Z., {et~al.} 2021, Astronomy and
  Computing, 36, 100487

\bibitem[{Mushotzky {et~al.}(2019)Mushotzky, Aird, Barger, Cappelluti, Chartas,
  Corrales, Eufrasio, Fabian, Falcone, Gallo, {et~al.}}]{mushotzky2019advanced}
Mushotzky, R.~F., Aird, J., Barger, A.~J., {et~al.} 2019, arXiv preprint
  arXiv:1903.04083

\bibitem[{{Muzzin} {et~al.}(2013){Muzzin}, {Marchesini}, {Stefanon}, {Franx},
  {McCracken}, {Milvang-Jensen}, {Dunlop}, {Fynbo}, {Brammer}, {Labb{\'e}}, \&
  et~al.}]{MuzzinMarchesiniStefanon_2013ApJ...777...18M}
{Muzzin}, A., {Marchesini}, D., {Stefanon}, M., {et~al.} 2013, \apj, 777, 18

\bibitem[{{Myers} {et~al.}(2023){Myers}, {Moustakas}, {Bailey}, {Weaver},
  {Cooper}, {Forero-Romero}, {Abolfathi}, {Alexander}, {Brooks}, {Chaussidon},
  \& et~al.}]{MyersMoustakasBailey_2023AJ....165...50M}
{Myers}, A.~D., {Moustakas}, J., {Bailey}, S., {et~al.} 2023, \aj, 165, 50

\bibitem[{{Nandra} {et~al.}(2013){Nandra}, {Barret}, {Barcons}, {Fabian}, {den
  Herder}, {Piro}, {Watson}, {Adami}, {Aird}, {Afonso}, \&
  et~al.}]{NandraBarretBarcons_2013arXiv1306.2307N}
{Nandra}, K., {Barret}, D., {Barcons}, X., {et~al.} 2013, arXiv e-prints,
  arXiv:1306.2307

\bibitem[{{Navarro} {et~al.}(1997){Navarro}, {Frenk}, \&
  {White}}]{NavarroFrenkWhite_1997ApJ...490..493N}
{Navarro}, J.~F., {Frenk}, C.~S., \& {White}, S. D.~M. 1997, \apj, 490, 493

\bibitem[{{Neumann} \& {Arnaud}(1999)}]{NeumannArnaud_1999A&A...348..711N}
{Neumann}, D.~M. \& {Arnaud}, M. 1999, \aap, 348, 711

\bibitem[{{Newsam} {et~al.}(1999){Newsam}, {McHardy}, {Jones}, \&
  {Mason}}]{NewsamMcHardyJones_1999MNRAS.310..255N}
{Newsam}, A.~M., {McHardy}, I.~M., {Jones}, L.~R., \& {Mason}, K.~O. 1999,
  \mnras, 310, 255

\bibitem[{{Nishimichi} {et~al.}(2019){Nishimichi}, {Takada}, {Takahashi},
  {Osato}, {Shirasaki}, {Oogi}, {Miyatake}, {Oguri}, {Murata}, {Kobayashi}, \&
  et~al.}]{NishimichiTakadaTakahashi_2019ApJ...884...29N}
{Nishimichi}, T., {Takada}, M., {Takahashi}, R., {et~al.} 2019, \apj, 884, 29

\bibitem[{{Nishizawa} {et~al.}(2018){Nishizawa}, {Oguri}, {Oogi}, {More},
  {Nishimichi}, {Nagashima}, {Lin}, {Mandelbaum}, {Takada}, {Bahcall}, \&
  et~al.}]{NishizawaOguriOogi_2018PASJ...70S..24N}
{Nishizawa}, A.~J., {Oguri}, M., {Oogi}, T., {et~al.} 2018, \pasj, 70, S24

\bibitem[{{Oppenheimer} \&
  {Schaye}(2013)}]{OppenheimerSchaye_2013MNRAS.434.1043O}
{Oppenheimer}, B.~D. \& {Schaye}, J. 2013, \mnras, 434, 1043

\bibitem[{{Oren} {et~al.}(2024){Oren}, {Sternberg}, {McKee}, {Faerman}, \&
  {Genel}}]{OrenSternbergMcKee_2024ApJ...974..291O}
{Oren}, Y., {Sternberg}, A., {McKee}, C.~F., {Faerman}, Y., \& {Genel}, S.
  2024, \apj, 974, 291

\bibitem[{{Pillepich} {et~al.}(2018){Pillepich}, {Nelson}, {Hernquist},
  {Springel}, {Pakmor}, {Torrey}, {Weinberger}, {Genel}, {Naiman}, {Marinacci},
  \& et~al.}]{PillepichNelsonHernquist_2018MNRAS.475..648P}
{Pillepich}, A., {Nelson}, D., {Hernquist}, L., {et~al.} 2018, \mnras, 475, 648

\bibitem[{{Planck Collaboration} {et~al.}(2020){Planck Collaboration},
  {Aghanim}, {Akrami}, {Ashdown}, {Aumont}, {Baccigalupi}, {Ballardini},
  {Banday}, {Barreiro}, {Bartolo}, \&
  et~al.}]{PlanckCollaborationAghanimAkrami_2020A&A...641A...6P}
{Planck Collaboration}, {Aghanim}, N., {Akrami}, Y., {et~al.} 2020, \aap, 641,
  A6

\bibitem[{{Poggianti} {et~al.}(2017){Poggianti}, {Jaff{\'e}}, {Moretti},
  {Gullieuszik}, {Radovich}, {Tonnesen}, {Fritz}, {Bettoni}, {Vulcani},
  {Fasano}, \& et~al.}]{PoggiantiJaffeMoretti_2017Natur.548..304P}
{Poggianti}, B.~M., {Jaff{\'e}}, Y.~L., {Moretti}, A., {et~al.} 2017, \nat,
  548, 304

\bibitem[{{Ponti} {et~al.}(2023{\natexlab{a}}){Ponti}, {Sanders}, {Locatelli},
  {Zheng}, {Zhang}, {Yeung}, {Freyberg}, {Dennerl}, {Comparat}, {Merloni}, \&
  et~al.}]{PontiSandersLocatelli_2023A&A...670A..99P}
{Ponti}, G., {Sanders}, J.~S., {Locatelli}, N., {et~al.} 2023{\natexlab{a}},
  \aap, 670, A99

\bibitem[{{Ponti} {et~al.}(2023{\natexlab{b}}){Ponti}, {Zheng}, {Locatelli},
  {Bianchi}, {Zhang}, {Anastasopoulou}, {Comparat}, {Dennerl}, {Freyberg},
  {Haberl}, \& et~al.}]{PontiZhengLocatelli_2023A&A...674A.195P}
{Ponti}, G., {Zheng}, X., {Locatelli}, N., {et~al.} 2023{\natexlab{b}}, \aap,
  674, A195

\bibitem[{{Pop} {et~al.}(2022){Pop}, {Hernquist}, {Nagai}, {Kannan},
  {Weinberger}, {Springel}, {Vogelsberger}, {Nelson}, {Pakmor}, {Pillepich}, \&
  et~al.}]{PopHernquistNagai_2022arXiv220511528P}
{Pop}, A.-R., {Hernquist}, L., {Nagai}, D., {et~al.} 2022, arXiv e-prints,
  arXiv:2205.11528

\bibitem[{{Popesso} {et~al.}(2024{\natexlab{a}}){Popesso}, {Biviano}, {Bulbul},
  {Merloni}, {Comparat}, {Clerc}, {Igo}, {Liu}, {Driver}, {Salvato}, \&
  et~al.}]{PopessoBivianoBulbul_2024MNRAS.527..895P}
{Popesso}, P., {Biviano}, A., {Bulbul}, E., {et~al.} 2024{\natexlab{a}},
  \mnras, 527, 895

\bibitem[{{Popesso} {et~al.}(2024{\natexlab{b}}){Popesso}, {Marini}, {Dolag},
  {Lamer}, {Csizi}, {Biffi}, {Robothan}, {Bravo}, {Biviano}, {Vladutesku-Zopp},
  \& et~al.}]{PopessoMariniDolag_2024arXiv241117120P}
{Popesso}, P., {Marini}, I., {Dolag}, K., {et~al.} 2024{\natexlab{b}}, arXiv
  e-prints, arXiv:2411.17120

\bibitem[{{Powell} {et~al.}(2022){Powell}, {Allen}, {Caglar}, {Cappelluti},
  {Harrison}, {Irving}, {Koss}, {Mantz}, {Oh}, {Ricci}, \&
  et~al.}]{PowellAllenCaglar_2022ApJ...938...77P}
{Powell}, M.~C., {Allen}, S.~W., {Caglar}, T., {et~al.} 2022, \apj, 938, 77

\bibitem[{{Powell} {et~al.}(2024){Powell}, {Krumpe}, {Coil}, \&
  {Miyaji}}]{PowellKrumpeCoil_2024A&A...686A..57P}
{Powell}, M.~C., {Krumpe}, M., {Coil}, A., \& {Miyaji}, T. 2024, \aap, 686, A57

\bibitem[{{Pratt} {et~al.}(2009){Pratt}, {Croston}, {Arnaud}, \&
  {B{\"o}hringer}}]{PrattCrostonArnaud_2009A&A...498..361P}
{Pratt}, G.~W., {Croston}, J.~H., {Arnaud}, M., \& {B{\"o}hringer}, H. 2009,
  \aap, 498, 361

\bibitem[{{Predehl} {et~al.}(2021){Predehl}, {Andritschke}, {Arefiev},
  {Babyshkin}, {Batanov}, {Becker}, {B{\"o}hringer}, {Bogomolov}, {Boller},
  {Borm}, \& et~al.}]{PredehlAndritschkeArefiev_2021A&A...647A...1P}
{Predehl}, P., {Andritschke}, R., {Arefiev}, V., {et~al.} 2021, \aap, 647, A1

\bibitem[{{Predehl} {et~al.}(2020){Predehl}, {Sunyaev}, {Becker}, {Brunner},
  {Burenin}, {Bykov}, {Cherepashchuk}, {Chugai}, {Churazov}, {Doroshenko}, \&
  et~al.}]{PredehlSunyaevBecker_2020Natur.588..227P}
{Predehl}, P., {Sunyaev}, R.~A., {Becker}, W., {et~al.} 2020, \nat, 588, 227

\bibitem[{{Refregier} {et~al.}(1997){Refregier}, {Helfand}, \&
  {McMahon}}]{RefregierHelfandMcMahon_1997ApJ...477...58R}
{Refregier}, A., {Helfand}, D.~J., \& {McMahon}, R.~G. 1997, \apj, 477, 58

\bibitem[{{Robotham} {et~al.}(2011){Robotham}, {Norberg}, {Driver}, {Baldry},
  {Bamford}, {Hopkins}, {Liske}, {Loveday}, {Merson}, {Peacock}, \&
  et~al.}]{RobothamNorbergDriver_2011MNRAS.416.2640R}
{Robotham}, A.~S.~G., {Norberg}, P., {Driver}, S.~P., {et~al.} 2011, \mnras,
  416, 2640

\bibitem[{{Ross} {et~al.}(2020){Ross}, {Bautista}, {Tojeiro}, {Alam}, {Bailey},
  {Burtin}, {Comparat}, {Dawson}, {de Mattia}, {du Mas des Bourboux}, \&
  et~al.}]{RossBautistaTojeiro_2020MNRAS.498.2354R}
{Ross}, A.~J., {Bautista}, J., {Tojeiro}, R., {et~al.} 2020, \mnras, 498, 2354

\bibitem[{{Rykoff} {et~al.}(2014){Rykoff}, {Rozo}, {Busha}, {Cunha},
  {Finoguenov}, {Evrard}, {Hao}, {Koester}, {Leauthaud}, {Nord}, \&
  et~al.}]{RykoffRozoBusha_2014ApJ...785..104R}
{Rykoff}, E.~S., {Rozo}, E., {Busha}, M.~T., {et~al.} 2014, \apj, 785, 104

\bibitem[{{Sabater} {et~al.}(2013){Sabater}, {Best}, \&
  {Argudo-Fern{\'a}ndez}}]{SabaterBestArgudo-Fernandez_2013MNRAS.430..638S}
{Sabater}, J., {Best}, P.~N., \& {Argudo-Fern{\'a}ndez}, M. 2013, \mnras, 430,
  638

\bibitem[{{Sabater} {et~al.}(2019){Sabater}, {Best}, {Hardcastle}, {Shimwell},
  {Tasse}, {Williams}, {Br{\"u}ggen}, {Cochrane}, {Croston}, {de Gasperin}, \&
  et~al.}]{SabaterBestHardcastle_2019A&A...622A..17S}
{Sabater}, J., {Best}, P.~N., {Hardcastle}, M.~J., {et~al.} 2019, \aap, 622,
  A17

\bibitem[{{Sanders} {et~al.}(2018){Sanders}, {Fabian}, {Russell}, \&
  {Walker}}]{SandersFabianRussell_2018MNRAS.474.1065S}
{Sanders}, J.~S., {Fabian}, A.~C., {Russell}, H.~R., \& {Walker}, S.~A. 2018,
  \mnras, 474, 1065

\bibitem[{{Schaye} {et~al.}(2015){Schaye}, {Crain}, {Bower}, {Furlong},
  {Schaller}, {Theuns}, {Dalla Vecchia}, {Frenk}, {McCarthy}, {Helly}, \&
  et~al.}]{SchayeCrainBower_2015MNRAS.446..521S}
{Schaye}, J., {Crain}, R.~A., {Bower}, R.~G., {et~al.} 2015, \mnras, 446, 521

\bibitem[{{Schaye} {et~al.}(2023){Schaye}, {Kugel}, {Schaller}, {Helly},
  {Braspenning}, {Elbers}, {McCarthy}, {van Daalen}, {Vandenbroucke}, {Frenk},
  \& et~al.}]{SchayeKugelSchaller_2023MNRAS.526.4978S}
{Schaye}, J., {Kugel}, R., {Schaller}, M., {et~al.} 2023, \mnras, 526, 4978

\bibitem[{{Schellenberger} \&
  {Reiprich}(2017)}]{SchellenbergerReiprich_2017MNRAS.469.3738S}
{Schellenberger}, G. \& {Reiprich}, T.~H. 2017, \mnras, 469, 3738

\bibitem[{{Schneider} {et~al.}(2022){Schneider}, {Freund}, {Czesla}, {Robrade},
  {Salvato}, \& {Schmitt}}]{SchneiderFreundCzesla_2022A&A...661A...6S}
{Schneider}, P.~C., {Freund}, S., {Czesla}, S., {et~al.} 2022, \aap, 661, A6

\bibitem[{{Schwope} {et~al.}(2024{\natexlab{a}}){Schwope}, {Kurpas}, {Baecke},
  {Knauff}, {St{\"u}tz}, {Tub{\'\i}n-Arenas}, {Standke}, {Anderson}, {Bauer},
  {Brandt}, \& et~al.}]{SchwopeKurpasBaecke_2024A&A...686A.110S}
{Schwope}, A., {Kurpas}, J., {Baecke}, P., {et~al.} 2024{\natexlab{a}}, \aap,
  686, A110

\bibitem[{{Schwope} {et~al.}(2024{\natexlab{b}}){Schwope}, {Knauff}, {Kurpas},
  {Salvato}, {Stelzer}, {St{\"u}tz}, \&
  {Tub{\'\i}n-Arenas}}]{SchwopeKnauffKurpas_2024A&A...690A.243S}
{Schwope}, A.~D., {Knauff}, K., {Kurpas}, J., {et~al.} 2024{\natexlab{b}},
  \aap, 690, A243

\bibitem[{{Seppi} {et~al.}(2022){Seppi}, {Comparat}, {Bulbul}, {Nandra},
  {Merloni}, {Clerc}, {Liu}, {Ghirardini}, {Liu}, {Salvato}, \&
  et~al.}]{SeppiComparatBulbul_2022A&A...665A..78S}
{Seppi}, R., {Comparat}, J., {Bulbul}, E., {et~al.} 2022, \aap, 665, A78

\bibitem[{{Seppi} {et~al.}(2024){Seppi}, {Comparat}, {Ghirardini}, {Garrel},
  {Artis}, {S{\'a}nchez}, {Liu}, {Clerc}, {Bulbul}, {Grandis}, \&
  et~al.}]{SeppiComparatGhirardini_2024A&A...686A.196S}
{Seppi}, R., {Comparat}, J., {Ghirardini}, V., {et~al.} 2024, \aap, 686, A196

\bibitem[{{Shreeram} {et~al.}(2024){Shreeram}, {Comparat}, {Merloni}, {Zhang},
  {Ponti}, {Nandra}, {ZuHone}, {Marini}, {Vladutescu-Zopp}, {Popesso}, \&
  et~al.}]{ShreeramComparatMerloni_2024arXiv240910397S}
{Shreeram}, S., {Comparat}, J., {Merloni}, A., {et~al.} 2024, arXiv e-prints,
  arXiv:2409.10397

\bibitem[{{Singh} {et~al.}(2024){Singh}, {Lau}, {Faerman}, {Stern}, \&
  {Nagai}}]{SinghLauFaerman_2024MNRAS.532.3222S}
{Singh}, P., {Lau}, E.~T., {Faerman}, Y., {Stern}, J., \& {Nagai}, D. 2024,
  \mnras, 532, 3222

\bibitem[{{Sinha} \& {Garrison}(2020)}]{SinhaGarrison_2020MNRAS.491.3022S}
{Sinha}, M. \& {Garrison}, L.~H. 2020, \mnras, 491, 3022

\bibitem[{{Soltan} {et~al.}(1997){Soltan}, {Hasinger}, {Egger}, {Snowden}, \&
  {Truemper}}]{SoltanHasingerEgger_1997A&A...320..705S}
{Soltan}, A.~M., {Hasinger}, G., {Egger}, R., {Snowden}, S., \& {Truemper}, J.
  1997, \aap, 320, 705

\bibitem[{{Strauss} {et~al.}(2002){Strauss}, {Weinberg}, {Lupton}, {Narayanan},
  {Annis}, {Bernardi}, {Blanton}, {Burles}, {Connolly}, {Dalcanton}, \&
  et~al.}]{StraussWeinbergLupton_2002AJ....124.1810S}
{Strauss}, M.~A., {Weinberg}, D.~H., {Lupton}, R.~H., {et~al.} 2002, \aj, 124,
  1810

\bibitem[{{Sunyaev} {et~al.}(2021){Sunyaev}, {Arefiev}, {Babyshkin},
  {Bogomolov}, {Borisov}, {Buntov}, {Brunner}, {Burenin}, {Churazov},
  {Coutinho}, \& et~al.}]{SunyaevArefievBabyshkin_2021A&A...656A.132S}
{Sunyaev}, R., {Arefiev}, V., {Babyshkin}, V., {et~al.} 2021, \aap, 656, A132

\bibitem[{{Tempel} {et~al.}(2016){Tempel}, {Kipper}, {Tamm}, {Gramann},
  {Einasto}, {Sepp}, \& {Tuvikene}}]{TempelKipperTamm_2016A&A...588A..14T}
{Tempel}, E., {Kipper}, R., {Tamm}, A., {et~al.} 2016, \aap, 588, A14

\bibitem[{{Tempel} {et~al.}(2018){Tempel}, {Kruuse}, {Kipper}, {Tuvikene},
  {Sorce}, \& {Stoica}}]{TempelKruuseKipper_2018A&A...618A..81T}
{Tempel}, E., {Kruuse}, M., {Kipper}, R., {et~al.} 2018, \aap, 618, A81

\bibitem[{{Tinker}(2021)}]{Tinker_2021ApJ...923..154T}
{Tinker}, J.~L. 2021, \apj, 923, 154

\bibitem[{{Tinker}(2022)}]{Tinker_2022AJ....163..126T}
{Tinker}, J.~L. 2022, \aj, 163, 126

\bibitem[{{Tinker} {et~al.}(2010){Tinker}, {Robertson}, {Kravtsov}, {Klypin},
  {Warren}, {Yepes}, \&
  {Gottl{\"o}ber}}]{TinkerRobertsonKravtsov_2010ApJ...724..878T}
{Tinker}, J.~L., {Robertson}, B.~E., {Kravtsov}, A.~V., {et~al.} 2010, \apj,
  724, 878

\bibitem[{{Truong} {et~al.}(2021){Truong}, {Pillepich}, {Nelson}, {Werner}, \&
  {Hernquist}}]{TruongPillepichNelson_2021MNRAS.508.1563T}
{Truong}, N., {Pillepich}, A., {Nelson}, D., {Werner}, N., \& {Hernquist}, L.
  2021, \mnras, 508, 1563

\bibitem[{{van den Bosch} {et~al.}(2013){van den Bosch}, {More}, {Cacciato},
  {Mo}, \& {Yang}}]{vandenBoschMoreCacciato_2013MNRAS.430..725V}
{van den Bosch}, F.~C., {More}, S., {Cacciato}, M., {Mo}, H., \& {Yang}, X.
  2013, \mnras, 430, 725

\bibitem[{{Vikhlinin} {et~al.}(2006){Vikhlinin}, {Kravtsov}, {Forman}, {Jones},
  {Markevitch}, {Murray}, \& {Van
  Speybroeck}}]{VikhlininKravtsovForman_2006ApJ...640..691V}
{Vikhlinin}, A., {Kravtsov}, A., {Forman}, W., {et~al.} 2006, \apj, 640, 691

\bibitem[{{Vladutescu-Zopp} {et~al.}(2023){Vladutescu-Zopp}, {Biffi}, \&
  {Dolag}}]{Vladutescu-ZoppBiffiDolag_2023A&A...669A..34V}
{Vladutescu-Zopp}, S., {Biffi}, V., \& {Dolag}, K. 2023, \aap, 669, A34

\bibitem[{{Waddell} {et~al.}(2024{\natexlab{a}}){Waddell}, {Buchner}, {Nandra},
  {Salvato}, {Merloni}, {Gauger}, {Boller}, {Seppi}, {Wolf}, {Liu}, \&
  et~al.}]{WaddellBuchnerNandra_2024arXiv240117306W}
{Waddell}, S. G.~H., {Buchner}, J., {Nandra}, K., {et~al.} 2024{\natexlab{a}},
  arXiv e-prints, arXiv:2401.17306

\bibitem[{{Waddell} {et~al.}(2024{\natexlab{b}}){Waddell}, {Nandra}, {Buchner},
  {Wu}, {Shen}, {Arcodia}, {Merloni}, {Salvato}, {Dauser}, {Boller}, \&
  et~al.}]{WaddellNandraBuchner_2024A&A...690A.132W}
{Waddell}, S.~G.~H., {Nandra}, K., {Buchner}, J., {et~al.} 2024{\natexlab{b}},
  \aap, 690, A132

\bibitem[{{Yang} {et~al.}(2005){Yang}, {Mo}, {van den Bosch}, \&
  {Jing}}]{YangMovandenBosch_2005MNRAS.356.1293Y}
{Yang}, X., {Mo}, H.~J., {van den Bosch}, F.~C., \& {Jing}, Y.~P. 2005, \mnras,
  356, 1293

\bibitem[{{Yeung} {et~al.}(2024){Yeung}, {Ponti}, {Freyberg}, {Dennerl}, {Liu},
  {Locatelli}, {Mayer}, {Sanders}, {Sasaki}, {Strong}, \&
  et~al.}]{YeungPontiFreyberg_2024A&A...690A.399Y}
{Yeung}, M. C.~H., {Ponti}, G., {Freyberg}, M.~J., {et~al.} 2024, \aap, 690,
  A399

\bibitem[{{Yuan} {et~al.}(2022){Yuan}, {Garrison}, {Hadzhiyska}, {Bose}, \&
  {Eisenstein}}]{YuanGarrisonHadzhiyska_2022MNRAS.510.3301Y}
{Yuan}, S., {Garrison}, L.~H., {Hadzhiyska}, B., {Bose}, S., \& {Eisenstein},
  D.~J. 2022, \mnras, 510, 3301

\bibitem[{{Zehavi} {et~al.}(2011){Zehavi}, {Zheng}, {Weinberg}, {Blanton},
  {Bahcall}, {Berlind}, {Brinkmann}, {Frieman}, {Gunn}, {Lupton}, \&
  et~al.}]{ZehaviZhengWeinberg_2011ApJ...736...59Z}
{Zehavi}, I., {Zheng}, Z., {Weinberg}, D.~H., {et~al.} 2011, \apj, 736, 59

\bibitem[{{Zehavi} {et~al.}(2005){Zehavi}, {Zheng}, {Weinberg}, {Frieman},
  {Berlind}, {Blanton}, {Scoccimarro}, {Sheth}, {Strauss}, {Kayo}, \&
  et~al.}]{ZehaviZhengWeinberg_2005ApJ...630....1Z}
{Zehavi}, I., {Zheng}, Z., {Weinberg}, D.~H., {et~al.} 2005, \apj, 630, 1

\bibitem[{{Zhang} {et~al.}(2023){Zhang}, {Behroozi}, {Volonteri}, {Silk},
  {Fan}, {Hopkins}, {Yang}, \&
  {Aird}}]{ZhangBehrooziVolonteri_2023MNRAS.518.2123Z}
{Zhang}, H., {Behroozi}, P., {Volonteri}, M., {et~al.} 2023, \mnras, 518, 2123

\bibitem[{{Zhang} {et~al.}(2024{\natexlab{a}}){Zhang}, {Ponti}, {Carretti},
  {Liu}, {Morris}, {Haverkorn}, {Locatelli}, {Zheng}, {Aharonian}, {Zhang}, \&
  et~al.}]{ZhangPontiCarretti_2024NatAs...8.1416Z}
{Zhang}, H.-S., {Ponti}, G., {Carretti}, E., {et~al.} 2024{\natexlab{a}},
  Nature Astronomy, 8, 1416

\bibitem[{{Zhang} {et~al.}(2024{\natexlab{b}}){Zhang}, {Comparat}, {Ponti},
  {Merloni}, {Nandra}, {Haberl}, {Locatelli}, {Zhang}, {Sanders}, {Zheng}, \&
  et~al.}]{ZhangComparatPonti_2024A&A...690A.267Z}
{Zhang}, Y., {Comparat}, J., {Ponti}, G., {et~al.} 2024{\natexlab{b}}, \aap,
  690, A267

\bibitem[{{Zhang} {et~al.}(2024{\natexlab{c}}){Zhang}, {Comparat}, {Ponti},
  {Merloni}, {Nandra}, {Haberl}, {Truong}, {Pillepich}, {Locatelli}, {Zhang},
  \& et~al.}]{ZhangComparatPonti_2024A&A...690A.268Z}
{Zhang}, Y., {Comparat}, J., {Ponti}, G., {et~al.} 2024{\natexlab{c}}, \aap,
  690, A268

\bibitem[{{Zhang} {et~al.}(2025){Zhang}, {Comparat}, {Ponti}, {Merloni},
  {Nandra}, {Haberl}, {Truong}, {Pillepich}, {Popesso}, {Locatelli}, \&
  et~al.}]{ZhangComparatPonti_2025A&A...693A.197Z}
{Zhang}, Y., {Comparat}, J., {Ponti}, G., {et~al.} 2025, \aap, 693, A197

\bibitem[{{Zhang} {et~al.}(2021){Zhang}, {Wang}, {Luo}, {Mo}, {Liang}, {Li},
  {Yang}, {Wang}, {Zhang}, {Hong}, \&
  et~al.}]{ZhangWangLuo_2021A&A...650A.155Z}
{Zhang}, Z., {Wang}, H., {Luo}, W., {et~al.} 2021, \aap, 650, A155

\bibitem[{{Zheng} {et~al.}(2024{\natexlab{a}}){Zheng}, {Ponti}, {Freyberg},
  {Sanders}, {Locatelli}, {Merloni}, {Strong}, {Sasaki}, {Comparat}, {Becker},
  \& et~al.}]{ZhengPontiFreyberg_2024A&A...681A..77Z}
{Zheng}, X., {Ponti}, G., {Freyberg}, M., {et~al.} 2024{\natexlab{a}}, \aap,
  681, A77

\bibitem[{{Zheng} {et~al.}(2024{\natexlab{b}}){Zheng}, {Ponti}, {Locatelli},
  {Sanders}, {Merloni}, {Becker}, {Comparat}, {Dennerl}, {Freyberg}, {Maitra},
  \& et~al.}]{ZhengPontiLocatelli_2024A&A...689A.328Z}
{Zheng}, X., {Ponti}, G., {Locatelli}, N., {et~al.} 2024{\natexlab{b}}, \aap,
  689, A328

\bibitem[{{Zheng} {et~al.}(2005){Zheng}, {Berlind}, {Weinberg}, {Benson},
  {Baugh}, {Cole}, {Dav{\'e}}, {Frenk}, {Katz}, \&
  {Lacey}}]{ZhengBerlindWeinberg_2005ApJ...633..791Z}
{Zheng}, Z., {Berlind}, A.~A., {Weinberg}, D.~H., {et~al.} 2005, \apj, 633, 791

\bibitem[{{Zheng} {et~al.}(2007){Zheng}, {Coil}, \&
  {Zehavi}}]{ZhengCoilZehavi_2007ApJ...667..760Z}
{Zheng}, Z., {Coil}, A.~L., \& {Zehavi}, I. 2007, \apj, 667, 760

\bibitem[{{Zhou} {et~al.}(2021){Zhou}, {Newman}, {Mao}, {Meisner}, {Moustakas},
  {Myers}, {Prakash}, {Zentner}, {Brooks}, {Duan}, \&
  et~al.}]{ZhouNewmanMao_2021MNRAS.501.3309Z}
{Zhou}, R., {Newman}, J.~A., {Mao}, Y.-Y., {et~al.} 2021, \mnras, 501, 3309

\bibitem[{{Zou} {et~al.}(2019){Zou}, {Gao}, {Zhou}, \&
  {Kong}}]{ZouGaoZhou_2019ApJS..242....8Z}
{Zou}, H., {Gao}, J., {Zhou}, X., \& {Kong}, X. 2019, \apjs, 242, 8

\bibitem[{{Zu} \& {Mandelbaum}(2015)}]{ZuMandelbaum_2015MNRAS.454.1161Z}
{Zu}, Y. \& {Mandelbaum}, R. 2015, \mnras, 454, 1161

\bibitem[{{Zu} \& {Mandelbaum}(2016)}]{ZuMandelbaum_2016MNRAS.457.4360Z}
{Zu}, Y. \& {Mandelbaum}, R. 2016, \mnras, 457, 4360

\end{thebibliography}

\appendix

\section{Galaxy mock catalog}
\label{sec:mocks}

With mock catalogs, we estimate the impact of possible systematic effects on the clustering measurement. 
We create mock catalogs using the Uchuu simulation \citep{IshiyamaPradaKlypin_2021MNRAS.506.4210I} and the UniverseMachine model \citep{BehrooziWechslerHearin_2019MNRAS.488.3143B, AungNagaiKlypin_2023MNRAS.519.1648A}. We create eight light cones following \citet{ComparatEckertFinoguenov_2020OJAp....3E..13C} that each spans 1/8th of the sky (by setting the observer at each corner of the simulated box). 

\subsection{Reference mock catalog}
We generate a reference mock catalog for each galaxy sample, applying the same selection as in the data but on the predicted galaxy population (i.e., using redshift and stellar mass). 
Since the UniverseMachine model is fitted upon different stellar mass functions, the predicted galaxy density as a function of stellar mass function is in excellent agreement with observations (Fig. \ref{fig:stellar:mass:density:functions}, \ref{fig:stellar:mass:functions}). 
Notably, the correlation functions predicted for each galaxy sample are close to the observed ones, see Fig. \ref{fig:wprp:final:mocks}. 
The fact that the measurements are close to the prediction of the UniverseMachine model (Fig. \ref{fig:wprp:final:mocks}) means that no large systematics have been missed in the measurement procedure.

\subsection{Impact of the photometric redshift on the correlation function}

We create mock catalogs where we degrade the redshift precision according to the performance of the photometric redshifts. We find that the quality of the redshift considered here does not cause significant distortions in the shape of the correlation function ($w^{\pi_{max}=100}_p(r_p)$) in the range $0.01<r_p[Mpc/h]<30$. The redshift uncertainty causes a constant loss of amplitude in the correlation function. For each sample (selections in redshift and stellar mass from the complete selection), we estimate the scatter of the photometric redshift.
We find that the performances for each sub-sample are better than for the complete sample with a scatter between 1 and 2 percent. Indeed, after discarding low mass, faint sources, and higher redshift sources, it is expected to perform better than the full sample (3 percent scatter). 
We find an expected loss in the correlation function amplitude between 2 and 7 percent. We use the tabulated value and apply a correction to the measurement for each sample (see $f_{obs}\in[0.93,0.98]$ in Fig. \ref{fig:photozbias:clustering:wprp} bottom rows of panels ).

\subsection{Impact of redshift space distortion}

With the mock catalog, we fit a model on the $w_p(r_p)$ ratio estimated in real space and redshift space. With that, we correct the observation from this shape bending of the signal. Since we are integrating up to $\pi_{max}=$100 $h^{-1}$Mpc on the line of sight, this is a minor correction for the maximum distance we fit to (log10(20-30 Mpc/h) = 1.6). The correction would be more prominent for smaller $\pi_{max}$ values. 
In \citet{vandenBoschMoreCacciato_2013MNRAS.430..725V}, they typically amount to 2\% (4, 10, 14) at R=4 (10, 25, 31) Mpc/h. We find similar values here 
(Fig. \ref{fig:rsd:correction}).
Overall, with the mock catalogs,    we derive a theoretical budget of a few percent corrections on each correlation function.

\section{Density of galaxies}
\label{sec:app:galaxy:density}

This appendix provides additional (large) figures and Tables related to the galaxy samples.

\begin{figure}
\centering
\includegraphics[width=0.95\columnwidth]{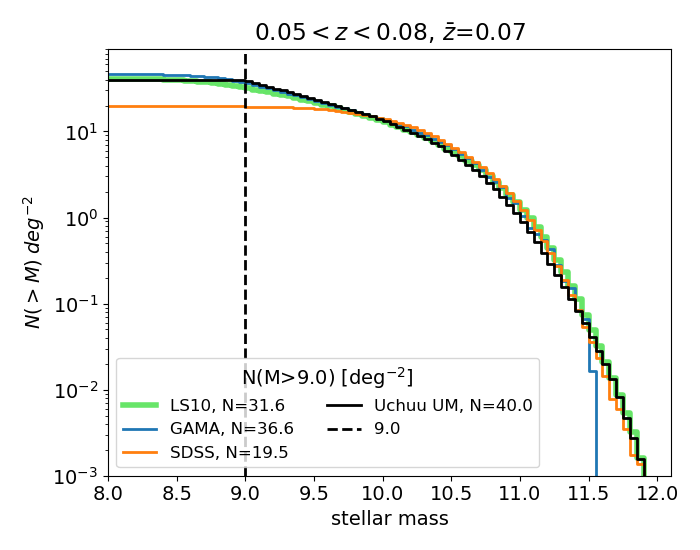}
\includegraphics[width=0.95\columnwidth]{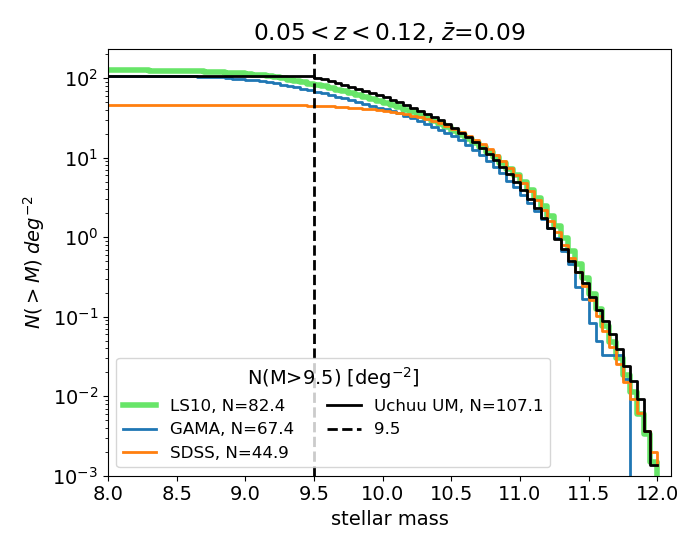}
\includegraphics[width=0.95\columnwidth]{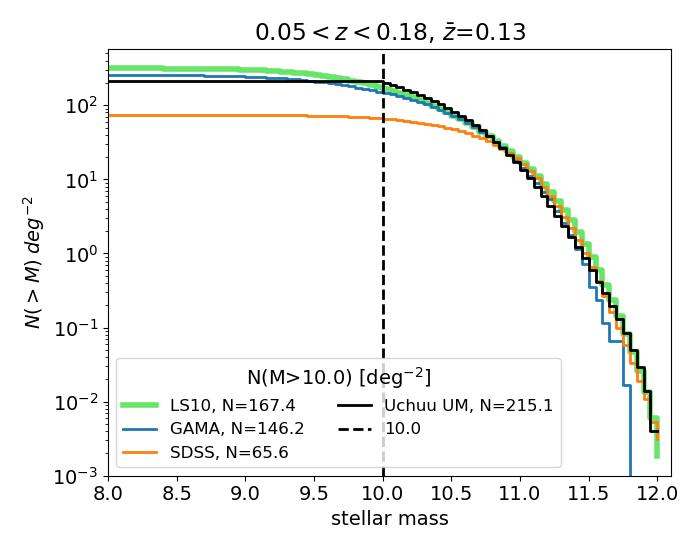}
\caption{Cumulative number density of galaxies per square degree as a function of stellar mass.}
\label{fig:stellar:mass:density:functions}
\end{figure}

\begin{figure}
\centering
\includegraphics[width=0.95\columnwidth]{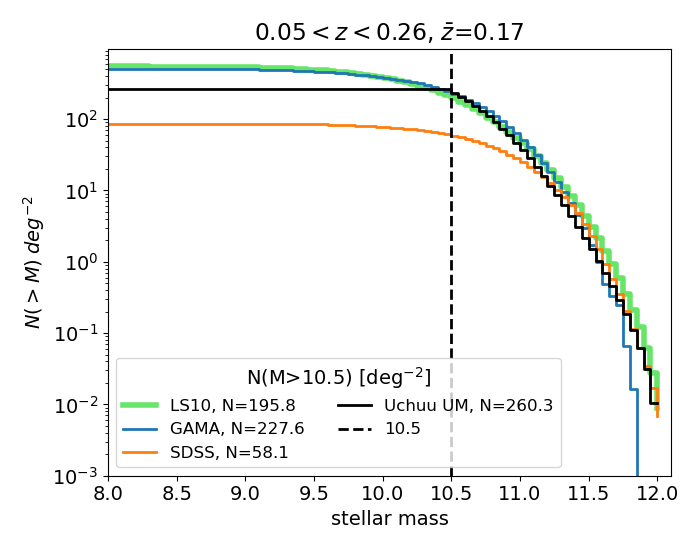}
\includegraphics[width=0.95\columnwidth]{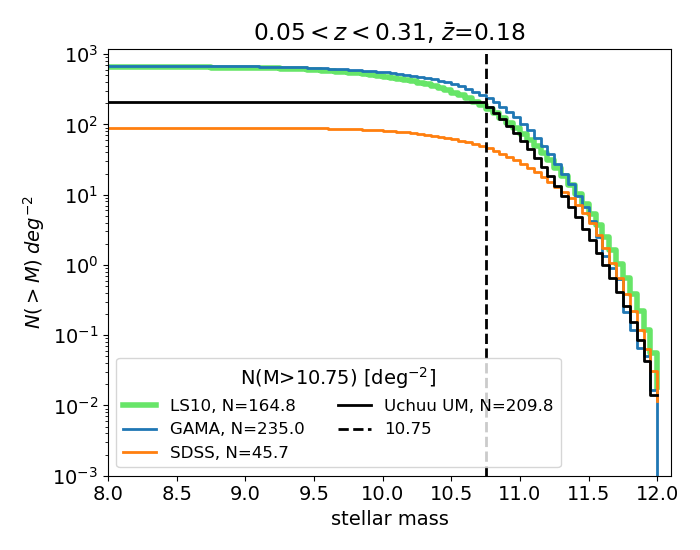}
\includegraphics[width=0.95\columnwidth]{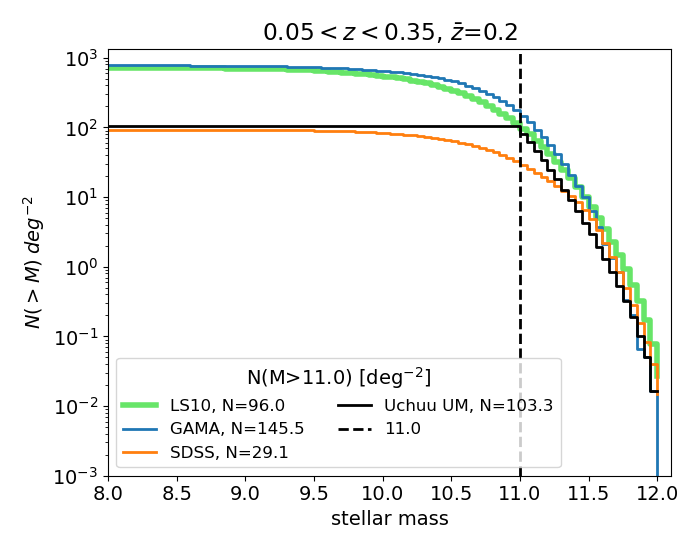}
\addtocounter{figure}{-1}\caption{Continued.}
\label{fig:stellar:mass:density:functions:2}
\end{figure}

\begin{figure}
\centering
\includegraphics[width=0.95\columnwidth]{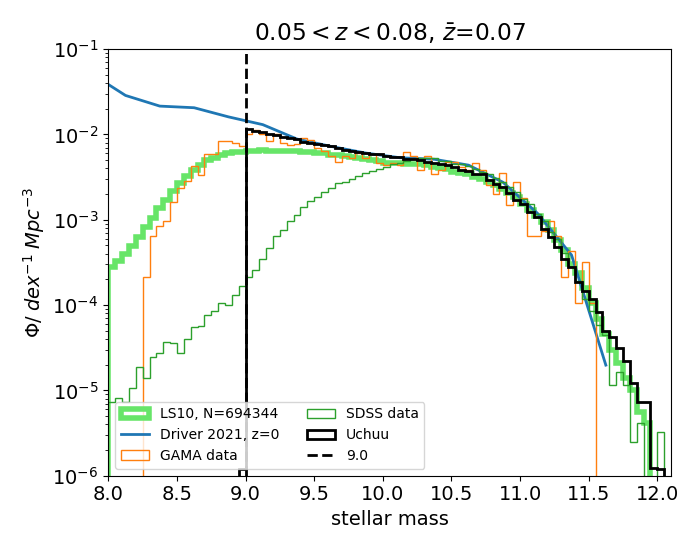}
\includegraphics[width=0.95\columnwidth]{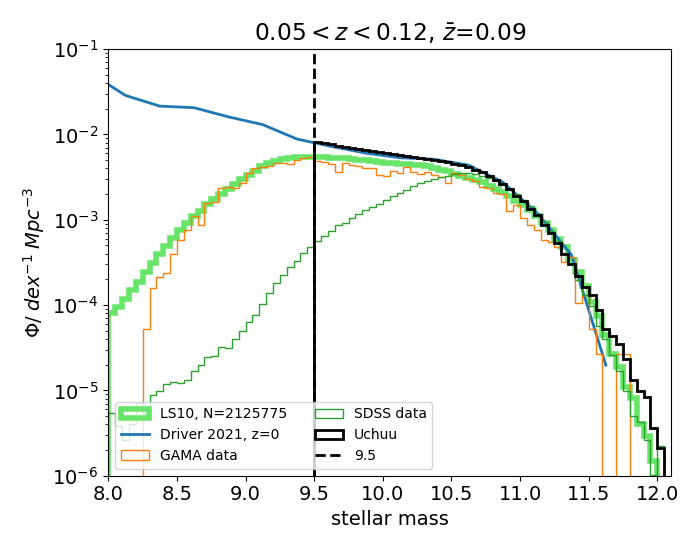}
\includegraphics[width=0.95\columnwidth]{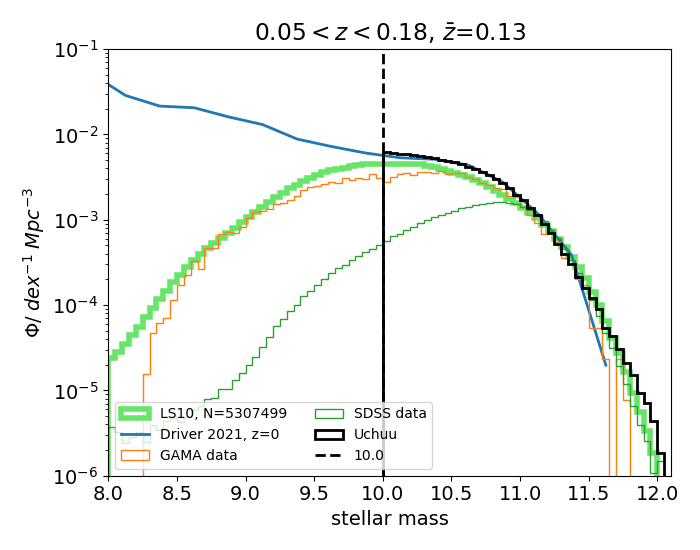}
\caption{Observed number density of galaxies per cubic Mpc as a function of stellar mass.}
\label{fig:stellar:mass:functions}
\end{figure}

\begin{figure}
\centering
\includegraphics[width=0.95\columnwidth]{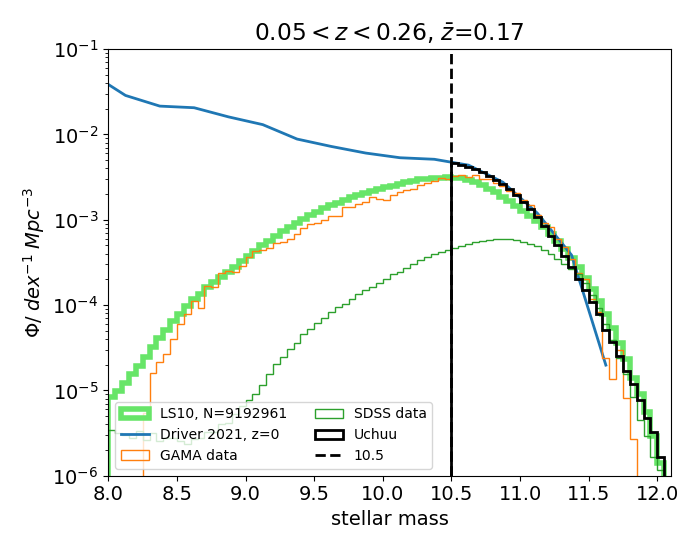}
\includegraphics[width=0.95\columnwidth]{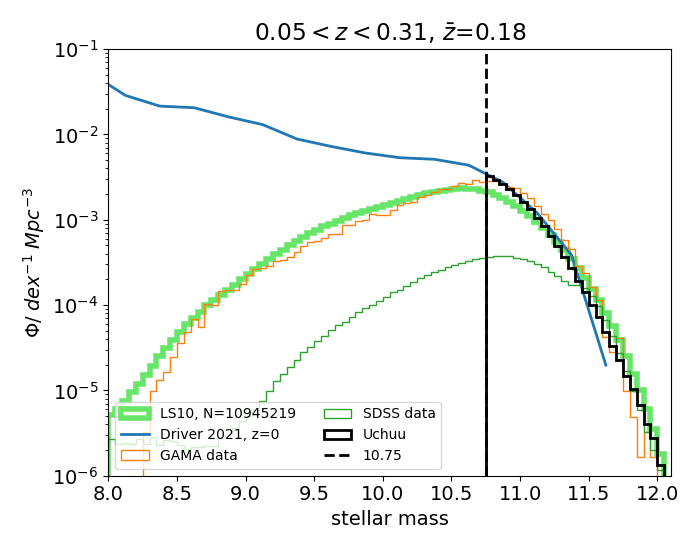}
\includegraphics[width=0.95\columnwidth]{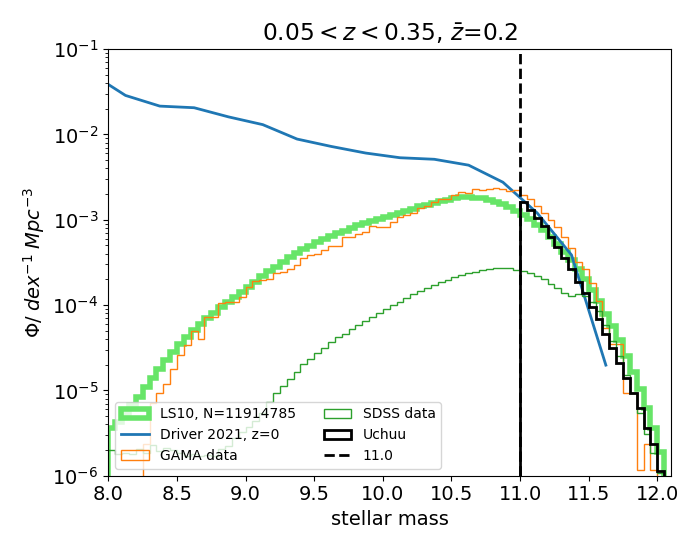}
\addtocounter{figure}{-1}\caption{Continued.}
\label{fig:stellar:mass:functions:2}
\end{figure}

\begin{table}
\caption{Number density of galaxies selected in LS10 compared to GAMA as a function of $r$-band magnitude.}
\label{tab:number:density:galaxies}
  \begin{center}
    \begin{tabular}{rr ll rr rr }
    \hline \hline
\multicolumn{2}{c}{$r$ mag} & \multicolumn{2}{c}{density [deg$^{-2}$]} & \multicolumn{1}{c}{ratio} \\
 min & max &  GAMA & LS10 & \\ 
\hline
16.0 & 16.5 & 13.8 & 14.8 & 0.929    \\ 
16.5 & 17.0 & 25.2 & 25.6 & 0.987    \\ 
17.0 & 17.5 & 51.4 & 52.6 & 0.978    \\ 
17.5 & 18.0 & 98.8 & 99.8 & 0.99     \\ 
18.0 & 18.5 & 192.9 & 188.0 & 1.026  \\ 
18.5 & 19.0 & 359.5 & 355.0 & 1.013  \\ 
19.0 & 19.5 & 627.3 & 629.3 & 0.997  \\ 
 \hline
    \end{tabular}
  \end{center}
\end{table}

\section{Correlation functions}
\label{sec:app:correlation:functions}
This appendix provides additional information, figures, and tables related to the correlation functions.

\subsection{Auto-correlation, field selection}
\label{subsec:app:XRAY:autocorr}
We use the X-ray event auto-correlation to determine where we measure the cross-correlation. 
We measure the auto-correlation of events on each eROSITA field (tiles of 3.6$\times$3.6 deg$^2$). 
The obtained distribution of correlation functions is generally well-behaved, albeit with several fields having outlying correlation functions. 
Indeed, due to bright Galactic foregrounds (resolved or unresolved, e.g., the eROSITA Bubbles \citep{PredehlSunyaevBecker_2020Natur.588..227P,ZhangPontiCarretti_2024NatAs...8.1416Z}, the angular correlation function may have a high power on degree scales. 
Most of these ($\sim50$ out of $\sim1600$) fields are at low galactic latitudes where our galaxy catalog does not extend. 
In these cases, the fraction of the correlation function due to point sources or diffuse emission is genuinely interesting but beyond the scope of this paper. 
In addition, a few outlier fields follow a flaring pattern that was not completely removed in a scan, so we remove these fields from the analysis.

\subsection*{Cross-correlation, field selection}
First, we remove fields that have a correlation function amplitude on large scales (in the two-halo term) at $\theta\sim0.6^\circ$ (corresponding to 6 (10) comoving Mpc at z=0.15 (0.25)) below (or above) the 1st (99th) percentile of the distribution of $w(\theta\sim0.6^\circ)$, corresponding heuristically to a $2.4\sigma$ clipping.
For a few more fields, the degree scale cross-correlation function is still a strong positive outlier, which we remove. 
For example, measuring $w(\theta\sim0.6^\circ)>0.15$ when the expectation is $w(\theta\sim0.6^\circ)\sim2\times 10^{-3}$ indicates a contamination by stars. 
These fields are at the edge of the survey, where the stellar density or extinction is high. 
We remove these fields from the analysis. 
A cut of the survey footprint using extinction and stellar density is equivalent. 

\begin{figure}
\centering
\includegraphics[width=0.95\columnwidth]{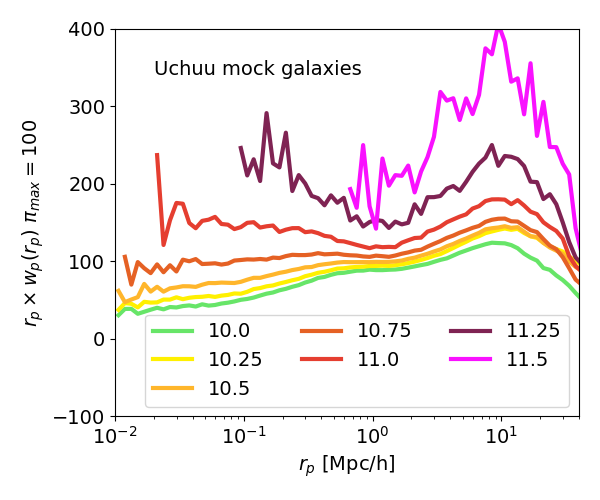}
\caption{Projected galaxy auto-correlation function ($w_p(r_p)$) predicted from the UCHUU mock catalogs on $\sim$5,000 deg$^2$.}
\label{fig:wprp:final:mocks}
\end{figure}

\begin{figure*}
\centering
\includegraphics[width=0.6\columnwidth]{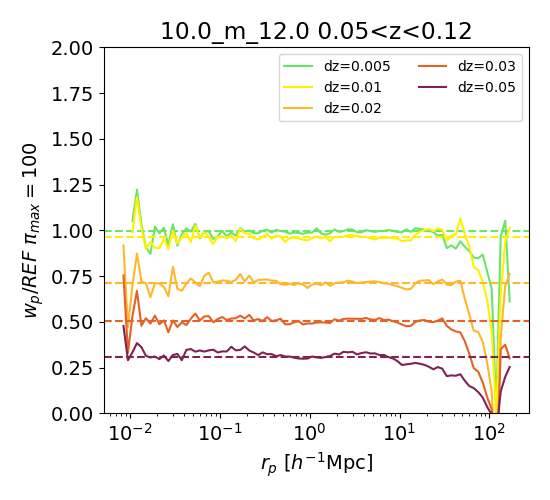}
\includegraphics[width=0.6\columnwidth]{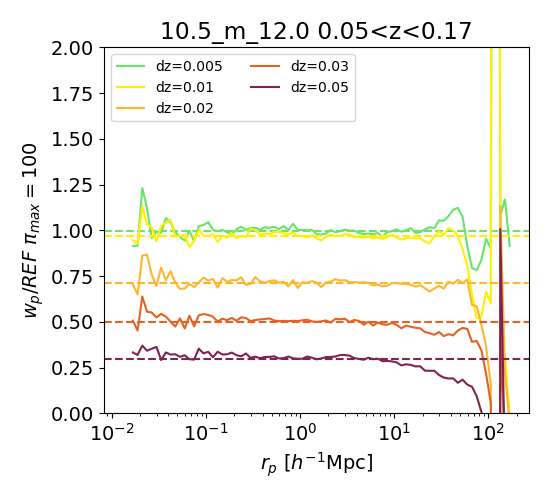}
\includegraphics[width=0.6\columnwidth]{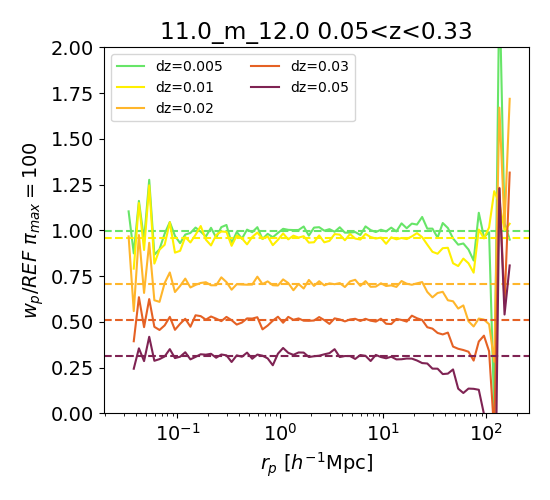}
\includegraphics[width=0.6\columnwidth]{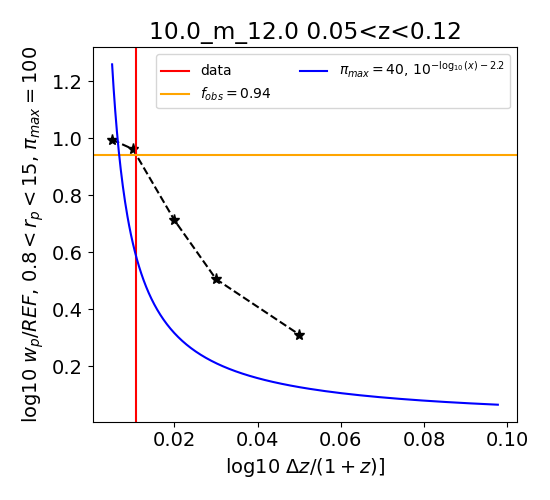}
\includegraphics[width=0.6\columnwidth]{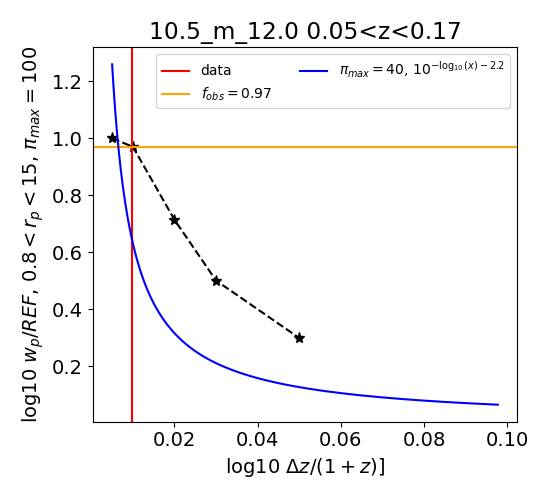}
\includegraphics[width=0.6\columnwidth]{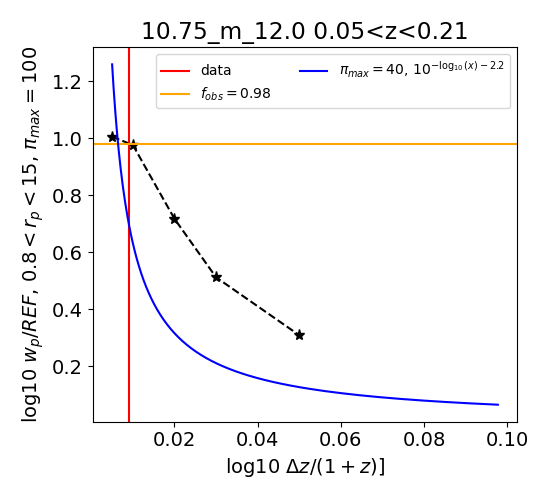}
\includegraphics[width=0.6\columnwidth]{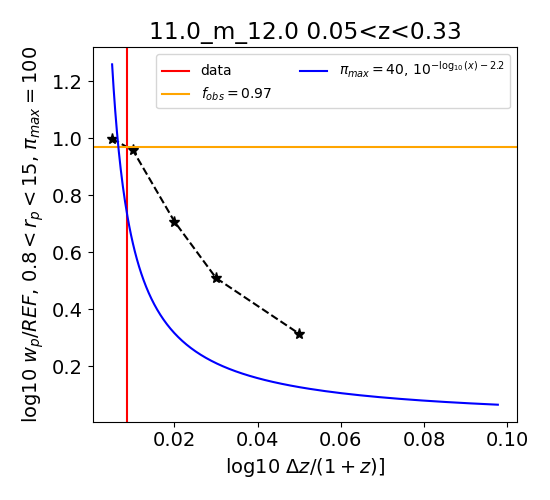}
\includegraphics[width=0.6\columnwidth]{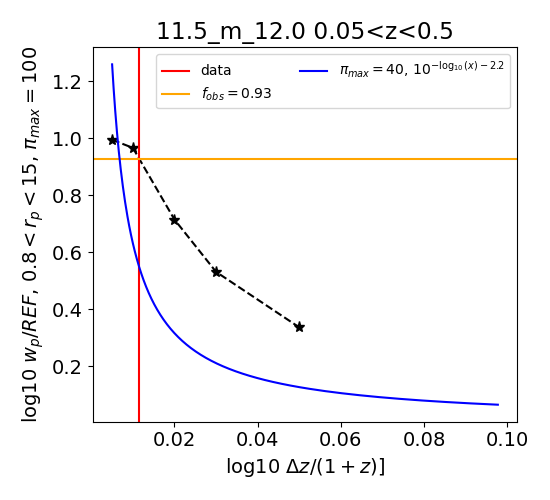}
\caption{Bias in amplitude when computing wprp in the presence of photometric errors for each sample considered here.}
\label{fig:photozbias:clustering:wprp}
\end{figure*}

\begin{figure*}
\centering
\includegraphics[width=0.6\columnwidth]{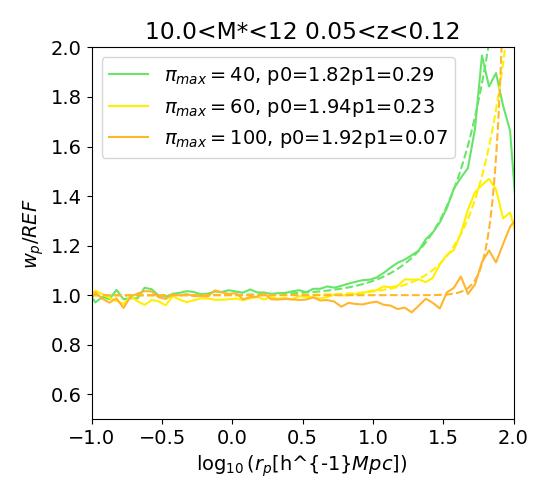}
\includegraphics[width=0.6\columnwidth]{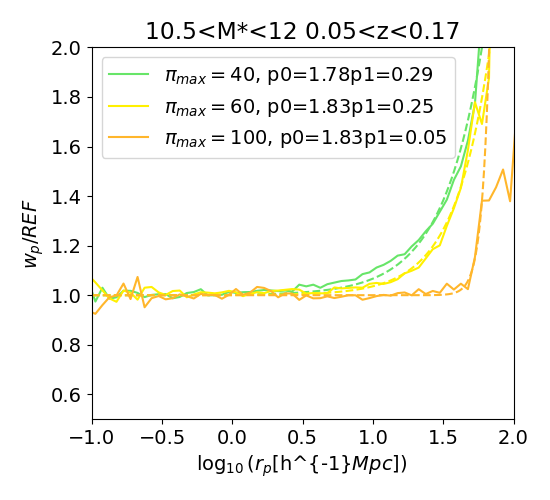}
\includegraphics[width=0.6\columnwidth]{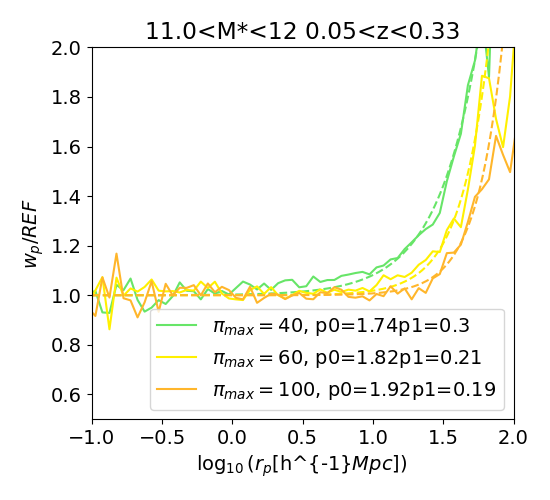}
\caption{Ratio of the real space over redshift space correlation function for all selections. data from Uchuu in solid lines and fitted model in dashed lines. The model is used to correct observations that we fit until 1.6 (40 Mpc/h)}
\label{fig:rsd:correction}
\end{figure*}

\begin{figure*}
\centering
\includegraphics[width=0.45\columnwidth]{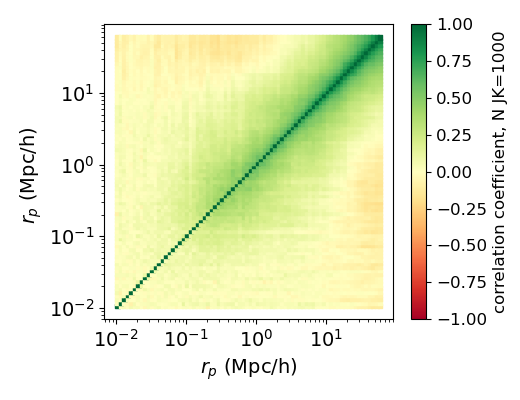}
\includegraphics[width=0.45\columnwidth]{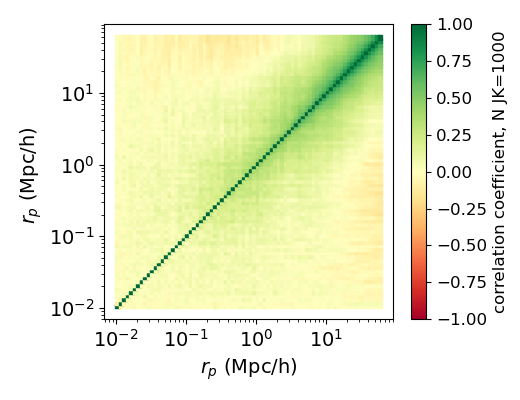}
\includegraphics[width=0.45\columnwidth]{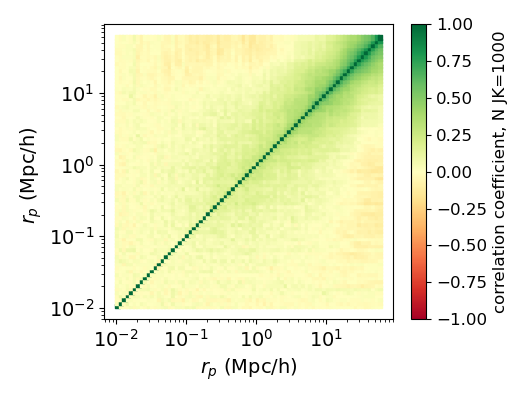}
\includegraphics[width=0.45\columnwidth]{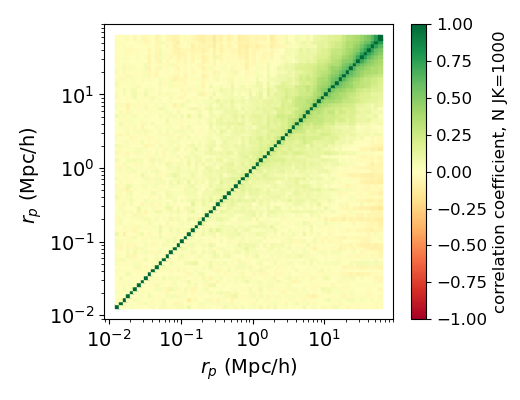}

\includegraphics[width=0.45\columnwidth]{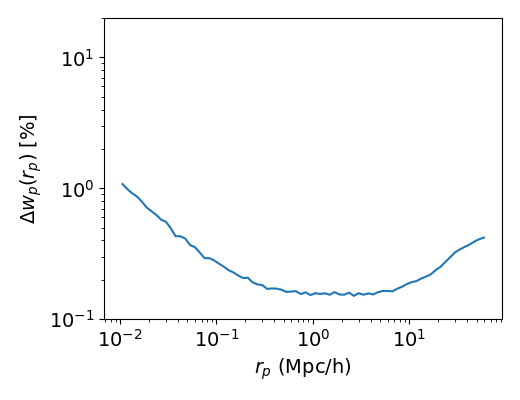}
\includegraphics[width=0.45\columnwidth]{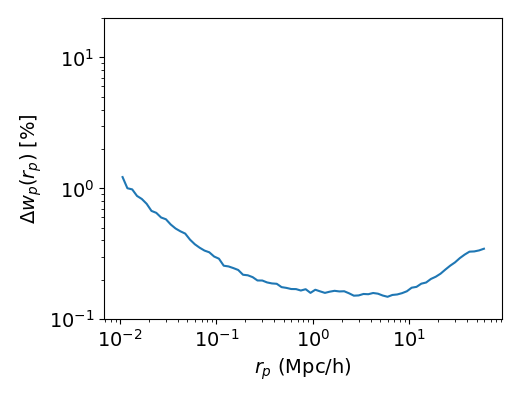}
\includegraphics[width=0.45\columnwidth]{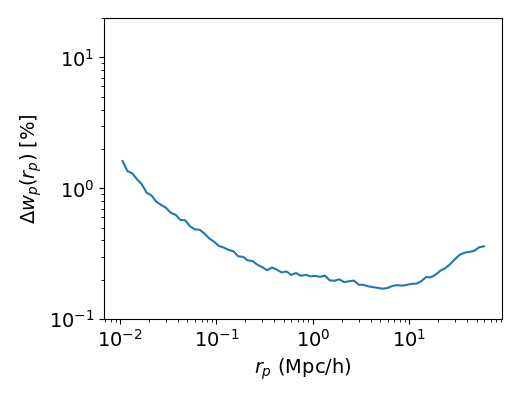}
\includegraphics[width=0.45\columnwidth]{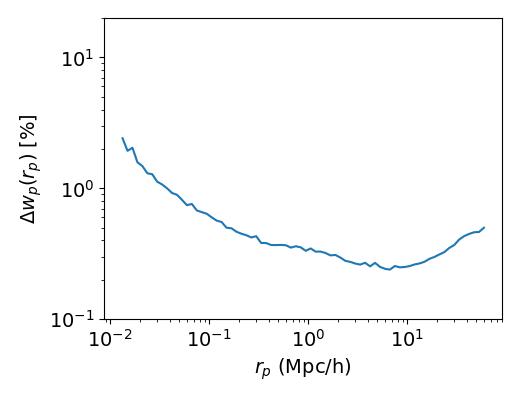}

\caption{correlation coefficient (top), square root of the diagonal of the covariance matrix (bottom) for M 10.0 10.5 (2nd column), 10.75 (3rd column), 11.0 (4th column).}
\label{fig:cvcm:10:105:1075:11}

\end{figure*}

\begin{figure*}
\centering
\includegraphics[width=1.5\columnwidth]{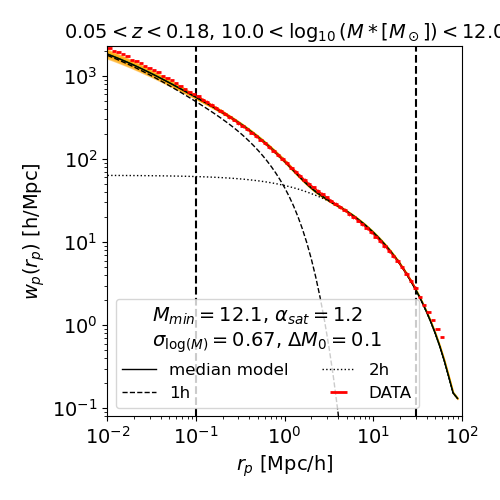}
\includegraphics[width=0.55\columnwidth]{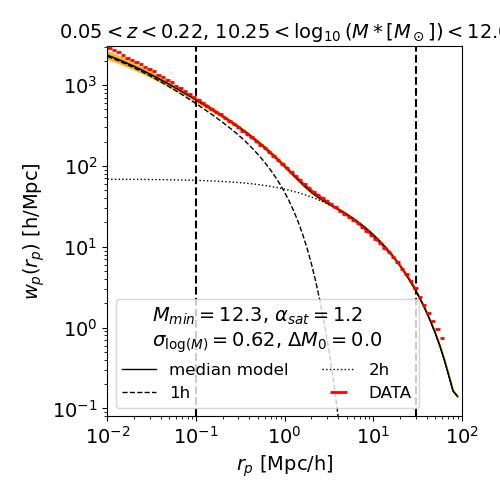}
\includegraphics[width=0.55\columnwidth]{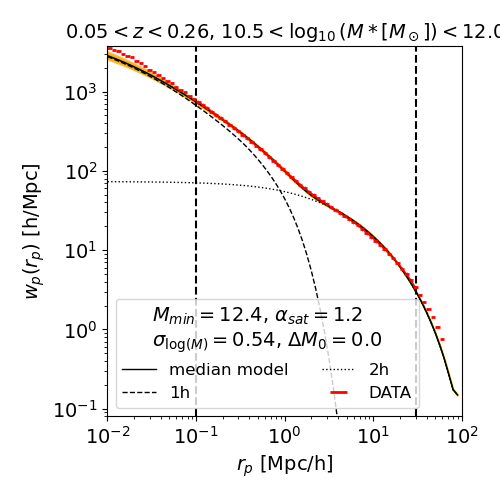}
\includegraphics[width=0.55\columnwidth]{wprp/LS10_VLIM_ANY_10.0_Mstar_12.0_0.05_z_0.18_N_2759238-wp-obs-mod.PNG}
\includegraphics[width=0.55\columnwidth]{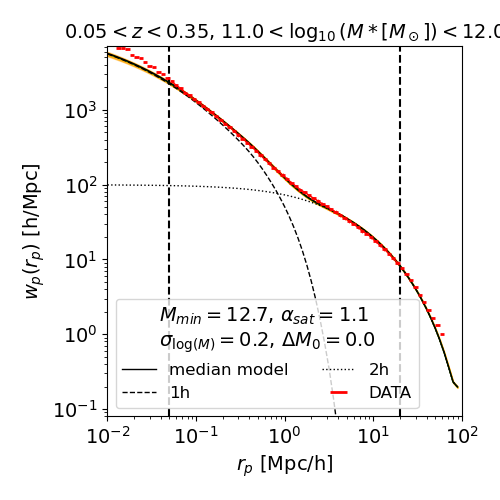}
\includegraphics[width=0.55\columnwidth]{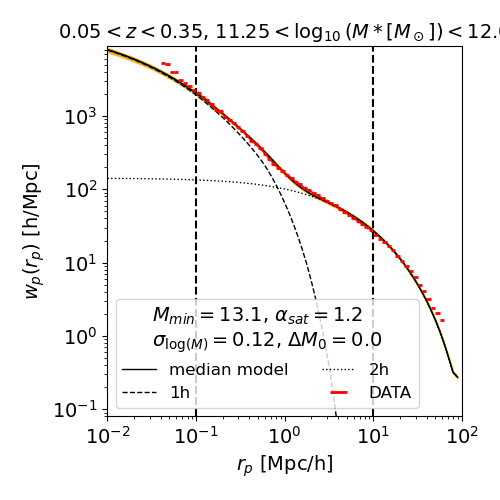}
\includegraphics[width=0.55\columnwidth]{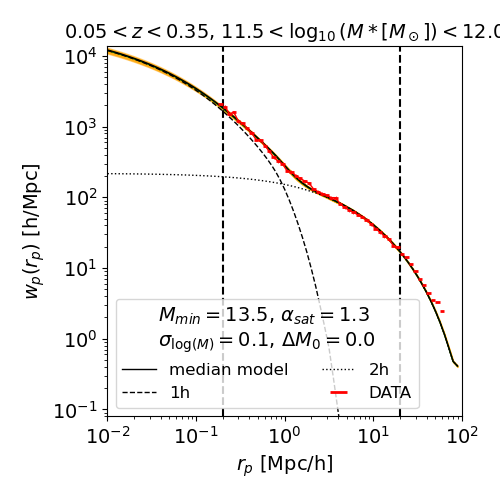}
\caption{Correlation function and best-fit HOD model for the samples considered. The models represent well the observations in the range 0.1-20 Mpc$/h$.}
\label{fig:wprpBFmodels:10:105:1075:11}
\end{figure*}

\begin{figure}
    \centering    \includegraphics[width=0.95\columnwidth]{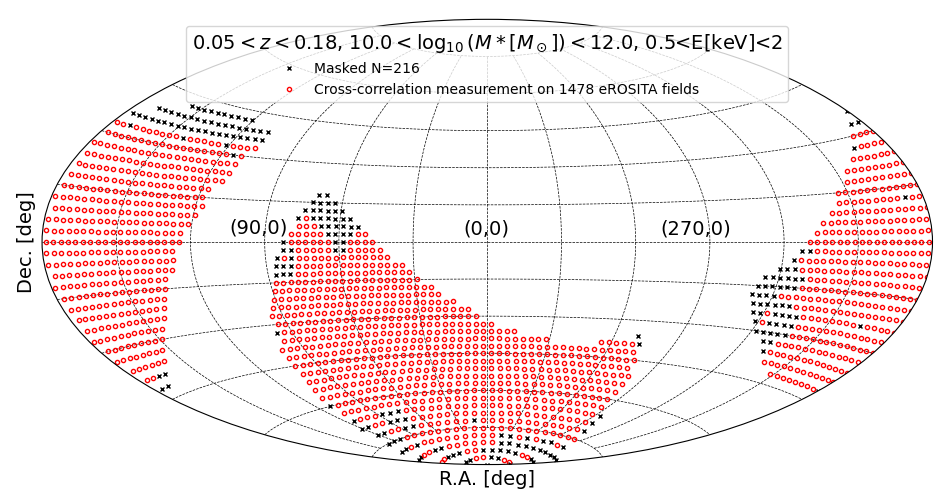}
    \caption{The footprint of the set of eROSITA fields used for the cross-correlation between X-rays and the sample with $\log_{10}(M^*/M_\odot)>10$ and $0.05<z<0.18$.}
    \label{fig:xcorr:footprint:M10}
\end{figure}

\begin{figure}
    \centering
    \includegraphics[width=0.9\columnwidth]{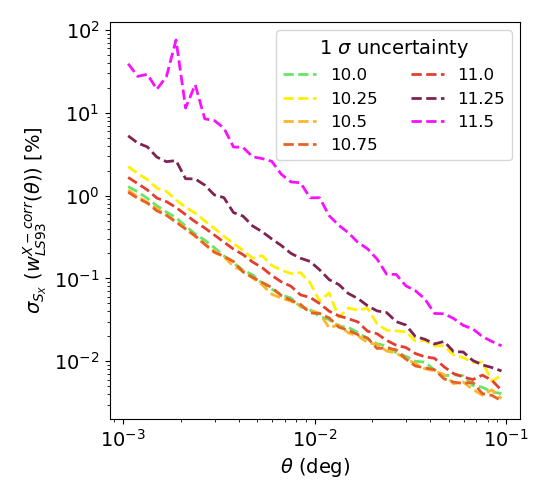}
    \includegraphics[width=0.9\columnwidth]{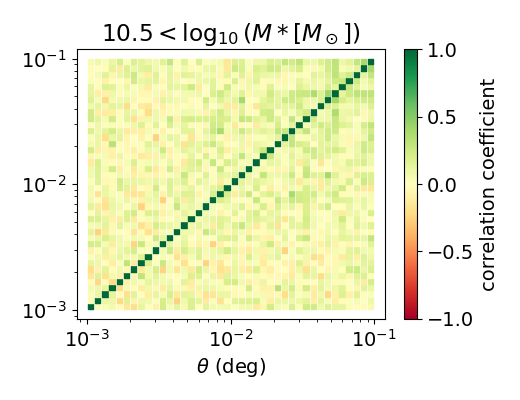}
    \caption{\textbf{Top}: Diagonal relative uncertainties on the cross-correlation. \textbf{Bottom}: Correlation coefficient for the sample with $\log_{10}(M^*/M_\odot)>10.5$.}
    \label{fig:xcorr:gal:evt:raw:uncertainties}
\end{figure}

\section{Selection of SF and QU}
\label{sec:app:SF:QU:selection}

In the Lephare runs made for this analysis, the SFR values inferred are (very) noisy, with only three or four broad bands.
They are not reliable enough to measure sSFR and split the sample into a set of quiescent and star-forming galaxies. 
Instead, we use a red-sequence model and corresponding colors \citep{IderChithamComparatFinoguenov_2020MNRAS.499.4768I,KlugeComparatLiu_2024A&A...688A.210K}. The typical scatter in the red sequence in the LS10 is 0.1 mag.

We define two boundaries in color $c_{RS}$ ($c_{BC}$) to select select red sequence (blue cloud) galaxies as a function of redshift : $g_{AB}-z_{AB}>RS(z)-c_{RS}$ ($g_{AB}-z_{AB}<RS(z)-c_{BC}$). 
In between, in the green valley, we have a mix of the two populations.  

We use the GAMA data, split into SF and QU with a log10 sSFR cut at -11 to optimize the boundary (best values for $c_{RS}$, $c_{BC}$) and to estimate completeness, purity and contamination of the obtained samples. 
In the GAMA-DR10 matched file, there are 46815 QU and 103160 SF galaxies with stellar mass larger than 9. Table \ref{tab:RS:thresholds} shows how the completeness and purity evolve as a function of the threshold value. 

The purity of the obtained RS (BC) sample varies between 59\% (68) and 69\% (97) and the completeness between 20\% (67) and 90\% (97). The trade offs to choose the best parameter to cut at is not obvious. For the RS selection the purity is never higher than 70\%. A choice could be to cut at 0.1 to obtain a completeness of 68\% and a similar purity, leaving less of the quarter of the selected objects to be SF contaminants. This is already a substantial number that will need to be modeled. 

For the BC selection with a $c_{BC}=0.34$ (0.22, 0.09) parameter, setting a similar completeness at 70.7\% (80.3, 90.2) gives a purity of 96.7\% (94.1, 84.5).
A ~5\% purity is perfectly acceptable, so we could strive for higher completeness and choose 0.22.

We chose $c_{RS}=0.1$ and $c_{BC}=0.22$ to split the sample into RS, BC and GV galaxies. This choice will depend on the application. There is a trend seen for the completeness (and to a lesser extent on the contamination) as a function of stellar mass that depends on the threshold chosen.

\begin{table}[]
    \centering
    \caption{Red sequence g-z color as a function of redshift.}
    \label{tab:rs:model}
    \begin{tabular}{cc}
    \hline \hline
redshift & g-z color DECam \\ \hline
0.05 & 1.42 \\
0.1 & 1.59 \\
0.15 & 1.76 \\
0.2 & 1.95 \\
0.25 & 2.15 \\
0.3 & 2.34 \\
0.35 & 2.54 \\
0.4 & 2.68 \\
0.23 & 2.83 \\
0.5 & 2.94 \\
0.55 & 3.06 \\
0.6 & 3.2 \\
\hline
    \end{tabular}
\end{table}

\begin{table*}[]
    \centering
        \caption{Red sequence selection parameter, estimated completeness, and purity.}

    \begin{tabular}{|c| c c c| c c c |}
\hline
 & \multicolumn{3}{|c|}{RS selection $g_{AB}-z_{AB}>RS(z)-c_{RS}$}&  \multicolumn{3}{|c|}{BC selection $g_{AB}-z_{AB}<RS(z)-c_{BC}$} \\
$c_{RS}$ & N$_{RS}$ ( QU / SF )  & completeness          & purity           & N$_{BC}$ ( QU / SF )  & completeness          & purity  \\    
$c_{BC}$ & selected       & QU/46815 ($\Delta$)   & QU/N$_{RS}$ ($\Delta$) & selected       & SF/103160 ($\Delta$)   & SF/N$_{BC}$ ($\Delta$)  \\
\hline
-0.09 & 7655 ( 4310 / 2671 ) & 9.2 ( 1.2 ) & 56.3 ( 1.4 ) & 147183 ( 42491 / 100448 ) & 97.4 ( -0.1 ) & 68.2 ( 0.3 ) \\

-0.08 & 8610 ( 4976 / 2846 ) & 10.6 ( 1.6 ) & 57.8 ( 1.5 ) & 146228 ( 41825 / 100273 ) & 97.2 ( -0.2 ) & 68.6 ( 0.4 ) \\

-0.07 & 9687 ( 5740 / 3045 ) & 12.3 ( 1.8 ) & 59.3 ( 1.4 ) & 145151 ( 41061 / 100074 ) & 97.0 ( -0.2 ) & 68.9 ( 0.4 ) \\

-0.06 & 10945 ( 6623 / 3266 ) & 14.1 ( 2.0 ) & 60.5 ( 1.1 ) & 143893 ( 40178 / 99853 ) & 96.8 ( -0.2 ) & 69.4 ( 0.5 ) \\

-0.05 & 12376 ( 7609 / 3540 ) & 16.3 ( 2.4 ) & 61.5 ( 1.2 ) & 142462 ( 39192 / 99579 ) & 96.5 ( -0.2 ) & 69.9 ( 0.5 ) \\

-0.04 & 14018 ( 8813 / 3797 ) & 18.8 ( 2.7 ) & 62.9 ( 1.4 ) & 140820 ( 37988 / 99322 ) & 96.3 ( -0.2 ) & 70.5 ( 0.7 ) \\

-0.03 & 15849 ( 10172 / 4081 ) & 21.7 ( 3.0 ) & 64.2 ( 1.2 ) & 138989 ( 36629 / 99038 ) & 96.0 ( -0.3 ) & 71.3 ( 0.8 ) \\

-0.02 & 17793 ( 11599 / 4382 ) & 24.8 ( 3.2 ) & 65.2 ( 0.9 ) & 137045 ( 35202 / 98737 ) & 95.7 ( -0.3 ) & 72.0 ( 0.8 ) \\

-0.01 & 19973 ( 13199 / 4746 ) & 28.2 ( 3.4 ) & 66.1 ( 0.9 ) & 134865 ( 33602 / 98373 ) & 95.4 ( -0.4 ) & 72.9 ( 1.0 ) \\

-0.0 & 22181 ( 14837 / 5102 ) & 31.7 ( 3.6 ) & 66.9 ( 0.7 ) & 132657 ( 31964 / 98017 ) & 95.0 ( -0.4 ) & 73.9 ( 1.0 ) \\

0.01 & 24596 ( 16604 / 5515 ) & 35.5 ( 3.9 ) & 67.5 ( 0.6 ) & 130242 ( 30197 / 97604 ) & 94.6 ( -0.4 ) & 74.9 ( 1.1 ) \\

0.02 & 27116 ( 18487 / 5961 ) & 39.5 ( 3.9 ) & 68.2 ( 0.5 ) & 127722 ( 28314 / 97158 ) & 94.2 ( -0.4 ) & 76.1 ( 1.1 ) \\

0.03 & 29642 ( 20284 / 6460 ) & 43.3 ( 3.9 ) & 68.4 ( 0.2 ) & 125196 ( 26517 / 96659 ) & 93.7 ( -0.5 ) & 77.2 ( 1.2 ) \\

0.04 & 32301 ( 22198 / 7004 ) & 47.4 ( 4.1 ) & 68.7 ( 0.2 ) & 122537 ( 24603 / 96115 ) & 93.2 ( -0.6 ) & 78.4 ( 1.2 ) \\

0.05 & 34908 ( 24068 / 7576 ) & 51.4 ( 3.9 ) & 68.9 ( 0.1 ) & 119930 ( 22733 / 95543 ) & 92.6 ( -0.6 ) & 79.7 ( 1.2 ) \\

0.06 & 37482 ( 25879 / 8172 ) & 55.3 ( 3.7 ) & \textbf{69.0} ( 0.0 ) & 117356 ( 20922 / 94947 ) & 92.0 ( -0.6 ) & 80.9 ( 1.2 ) \\

0.07 & 39908 ( 27505 / 8802 ) & 58.8 ( 3.5 ) & 68.9 ( -0.1 ) & 114930 ( 19296 / 94317 ) & 91.4 ( -0.6 ) & 82.1 ( 1.2 ) \\

0.08 & 42326 ( 29122 / 9445 ) & 62.2 ( 3.5 ) & 68.8 ( -0.1 ) & 112512 ( 17679 / 93674 ) & 90.8 ( -0.6 ) & 83.3 ( 1.2 ) \\

0.09 & 44766 ( 30739 / 10119 ) & 65.7 ( 3.2 ) & 68.7 ( -0.2 ) & 110072 ( 16062 / 93000 ) & \textbf{90.2} ( -0.6 ) & 84.5 ( 1.1 ) \\

0.1 & 47001 ( 32135 / 10842 ) & \textbf{68.6} ( 2.9 ) & 68.4 ( -0.3 ) & 107837 ( 14666 / 92277 ) & 89.5 ( -0.8 ) & 85.6 ( 1.0 ) \\

0.11 & 49178 ( 33470 / 11591 ) & 71.5 ( 2.8 ) & 68.1 ( -0.4 ) & 105660 ( 13331 / 91528 ) & 88.7 ( -0.8 ) & 86.6 ( 1.0 ) \\

0.12 & 51226 ( 34693 / 12312 ) & 74.1 ( 2.5 ) & 67.7 ( -0.4 ) & 103612 ( 12108 / 90807 ) & 88.0 ( -0.7 ) & 87.6 ( 1.0 ) \\

0.13 & 53194 ( 35806 / 13082 ) & 76.5 ( 2.2 ) & 67.3 ( -0.4 ) & 101644 ( 10995 / 90037 ) & 87.3 ( -0.8 ) & 88.6 ( 0.9 ) \\

0.14 & 54999 ( 36781 / 13836 ) & 78.6 ( 2.0 ) & 66.9 ( -0.4 ) & 99839 ( 10020 / 89283 ) & 86.5 ( -0.8 ) & 89.4 ( 0.9 ) \\

0.15 & 56843 ( 37737 / 14654 ) & \textbf{80.6} ( 1.9 ) & 66.4 ( -0.5 ) & 97995 ( 9064 / 88465 ) & 85.8 ( -0.8 ) & \textbf{90.3} ( 0.8 ) \\

0.16 & 58472 ( 38557 / 15423 ) & 82.4 ( 1.7 ) & 65.9 ( -0.5 ) & 96366 ( 8244 / 87696 ) & 85.0 ( -0.8 ) & 91.0 ( 0.6 ) \\

0.17 & 59984 ( 39263 / 16186 ) & 83.9 ( 1.4 ) & 65.5 ( -0.5 ) & 94854 ( 7538 / 86933 ) & 84.3 ( -0.8 ) & 91.6 ( 0.6 ) \\

0.18 & 61456 ( 39902 / 16984 ) & 85.2 ( 1.2 ) & 64.9 ( -0.5 ) & 93382 ( 6899 / 86135 ) & 83.5 ( -0.8 ) & 92.2 ( 0.6 ) \\

0.19 & 62827 ( 40444 / 17767 ) & 86.4 ( 1.1 ) & 64.4 ( -0.6 ) & 92011 ( 6357 / 85352 ) & 82.7 ( -0.8 ) & 92.8 ( 0.5 ) \\

0.2 & 64218 ( 40971 / 18594 ) & 87.5 ( 1.1 ) & 63.8 ( -0.6 ) & 90620 ( 5830 / 84525 ) & 81.9 ( -0.8 ) & 93.3 ( 0.5 ) \\

0.21 & 65531 ( 41455 / 19400 ) & 88.6 ( 1.0 ) & 63.3 ( -0.5 ) & 89307 ( 5346 / 83719 ) & 81.2 ( -0.8 ) & 93.7 ( 0.4 ) \\

0.22 & 66789 ( 41852 / 20236 ) & 89.4 ( 0.8 ) & 62.7 ( -0.6 ) & 88049 ( 4949 / 82883 ) & \textbf{80.3} ( -0.9 ) & 94.1 ( 0.4 ) \\

0.23 & 68002 ( 42202 / 21079 ) & \textbf{90.1} ( 0.8 ) & 62.1 ( -0.6 ) & 86836 ( 4599 / 82040 ) & 79.5 ( -0.8 ) & 94.5 ( 0.4 ) \\

0.24 & 69183 ( 42533 / 21918 ) & 90.9 ( 0.7 ) & 61.5 ( -0.6 ) & 85655 ( 4268 / 81201 ) & 78.7 ( -0.8 ) & 94.8 ( 0.3 ) \\

0.25 & 70284 ( 42827 / 22718 ) & 91.5 ( 0.6 ) & 60.9 ( -0.6 ) & 84554 ( 3974 / 80401 ) & 77.9 ( -0.8 ) & 95.1 ( 0.3 ) \\

0.26 & 71379 ( 43120 / 23501 ) & 92.1 ( 0.5 ) & 60.4 ( -0.6 ) & 83459 ( 3681 / 79618 ) & 77.2 ( -0.8 ) & 95.4 ( 0.2 ) \\

0.27 & 72449 ( 43355 / 24325 ) & 92.6 ( 0.5 ) & 59.8 ( -0.6 ) & 82389 ( 3446 / 78794 ) & 76.4 ( -0.8 ) & 95.6 ( 0.2 ) \\

0.28 & 73513 ( 43576 / 25160 ) & 93.1 ( 0.4 ) & 59.3 ( -0.5 ) & 81325 ( 3225 / 77959 ) & 75.6 ( -0.8 ) & 95.9 ( 0.2 ) \\

0.29 & 74513 ( 43743 / 25987 ) & 93.4 ( 0.4 ) & 58.7 ( -0.5 ) & 80325 ( 3058 / 77132 ) & 74.8 ( -0.8 ) & 96.0 ( 0.1 ) \\

0.3 & 75491 ( 43908 / 26795 ) & 93.8 ( 0.4 ) & 58.2 ( -0.6 ) & 79347 ( 2893 / 76324 ) & 74.0 ( -0.8 ) & 96.2 ( 0.2 ) \\

0.31 & 76476 ( 44078 / 27606 ) & 94.2 ( 0.4 ) & 57.6 ( -0.6 ) & 78362 ( 2723 / 75513 ) & 73.2 ( -0.8 ) & 96.4 ( 0.1 ) \\

0.32 & 77476 ( 44234 / 28448 ) & 94.5 ( 0.3 ) & 57.1 ( -0.6 ) & 77362 ( 2567 / 74671 ) & 72.4 ( -0.8 ) & 96.5 ( 0.1 ) \\

0.33 & 78446 ( 44359 / 29287 ) & 94.8 ( 0.2 ) & 56.5 ( -0.6 ) & 76392 ( 2442 / 73832 ) & 71.6 ( -0.9 ) & 96.6 ( 0.1 ) \\

0.34 & 79438 ( 44464 / 30170 ) & 95.0 ( 0.2 ) & 56.0 ( -0.6 ) & 75400 ( 2337 / 72949 ) & \textbf{70.7} ( -0.8 ) & 96.7 ( 0.1 ) \\

0.35 & 80390 ( 44560 / 31024 ) & 95.2 ( 0.2 ) & 55.4 ( -0.6 ) & 74448 ( 2241 / 72095 ) & 69.9 ( -0.8 ) & 96.8 ( 0.1 ) \\

0.36 & 81348 ( 44675 / 31866 ) & 95.4 ( 0.2 ) & 54.9 ( -0.5 ) & 73490 ( 2126 / 71253 ) & 69.1 ( -0.8 ) & 97.0 ( 0.1 ) \\

0.37 & 82255 ( 44764 / 32683 ) & 95.6 ( 0.2 ) & 54.4 ( -0.5 ) & 72583 ( 2037 / 70436 ) & 68.3 ( -0.8 ) & 97.0 ( 0.0 ) \\

0.38 & 83224 ( 44847 / 33567 ) & 95.8 ( 0.2 ) & 53.9 ( -0.5 ) & 71614 ( 1954 / 69552 ) & 67.4 ( -0.9 ) & 97.1 ( 0.1 ) \\

\hline
    \end{tabular}
    \label{tab:RS:thresholds}
\end{table*}

\end{document}